\documentclass[12pt]{article}         %
\usepackage{times}                               %
\usepackage{epsfig}
\usepackage{chicagor}
\usepackage{amssymb}

\newtheorem{THEOREM}{Theorem}[section]
\newenvironment{theorem}{\begin{THEOREM} \hspace{-.85em} {\bf :} }%
                        {\end{THEOREM}}
\newtheorem{LEMMA}[THEOREM]{Lemma}
\newenvironment{lemma}{\begin{LEMMA} \hspace{-.85em} {\bf :} }%
                      {\end{LEMMA}}
\newtheorem{COROLLARY}[THEOREM]{Corollary}
\newenvironment{corollary}{\begin{COROLLARY} \hspace{-.85em} {\bf :} }%
                          {\end{COROLLARY}}
\newtheorem{PROPOSITION}[THEOREM]{Proposition}
\newenvironment{proposition}{\begin{PROPOSITION} \hspace{-.85em} {\bf :} }%
                            {\end{PROPOSITION}}
\newtheorem{DEFINITION}[THEOREM]{Definition}
\newenvironment{definition}{\begin{DEFINITION} \hspace{-.85em} {\bf :} \rm}%
                            {\end{DEFINITION}}
\newtheorem{CLAIM}[THEOREM]{Claim}
\newenvironment{claim}{\begin{CLAIM} \hspace{-.85em} {\bf :} \rm}%
                            {\end{CLAIM}}
\newtheorem{EXAMPLE}[THEOREM]{Example}
\newenvironment{example}{\begin{EXAMPLE} \hspace{-.85em} {\bf :} \rm}%
                            {\end{EXAMPLE}}
\newtheorem{REMARK}[THEOREM]{Remark}
\newenvironment{remark}{\begin{REMARK} \hspace{-.85em} {\bf :} \rm}%
                            {\end{REMARK}}

\newcommand{\thm}{\begin{theorem}}
\newcommand{\lem}{\begin{lemma}}
\newcommand{\pro}{\begin{proposition}}
\newcommand{\dfn}{\begin{definition}}
\newcommand{\rem}{\begin{remark}}
\newcommand{\xam}{\begin{example}}
\newcommand{\cor}{\begin{corollary}}
\newcommand{\prf}{\noindent{\bf Proof:} }
\newcommand{\ethm}{\end{theorem}}
\newcommand{\elem}{\end{lemma}}
\newcommand{\epro}{\end{proposition}}
\newcommand{\edfn}{\bbox\end{definition}}
\newcommand{\erem}{\bbox\end{remark}}
\newcommand{\exam}{\bbox\end{example}}
\newcommand{\ecor}{\end{corollary}}
\newcommand{\eprf}{\bbox\vspace{0.1in}}
\newcommand{\beqn}{\begin{equation}}
\newcommand{\eeqn}{\end{equation}}

\newcommand{\bbox}{\vrule height7pt width4pt depth1pt}

\newcommand{\clm}{\begin{claim}}
\newcommand{\eclm}{\end{claim}}

\newcommand{\sat}{\models}

\newcommand{\union}{\cup}
\newcommand{\inter}{\cap}

\renewcommand{\phi}{\varphi}

\newcommand{\A}{{\cal A}}

\newcommand{\D}{{\cal D}}

\newcommand{\F}{{\cal F}}

\newcommand{\I}{{\cal I}}

\newcommand{\K}{{\cal K}}
\newcommand{\M}{{\cal M}}

\renewcommand{\P}{{\cal P}}
\newcommand{\Q}{{\cal Q}}
\newcommand{\R}{{\cal R}}
\newcommand{\T}{{\cal T}}

\newcommand{\W}{{\cal W}}

\newcommand{\<}{\langle}
\renewcommand{\>}{\rangle}

\newcommand{\ol}{\setlength{\itemsep}{0pt}\begin{enumerate}}
\newcommand{\eol}{\end{enumerate}\setlength{\itemsep}{-\parsep}}
\newcommand{\ul}{\setlength{\itemsep}{0pt}\begin{itemize}}
\newcommand{\dl}{\setlength{\itemsep}{0pt}\begin{description}}
\newcommand{\edl}{\end{description}\setlength{\itemsep}{-\parsep}}
\newcommand{\eul}{\end{itemize}\setlength{\itemsep}{-\parsep}}

\newcommand{\commentout}[1]{}

\newcommand{\bi}{\begin{itemize}}
\newcommand{\ei}{\end{itemize}}
\newcommand{\be}{\begin{enumerate}}
\newcommand{\ee}{\end{enumerate}}

\newenvironment{oldthm}[1]{\par\noindent{\bf Theorem #1:} \em \noindent}{\par}
\newenvironment{oldlem}[1]{\par\noindent{\bf Lemma #1:} \em \noindent}{\par}
\newenvironment{oldcor}[1]{\par\noindent{\bf Corollary #1:} \em \noindent}{\par}
\newenvironment{oldpro}[1]{\par\noindent{\bf Proposition #1:} \em \noindent}{\par}
\newcommand{\othm}[1]{\begin{oldthm}{\ref{#1}}}
\newcommand{\eothm}{\end{oldthm} \medskip}
\newcommand{\olem}[1]{\begin{oldlem}{\ref{#1}}}
\newcommand{\eolem}{\end{oldlem} \medskip}
\newcommand{\ocor}[1]{\begin{oldcor}{\ref{#1}}}
\newcommand{\eocor}{\end{oldcor} \medskip}
\newcommand{\opro}[1]{\begin{oldpro}{\ref{#1}}}
\newcommand{\eopro}{\end{oldpro} \medskip}

\newcommand{\Pltriv}{\Pl_{\mathrm{triv}}}
\newcommand{\PR}{{\cal PR}}
\newcommand{\PL}{{\cal PL}}
\newcommand{\PT}{{\cal PT}}
\newcommand{\RT}{{\cal RT}}
\newcommand{\hide}[1]{}
\newcommand{\mc}{\mathcal}
\newcommand{\mb}{\mathbf}
\newcommand{\lles}{\mbox{LLES}}
\newcommand{\interleave}{\mbox{\it interleave}}
\renewcommand{\S}{{\cal S}}
\renewcommand{\SS}{\mbox{{\it SS}}}
\renewcommand{\mid}{\, | \,}
\newcommand{\strat}{\mathit{strat}}

\newcommand{\goesto}[1]{\stackrel{#1}{\rightarrow}}

\newcommand{\Pl}{\mbox{Pl}}

\newcommand{\di}{\mbox{$\Diamond\hskip -5pt\mbox{-}\hskip 3pt$}}
\renewcommand{\L}{{\cal L}}
\newcommand{\INIT}{\mathit{INIT}}
\newcommand{\Dom}{\mathrm{Dom}}

\begin{document}

\title{Secrecy in Multiagent Systems%
\thanks{Authors supported in part by NSF under grant IRI-96-25901 and
IIS-0090145, by ONR under grants  N00014-00-1-03-41 and
N00014-01-10-511, and by the DoD Multidisciplinary University Research
Initiative (MURI) program administered by the ONR under
grant N00014-01-1-0795.
Kevin O'Neill was also supported in part by a graduate fellowship from the
National Science and Engineering Research Council of Canada.
A preliminary version of this paper appeared at the 15th IEEE Computer Security
Foundations Workshop, 2002.
}}

\author{Joseph Halpern \& Kevin O'Neill\\
Department of Computer Science\\
Cornell University\\
\{halpern,oneill\}@cs.cornell.edu\\
}

\maketitle
\thispagestyle{empty}

\begin{abstract}
We introduce a general framework for reasoning about
secrecy and privacy
requirements in multiagent systems. 
Our definitions extend earlier definitions of secrecy and
nondeducibility given by Shannon and Sutherland.
Roughly speaking,
one agent maintains secrecy with respect to another 
if the second agent cannot rule out any possibilities for the
behavior or state of the first agent. 
We show that the framework can
handle probability and nondeterminism
in a clean way, is useful for reasoning about 
asynchronous systems as well as synchronous systems, and
suggests generalizations of secrecy that may be useful for dealing with
issues such as resource-bounded reasoning. We also show that
a number of well-known attempts to characterize the absence of
information flow are
special cases of our definitions of secrecy.
\end{abstract}

\section{Introduction}\label{sec:intro}

In the past two decades there have been many attempts to define what it
means for a system to be perfectly secure, in the sense that one group of
agents is unable to deduce anything at all about the behavior of another.
More generally, many papers in computer science have, in a variety of
different settings, defined properties of ``secrecy'' or ``privacy'' and
have discussed techniques for achieving these properties. In the 
computer-security literature, early definitions of ``perfect security''
were based 
on two different intuitions. \emph{Noninterference} \cite{GM82} attempted
to capture the intuition that a classified agent is unable to interfere
with an unclassified agent, while \emph{nondeducibility}
\cite{sutherland86} attempted to capture the intuition that an unclassified
agent is unable to deduce anything about the state of a classified agent.
Others definitions have involved a notion of ``information flow'', and
taken a system to be secure if it is impossible for information to flow
from a classified user to an unclassified user. With these basic ideas in
mind, definitions of security have been provided for a wide variety of
system models, including semantic models that encode all possible
input/output behaviors of a computing system and language-based models
that deal with process algebras and with more traditional constructs such
as imperative programming languages. (Focardi and
Gorrieri~\citeyear{focardi01} provide a classification of security
properties expressed using process algebras; Sabelfeld and
Myers~\citeyear{SabelfeldMyers03} give a survey of language-based
techniques.)

Sutherland's definition of nondeducibility was based on a simple idea:
a system can be described as a set of ``worlds'' that encode the
local states of classified and unclassified users, and security is
maintained if classified and unclassified states are 
independent in the
sense that an unclassified user can never totally rule out any classified state
based on his own local state. 
As we shall see, nondeducibility
is closely related to Shannon's~\citeyear{Shannon49}
probabilistic definition of secrecy in the context of cryptography,
which
requires 
classified and unclassified events 
to be
(probabilistically) independent.
In other words, the unclassified
agent's posterior probability of a classified event should be the same as
his prior probability of that event before he began interacting with the
system.

Definitions of noninterference based on Goguen and Meseguer's early work
are quite different in flavor from the definitions of Shannon and
Sutherland.  Typically, they represent the system as a set of input/output
traces, and deem the system secure if the set of traces is closed under
operations that add or remove classified events. 
Variants of this idea
have been proposed to deal with issues such as
verification, system composition, timing attacks, and so on. 
Although these definitions have been useful for solving a variety of
technical problems, the 
complexity of some 
of
this work has, in our view, obscured the simplicity
of earlier definitions based on the notion of independence. While
nondeducibility has been criticized for its inability to deal with a
variety of security concerns,
we claim that the 
basic idea captures notions of secrecy and privacy in an elegant and
useful way.

In this paper we define secrecy
in terms of an agent's knowledge,
using
the ``runs-and-systems'' framework~\cite{FHMV}.
Our definitions can be viewed as generalizing the notion
of nondeducibility to systems in which 
agents interact with each other over time.
The runs and systems framework generalizes the standard 
input/output trace models that 
have
been used in many definitions of
noninterference. 
The
trace-based approach has been concerned primarily with the input and
output values exchanged as a user or observer interacts with the system.
Thus, with a trace-based approach, it is possible to define secrecy only
for systems 
that can be characterized by observable input and output events. This is
insufficient for modeling a variety of interesting systems. As Focardi and
Gorrieri \citeyear{focardi01} point out, for example, it is difficult to
deal with issues such as deadlock using a purely trace-based approach. It
is also difficult to represent an agent's notion of time in systems that
may exhibit differing degrees of synchrony. As we shall see, the added
generality of the runs and systems approach lets us deal with these
issues in a straightforward way.

Many frameworks for reasoning about secrecy and information flow 
have assumed, often implicitly, a very coarse notion
of uncertainty.  Either an agent knows, with certainty, that some fact is
true, or she does not; a definition of secrecy (with respect to some
agent) amounts to a characterization of which facts the agent must not
know, or which facts she must think are possible. Indeed, this is
precisely the intuition that we make precise in Section~\ref{sec:k+sec}.
In the literature, such definitions are called {\em possibilistic},
because they consider only what 
agents consider possible or
impossible. In practice, however, such a coarse-grained notion of
uncertainty is simply too 
weak; it 
is easy to concoct examples where one
agent has possibilistic secrecy, but where intuition suggests that secrecy
is not maintained. We extend our definitions of secrecy to incorporate
probability, a much more fine-grained notion of uncertainty. Just as
Shannon's definitions of secrecy can be viewed as a probabilistic
strengthening of Sutherland's definition of nondeducibility, our
definitions of probabilistic secrecy generalize the possibilistic
definitions we give. In fact, there is a sense in which they are
the same definitions, except with a different measure of uncertainty---a
point made precise when we generalize to {\em plausibilistic secrecy\/} in
Section~\ref{sec:plaus}.

Our approach has an additional advantage: it enables us to provide 
syntactic characterizations of secrecy, using 
a logic that includes modal operators for reasoning about
knowledge and probability. We discuss what it means for a
fact to ``depend on'' the state of an agent 
and show that secrecy can be characterized as
the requirement that unclassified agents never know any fact 
that depends on the state of a classified agent. (In the
probabilistic case, 
the requirement is that unclassified agents must think that any such fact
is equally likely at all points of the system.) This knowledge-based
characterization lets us make precise the connection between secrecy (of
the classified agent with respect to the unclassified agent) and the
notion of a ``secret'', i.e., a fact about the system that an agent is not
allowed to know. 
This syntactic approach also opens the door to natural generalizations
of information-flow properties
that require secrecy for only some facts, as well as allowing us to
consider notions of secrecy based on more computational notions of
knowledge, which may be more appropriate for resource-bounded agents.

As we show in Section~\ref{sec:rw}, our approach provides insight into a
number of other definitions of secrecy, privacy, and noninterference
that have been proposed in the literature.  We illustrate this point by
considering {\em separability} \cite{mclean94}, {\em generalized
noninterference}~\cite{mclean94}, {\em 
nondeducibility on strategies}~\cite{Wittbold&Johnson}, and {\em probabilistic
noninterference}~\cite{gray98}.
One of our goals in this section, obviously, is to convince the
reader that our definitions are in fact as general as we claim they are.
More importantly, we hope that providing a unified framework for
comparing definitions of secrecy will facilitate the
cross-fertilization of ideas.

The rest of the paper is organized as follows. Section~\ref{sec:review}
reviews the multiagent systems framework and the definition of knowledge
in multiagent systems.  
In Section~\ref{sec:nonprobsec} we define secrecy and relate it to
Sutherland's notion of nondeducibility. We also consider syntactic
definitions of secrecy using a logic of knowledge.
Section~\ref{sec:prob} considers probabilistic secrecy, while
Section~\ref{sec:plaus} considers plausibilistic secrecy. 
In Section~\ref{sec:rw} we compare our definitions with others that have
been given in the security literature.  We conclude in
section~\ref{sec:conclusion}.  Most proofs are deferred to the appendix.

\section{Knowledge and Multiagent Systems}\label{sec:review}

A multiagent system consists of $n$ agents, each of whom is in some
{\em local state\/} at a given point in time. We assume that an agent's
local
state encapsulates all the information to which 
she has
access. In 
a security setting the local state of an agent might
include 
initial information regarding keys, the messages she has
sent and received, and perhaps the reading of a clock. The basic
framework makes no assumptions about the precise nature of the local
state.

We can view the whole system as being in some {\em global state},
which
is a tuple consisting of the local state of each agent and the
state of the environment, where the environment consists of everything
relevant to the system that is not contained in the state of
the agents. Thus, a global state has the form $(s_e, s_1,\ldots, s_n)$,
where $s_e$ is the state of the environment and $s_i$ is agent $i$'s
state,
for $i = 1, \ldots , n$. 

A {\em run\/} is a function from time to global states. Intuitively,
a run is a complete description of what happens over time in one
possible execution of the system. A {\em point\/} is a pair $(r, m)$
consisting
of a run $r$ and a time $m$. For simplicity, we take time to range over
the natural numbers.
At a point $(r, m)$, the system is in some global state $r(m)$. If $r(m)
= (s_e, s_1,
\ldots , s_n)$, then we take $r_i(m)$ to be $s_i$, agent $i$'s local
state at the point $(r, m)$.  
Formally, a {\em system\/} consists of a set of runs (or
executions).   
Let $\PT(\R)$ denote the points in a system $\R$.  

Given a system $\R$, let $\K_i(r,m)$ be the set of points in $\PT(\R)$ 
that $i$ thinks are possible at $(r,m)$, i.e.,
\[ \K_i(r,m) = \{ (r',m') \in \PT(\R) : r'_i(m') = r_i(m) \}. \]
The set $\K_i(r,m)$ is often called an {\em $i$-information set}
because, intuitively, it corresponds to the system-dependent 
information encoded in $i$'s local state at the point $(r,m)$.

A natural question to ask is where these runs come from. 
While the framework itself does not deal with this issue, 
in practice, we are interested in systems where the runs are
generated by a simple set of rules, such as a communication or security
protocol, a program written in some programming language, or a 
process described in a concurrent process language. Translating such
rules to a set of runs is not always straightforward,
but doing so is often useful inasmuch as
it forces us to think carefully about what features of
the system are essential and relevant to the safety or correctness
issues that we are interested in.
This, in turn, determines the form of the local states.

To reason formally about secrecy in multiagent
systems, 
we use a logic of
knowledge and time. Starting with a set $\Phi$ of primitive
propositions, we close off under negation, conjunction,
the modal operators $K_i$ for $i = 1, \ldots, n$, and $\di$.  
In the context of security protocols, the set $\Phi$ might consist of
primitive propositions corresponding to facts such as
``the key is $n$'' or ``agent $A$ sent the message $m$ to $B$''. 
As usual, $K_i \phi$ means that  agent $i$ knows $\phi$;
$K_i \phi$
at a point $(r,m)$ if $\phi$ is true at all points in $\K_i(r,m)$.
Finally, $\di \phi$ is true at a
point $(r,m)$ if $\phi$ is true at some point on run $r$ (either before,
at, or after time $m$).  While it is, of course, possible to define
other temporal operators, the $\di$ operator will prove particularly
useful in our definitions.

We use the standard approach \cite{FHMV} to give semantics to this
language.  An interpreted system $\I$ consists of a pair $(\R,
\pi)$, where $\R$ is a system and $\pi$ is an interpretation for the
primitive
propositions in $\Phi$ that assigns truth values to the primitive
propositions at the global states. Thus, for every $p\in\Phi$ and global
state $s$ that arises in $\R$, we have $(\pi(s))(p)\in \{\bf true, \bf
false\}$. Of course, 
$\pi$ also induces an interpretation over the points in  $\PT(\R)$: simply
take $\pi(r, m)$ to be $\pi(r(m))$. 
We 
now define what it means for a formula $\phi$ to be true at a
point
$(r, m)$ in an interpreted system $\I$, written $(\I, r, m)\sat\phi$, by
induction on the structure of formulas:
\begin{itemize}
\item $(\I,r,m) \sat p$ iff $(\pi(r,m))(p) = {\bf true}$;
\item $(\I,r,m) \sat \phi \land \psi$ iff $(\I,r,m) \sat \phi$ and 
$(\I,r,m) \sat \psi$;
\item $(\I,r,m) \sat \neg \phi$ iff $(\I,r,m) \not\sat \phi$;
\item $(\I, r, m) \sat K_i\phi ~\mbox{iff}~ (\I , r', m') \sat\phi$ for
all $(r',m') \in \K_i(r,m)$;
\item $(\I,r,m) \sat \di \phi ~\mbox{iff there exists}~ n ~\mbox{such that}~
(\I,r,n) \sat \phi$.
\end{itemize}
As usual, we say that $\phi$ is {\em valid in $\I$\/} and write $\I \sat
\phi$ if $(\I,r,m) \sat \phi$ for all points 
$(r,m)$ in $\I$; 
similarly, $\phi$ is {\em satisfiable in $\I$\/} if $(\I,r,m) \sat \phi$
for some point $(r,m)$ in $\I$.
We abbreviate $\neg K_i \neg \phi$ as $P_i \phi$.  We read $P_i \phi$ as
``(according to agent $i$) $\phi$ is possible''.  Note that 
$(\I,r,m) \sat P_i \phi$ if there exists 
a
point $(r',m') \in 
\K_i(r,m)$ such that $(\I,r',m') \sat \phi$.

The systems framework lets us express in a natural way some standard
assumptions about systems.  For example, 
we can reason about {\em synchronous} systems, where agents always know
the time. 
Formally, $\R$ is
synchronous if, for all agents $i$ and points $(r,m)$ and $(r',m')$, if
$r_i(m) = r'_i(m')$, then $m = m'$.%

Another standard assumption (implicitly made in almost all
systems models
considered in the security literature) is that agents have {\em perfect
recall}.  Roughly speaking, an agent with perfect recall 
can reconstruct his complete local history.
In synchronous systems, 
for example,
an agent's local state
changes with every tick of the external clock, so agent~$i$'s having
perfect recall implies that the sequence
$\langle r_i(0),\ldots,r_i(m)\rangle$ must be encoded in~$r_i(m+1)$.
To formalize this intuition, let
{\em agent~$i$'s local-state sequence at the point $(r,m)$\/}
be the sequence of local states she has gone through in run~$r$ up
to time~$m$, without consecutive repetitions.  Thus, if from time 0
through time~4 in run~$r$ agent~$i$
has gone through the sequence $\< s_i,s_i,s_i',s_i,s_i \>$ of
local states,
where $s_i \ne s_i'$,
then her local-state sequence at $(r,4)$ is $\< s_i,s_i',s_i \>$.
Intuitively, an agent has perfect recall if her current local
state encodes her local-state sequence.
More formally, we say that
{\em agent~$i$ has perfect recall
in system~$\R$\/}\index{perfect recall}
if, at all points $(r,m)$ and $(r',m')$ in~$\PT(\R)$,
if $(r',m') \in \K_i(r,m)$, then
agent~$i$ has the same local-state sequence at both $(r,m)$ and
$(r',m')$.  Thus,
agent~$i$ has perfect recall if she ``remembers'' her local-state
sequence at all times.  
It is easy to check that perfect recall has the following key property:
if $(r',m_1') \in \K_i(r,m_1)$, then 
for all $m_2 \le m_1$, there exists 
$m_2' \le m_1'$ such that 
$(r',m_2') \in \K_i(r,m_2)$.
(See \cite{FHMV} for more discussion of this definition.)

\section{Secrecy in Nonprobabilistic Systems}\label{sec:nonprobsec}

\subsection{Defining Secrecy}\label{sec:secrecy}

In this section, we give abstract definitions of secrecy and 
motivate these definitions using the runs
and systems model. 
Roughly speaking, we define secrecy so as 
to ensure that low-level agents do not know anything about the
state of high-level agents. 
In Section~\ref{sec:k+sec}, we formalize these intuitions using 
the epistemic logic of Section~\ref{sec:review}.

The strongest notion of secrecy that we consider 
in this section
is the requirement
that a low-level agent, based on her local state, should never be able
to determine anything about the local state of the high-level agent.
More specifically, the low-level agent should never be able to rule
out any possible high-level state.
In terms of knowledge, this means that the low-level agent must
never know that some high-level state is incompatible with her
current low-level state.
To ensure
that the low-level agent $L$ is not able to rule out any possible 
high-level states, we insist that 
every low possible low-level state is compatible with every possible
high-level state.

\dfn\label{dfn:totsecrecy}
Agent $j$ maintains {\em total secrecy\/} with respect to $i$ in system
$\R$ if, for
all points $(r,m)$ and $(r',m')$ in 
$\PT(\R)$, 
$\K_i(r,m) \inter \K_j(r',m') \ne \emptyset$.
\edfn
Note that if we take $i$ to be the low-agent $L$ and $j$ to be the
high-level
agent $H$, then 
this definition just formalizes the informal definition given above. At the
point $(r,m)$, $L$ cannot rule out any possible local state of $H$.

Total secrecy is a strong property. 
For almost any imaginable system, it is, in fact, too strong to be useful.
There are two important respects in which it is too strong.
The first respect has to do with the fact that total secrecy protects
{\em everything\/} about the state of the high-level agent.
In some systems, we might want only some part of the high-level
agent's state to be kept secret from the low-level agent. For example,
we might want the high-level agent to be able to see the state of the
low-level agent, in which case our definitions are too strong because
they rule out any 
correlation
between the states of the high-level and
low-level agents. We can correct this situation by 
extracting from $H$'s state the information that is relevant and
ensuring
that the relevant part of $H$'s state is kept secret.

\dfn
A {\em $j$-information function\/} on $\R$ is a function $f$
from $\PT(\R)$ to some range that depends only on $j$'s local state; that
is $f(r,m) = f(r',m')$ if $r_j(m) = r'_j(m')$.
\edfn
Thus, for example, if $j$'s local state 
at any point $(r,m)$ includes both its input and
output, $f(r,m)$ could be just the output component of $j$'s local state.

\dfn If $f$ is a $j$-information function, agent
$j$ maintains {\em total $f$-secrecy\/} with respect to $i$ in system
$\R$ if, 
for all  points $(r,m)$ and values $v$ in the range of $f$, 
$\K_i(r,m) \inter f^{-1}(v) \ne \emptyset$
(where $f^{-1}(v)$ is simply the preimage of $v$,
that is, all points $(r,m)$ such that $f(r,m) = v$).
\edfn
Of course, if $f(r,m) = r_j(m)$, then $f^{-1}(r'_j(m')) = \K_j(r',m')$,
so total secrecy is a special case of total $f$-secrecy.

Total $f$-secrecy is a special case of {\em nondeducibility}, introduced
by Sutherland~\citeyear{sutherland86}. 
Sutherland considers ``abstract'' systems that are characterized by a
set $W$ of worlds.  He focuses on two agents, whose views are
represented by information functions $g$ and $h$ on $W$.
Sutherland says that {\em no information flows from $g$ to $h$\/}
if, for all worlds $w, w' \in W$, there exists
some world $w'' \in W$ such that $g(w'') = g(w)$ and $h(w'') = h(w')$.
This notion is often called 
{\em nondeducibility (with respect to $g$ and $h$)} in the literature.
To see how 
total $f$-secrecy
is a special case of 
nondeducibility,
let $W = \PT(\R)$, the set of all points of the system. Given a point 
$(r,m)$, let $g(r,m) = r_i(m)$.
Then total $f$-secrecy is equivalent to nondeducibility 
with respect to $g$ and $f$.

Note that nondeducibility is symmetric:~no information flows from
$g$ to $h$ iff no information flows from $h$ to $g$.
Since most standard noninterference properties focus only on
protecting the state of some high-level agent, symmetry appears
to suggest that if the actions of a high-level
agent are kept secret from a low-level agent, then the actions of a
low-level agent must also be kept secret from the high-level agent.
Our definitions help 
to
clarify this issue.
Total secrecy as we have defined it is indeed symmetric:~$j$
maintains total secrecy with respect to $i$ iff $i$ maintains total
secrecy with respect to $j$. However,
total $f$-secrecy is not symmetric  in general.
If $j$ maintains total $f$-secrecy with respect to $i$, it may
not even make sense to talk about $i$ maintaining total $f$-secrecy with
respect to $j$, because $f$ may not be an $i$-information function.
Thus, although $f$-secrecy is an instantiation of nondeducibility (with
respect to an appropriate $g$ and $h$),
the symmetry at the level of $g$ and $h$ does not translate
to symmetry 
at the level of $f$-secrecy, which is where it matters.
While $f$-secrecy is useful conceptually, it is essentially a trivial
technical generalization of the basic notion of secrecy, because for any agent
$j$ and $j$-information function $f$, we can reason about a new agent
$j_f$ whose local state at any point $(r,m)$ is $r_{j_f}(m) = f(r_j,m)$.
Therefore, every theorem we prove involving secrecy
holds for $f$-secrecy as well.
For this reason, and to simplify the definitions 
given
in the remainder
of the paper, we ignore information functions, and deal only with
secrecy of one agent with respect to another.
We remark that this ability to ``create'' new agents, by identifying an
agent with a function on global states, turns out to be quite useful,
since our definitions hold without change for any agent created this way.

The second respect in which total secrecy is too strong involves {\em time}.
To understand the issue,
consider synchronous systems (as defined in Section~\ref{sec:review}). 
In such systems, the 
low-level agent knows 
the
time and knows that the high-level agent
knows it too.  Thus, the low-level agent can rule out all
high-level states except those
that occur at the current time. Even in semisynchronous
systems, where agents know the time to within some tolerance $\epsilon$,
total secrecy is impossible, because low-level agents can rule out
high-level states that occur only in the distant past or future.

We now present two ways of resolving this problem. The first way
weakens total secrecy by considering 
runs, 
instead of points.
Total secrecy (of $j$ with respect to $i$) says that at 
at all times, agent $i$ must consider all state of $j$ to be
(currently) possible.
A weaker version of total secrecy says that 
at all times, $i$ must consider it possible that every possible state
of $j$ either occurs at that time, or at some point in the past or future.
We formalize this in the following definition. 
Given a set $U$ of points, let $\R(U)$
consist of the runs in $\R$ going through a point in $U$.  That is, 
$\R(U) = \{r \in \R: (r,m) \in U \mbox{ for some $m$}\}.$

\dfn\label{dfn:weaktotalsec}
Agent $j$ maintains {\em run-based secrecy} with respect to $j$ in
system $\R$ if, for all points $(r,m)$ and $(r',m')$ in $\PT(\R)$,
$\R(\K_i(r,m)) \inter \R(\K_j(r',m')) \ne \emptyset$.
\edfn

It is easy to check that $j$ maintains run-based secrecy with respect to $j$ in
system $\R$ iff for all points $(r,m)$ and $(r',m')$ in $\PT(\R)$,
there exists a run $r''$ and times $n$ and $n'$
such that $r_i''(n) = r_i(m)$ and $r_j''(n') = r_j'(m')$.
To relate the formal definition to 
its informal motivation,
note that every state of $j$ that occurs in the system has the form
$r'_j(m')$ for 
some point $(r',m')$. Suppose that $i$'s state is $r_i(m)$. 
If there exists a point $(r'',n'')$ such that
$r''_i(n'') = r_i(m)$ and $r''_j(n'') = r_j'(m')$, agent $i$ considers it
possible that $j$ currently has state $r_j'(m')$.  If instead $r''_j(n)
= r_j'(m')$ for $n < n''$, then $i$ currently considers it possible that
$j$ was in state $r'_j(m')$ at some point in the past; similarly, if 
$n > n''$,
then $i$
thinks that $j$  
could
be in state $r'_j(m')$ at some point in the future.
Note that total secrecy implies run-based secrecy, 
but the converse is not necessarily true (as shown in
Example \ref{xam:no_ind}).
While run-based secrecy is still a very strong security property, it
seems much more reasonable than total secrecy. 

\commentout{
The second respect in which total secrecy is too strong is not captured
by Sutherland's definition in any obvious way.  To understand the issue,
consider synchronous systems (as defined in Section~\ref{sec:review}). 
In such systems, the low-level
agent can always be certain that the high-level agent knows what time it
is, and can thus rule out {\em all} high-level states except those
that occur at the current time. Even in semisynchronous
systems, where agents know the time to within some tolerance $\epsilon$,
total secrecy is impossible, because low-level agents can rule out
high-level states that occur only in the distant past or future.
}
The second way to 
weaken
total secrecy is to relax the requirement that
the low-level agent cannot rule out any possible high-level states.
We make this formal as follows.  
\dfn
An {\em $i$-allowability function\/} on $\R$ is a function $C$ 
from  $\PT(\R)$ to subsets of $\PT(\R)$  such that 
$\K_i(r,m) \subseteq C(r,m)$
for all $(r,m) \in \PT(\R)$.
\edfn
Intuitively,
$\PT(\R) - C(r,m)$ is the set of points that $i$ is allowed to ``rule
out''
at the point $(r,m)$.  It seems reasonable 
to insist that the points that $i$ considers
possible at $(r,m)$ 
not be ruled out,
which is why we require that $\K_i(r,m) \subseteq C(r,m)$.

\dfn
If $C$ is an $i$-allowability function,
then $j$ {\em maintains $C$-secrecy with respect to $i$\/} if,
for all points $(r,m) \in \PT(\R)$ 
and $(r',m') \in C(r,m)$,
we have
$\K_i(r,m) \inter \K_j(r',m') \ne \emptyset$.
\edfn
If $C(r,m) = \PT(\R)$ for all points $(r,m) \in \PT(\R)$, 
then $C$-secrecy reduces to total secrecy.  
In general, allowability functions give a generalization of secrecy that is
orthogonal to information functions.  

Synchrony can be captured by the allowability function
$S(r,m) = \{ (r',m) : r' \in \R\}$.
Informally, this says that agent $i$ is allowed to know what time it is.
We sometimes call $S$-secrecy {\em synchronous secrecy}.  
In synchronous systems, synchronous secrecy has a simple characterization.
\pro\label{pro:S-secrecy}
Agent $j$ maintains synchronous secrecy with
respect to $i$ in a synchronous system $\R$ iff, for all runs $r, r' \in
\R$ and times $m$, 
we have that 
$\K_i(r,m) \inter \K_j(r',m) \ne
\emptyset$.  
\epro

\prf
This follows trivially from the definitions.
\eprf

In synchronous input/output trace systems,
synchronous secrecy is essentially equivalent to the standard notion of 
{\em separability}~\cite{mclean94}.
(Total secrecy 
can be viewed as an
asynchronous version of separability.
See Section~\ref{sec:sts}
for
further discussion of this issue.)
The security literature has typically
focused on either synchronous systems or completely asynchronous systems.
One advantage of our framework is that we
can easily model both of these extreme cases, as well as being able to
handle in-between cases,
which do not seem to have been considered 
up to now.
Consider a semisynchronous system where agents know the
time to within a tolerance of $\epsilon$.  
At time 5, for example, an agent
knows that the true time is in the interval
$[5 - \epsilon, 5 + \epsilon]$.  This corresponds to the allowability function
$\SS(r,m) = \{(r',m') : |m-m'| \le \epsilon\}$, for the appropriate
$\epsilon$. 
We believe that any attempt to define security for semisynchronous
systems will require something like 
allowability functions. 
\commentout{
Because allowability functions are so general, secrecy with respect to an
allowability function is not generally a symmetric condition. However,
two allowability functions $C_i, C_j$ may themselves
symmetric, so that $(r,m) \in C_j(r',m')$ iff $(r',m') \in C_i(r,m)$.
In this case, it is easy to check that $j$ maintains $C_i$-secrecy with 
respect to $i$ if and only $i$ maintains $C_j$-secrecy with respect to $j$.
Note that the synchronous and semisynchronous allowability functions that
we just defined are both symmetric.
}
The notions of run-based secrecy and 
$C$-secrecy are
distinct, in the sense that there are systems where run-based secrecy
holds 
and $C$-secrecy does not, and other
systems where $C$-secrecy holds but run-based secrecy does not.
If agents do not have perfect recall, we may have synchronous
secrecy without having run-based secrecy,
and if the system is asynchronous, we may have run-based secrecy
without having synchronous secrecy. 
(See Appendix \ref{app:examples} for examples.)
On the other hand, there are contexts in which
the two approaches capture the same intuitions.  Consider 
our definition of synchronous secrecy.
While synchronous secrecy may seem like a reasonable condition, 
intuition might at first seem to suggest that
it goes too far in weakening total secrecy.
Informally, $j$ maintains {\em total} secrecy with respect to $i$ if $i$ never
learns anything not only about $j$'s current 
state, but also his possible future and future states. Synchronous
secrecy seems to say only that $i$ never learns anything about $j$'s 
state {\em at the current time}.  
However, when agents have perfect recall,
it turns out that synchronous secrecy implies run-based secrecy, thus
addressing this concern. 

To make this precise
for a more general class of allowability functions,
we need the following
definition,
which captures the intuition that an allowability function depends only
on timing.
Given any two runs, we want the allowability function to map points
on the first run to contiguous, nonempty sets of points on the second
run in a way 
that respects the ordering of points on the first run, 
and covers all points on the second run. %

\dfn 
An allowability function $C$ {\em depends only on timing\/} if 
it satisfies the following three conditions: (a) for all runs $r, r' \in \R$,
and all times $m'$, there exists $m$ such that $(r',m') \in C(r,m)$; 
(b) if $(r',m') \in C(r,m)$, and $n \geq m$ (resp. $n \leq m$), 
there exists $n' \geq m'$ (resp. $n' \leq m'$) such that $(r',n') \in C(r,n)$;
(c) if $(r',n_1) \in C(r,m)$, $(r',n_2) \in C(r,m)$, and $n_1 \leq m' \leq n_2$,
then $(r',m') \in C(r,m)$.
\commentout{
it satisfies the following three conditions:
(a) for all points $(r,m)$ and runs
$r'$ there exists a time $m'$ such that $(r,m) \in C(r',m')$,
(b) for all points $(r,m)$ and runs $r'$ there exists a time $m'$ such
that $(r',m') \in C(r,m)$, and 
(c) if $(r',m_1') \in C(r,m_1)$, 
$(r',m_2') \in C(r,m_2)$, $m_1 \le m_2$ and $m_2' \le m_1'$, then
$C(r,m_1) = C(r,m_2)$.  
}
\edfn
It is easy to
check that both synchronous and semi-synchronous allowability functions
depend only on timing.
We now show that $C$-secrecy implies run-based secrecy if $C$ depends
only on timing.

\pro{\label{pro:timing}}
If $\R$ is a system where $i$ and $j$ have perfect recall, 
$C$ depends only on timing, and
$j$ maintains $C$-secrecy with respect to $i$, then
$j$ maintains run-based secrecy with respect to $i$.
\epro

In synchronous systems with perfect recall, 
synchronous secrecy and run-based secrecy 
agree.
This
reinforces our
claim that both definitions are natural, useful weakenings of 
total secrecy.

\pro\label{pro:sync<->weak}
If $\R$ is a synchronous system where both $i$ and $j$ have
perfect recall, then agent $j$ maintains synchronous
secrecy with respect to $i$ iff $j$ maintains 
run-based secrecy with respect to $i$.
\epro

The requirement in Proposition~\ref{pro:sync<->weak} that both agents
have perfect recall is necessary; see Example \ref{xam:need_recall} 
for details.
Without
perfect recall, two things can go wrong. 
First,
if $i$ does not have perfect
recall, she might be able to determine 
at time $n$ 
what $j$'s state is going to be 
at some future time $n' > n$, but then
forget about it by time $n'$, so that 
$j$ maintains synchronous secrecy with respect to $i$, but not 
run-based secrecy.   Second, if $j$ does not have perfect
recall, $i$ 
might learn something about $j$'s state in the past, 
but $j$ might still maintain synchronous secrecy with respect to $i$
because $j$ has forgotten this information by the time $i$ learns it.
These examples suggest that secrecy is perhaps not 
as
interesting 
when
agents can forget things that have happened in the past.
Intuitively, we should be proving secrecy under the assumption of perfect recall,
rather than trusting that agents will forget important facts whenever we want them to.
\commentout{
A final way to extend $f$-$C$-secrecy is an obvious extension to 
multiple agents.
So far we have considered only pairwise security.  It is natural to ask
what it means for $j$ to maintain secrecy with respect to some set of
agents.  There are several possible definitions.  One is simply that $j$
maintains secrecy with respect to each agent in the set (according to
one of the definitions above).  A second captures the intuition that $j$
maintains secrecy even if all the agents in the set could put their
information together.  (This corresponds to the notion of distributed
knowledge \cite{FHMV}.)  For example
we can define total secrecy with respect to a group of agents as
follows.

\dfn
Agent $j$ maintains {\em total secrecy\/} with respect to the set $A$ of
agents in system $\R$ if 
$\inter_{i \in A \union \{j\}} \K_i(r,m) \ne \emptyset$
for every point $(r,m) \in \PT(\R)$.
\edfn
There are similar variants 
for the other definitions considered in this section.

We conclude this section by examining the extent to which
$f$-$C$-secrecy can be captured in Sutherland's original framework.
The notion of total $f$-secrecy is clearly a special case of
Sutherland's notion of nondeducibility.  We suggested earlier that 
$f$-$C$-secrecy is a generalization that cannot be captured in
Sutherland's framework in any obvious way.  However, as we now show, it
can in a precise sense be captured in his framework, albeit in an
awkward way.

Given a system $\R$, an allowability function $C$, and a $j$-information
function $f$, we can
show that there exists a set of worlds $W_{\R,C}$
(depending only on $\R$ and $C$) and information functions $g$ and $h$
on $W_{\R,C}$ such that $j$ maintains $f$-$C$-secrecy in $\R$ 
iff no
information flows from $g$ to $h$ in $W_{\R,C}$.  
Actually, this result is almost trivially true.
We can take a fixed set of worlds
$W'$ and two pairs of information functions $(g,h)$ and $(g',h')$ on
$W'$, such that no information flows from $g$ to $h$ and information
does flow from $g'$ to $h'$.  We then map $(\R,C,f)$ to $(W',g,h)$ if
$j$ preserves $f$-$C$-secrecy in 
$R$; otherwise we map it to
$(W',g',h')$.
To make this statement nontrivial, we require that the mappings satisfy
a uniformity condition.  As the following result shows, this too can be done.

\pro 
\label{pro:finiteND}
Given a system $\R$, an $i$-allowability function $C$, and a
$j$-information function $f$, there exists
a set $W_{\R,C}$ of worlds and information functions $g$ and $h$ on $W_{\R,C}$  
such that $j$ maintains $f$-$C$-secrecy with respect to $i$
in $\R$ iff information does not flow from $g$ to $h$ in $W_{\R,C}$.
Moreover, the mapping from $(\R,C,f)$ to $(W_{\R,C},g,h)$ is uniform in
the sense that if $\R' \subseteq \R$, and $f'$ and $C'$ are the
restrictions of $f$ and $C$ to $\R'$, then $W_{\R',C'} \subseteq
W_{\R,C}$ and $j$ maintains $f'$-$C'$ secrecy with respect to $i$ in
$\R'$ iff no information flows from $g'$ to $h'$ in $W_{\R',C'}$, where
$g'$  and $h'$ are the restrictions of $g$ and $h$ to $W_{\R',C'}$.
\epro

\prf Let 
\[ W_{\R,C} = \{((r',m'), C(r,m)) : (r',m') \in C(r,m)\}, \]  
let
$g((r',m'),C(r,m)) = (r'_i(m),C(r,m))$ and let $h((r',m'),C(r,m)) =
(f(r,m),C(r,m))$.  It is straightforward to check that $j$
maintains $f$-$C$-secrecy with respect to $i$ in $\R$ iff  
information does not flow from $g$ to $h$ in $W_{\R,C}$, and that the
map is uniform as desired.  We leave details to the reader.
\eprf
} %

\subsection{Characterizing Secrecy Syntactically}\label{sec:k+sec}

Our definition of secrecy is 
semantic; it is given in terms of the local states of the agents.  
As we shall see, it is helpful to reason syntactically about secrecy,
using the logic of knowledge 
discussed in Section~\ref{sec:review}.
Our goal in this section is to characterize 
secrecy in terms of the knowledge---or more precisely, the lack
of knowledge---of the agent with respect to whom secrecy is maintained. 
To this end, we show that the state of an agent $j$ is kept secret 
from an agent $i$ exactly if $i$ does not know any formulas 
that depend only on the state of $j$,
or, dually, if $i$ always thinks that any formula that depends
on the state of $j$ is currently possible.

For this characterization, we 
use the modal logic of knowledge described in Section \ref{sec:review}.
But first, we need 
to define what it means for
a formula to depend on the local state of a particular agent. Given
an agent $j$, a formula $\phi$ is {\em $j$-local} in an interpreted
system
$\I$ if, 
for all points $(r,m)$ and 
$(r',m')$ such that $r_j(m) = r'_j(m')$, $(\I,r,m) \sat \phi$ iff 
$(\I,r',m') \sat \phi$. 
It is easy to check that $\phi$ is $j$-local in $\I$ iff 
$\I \sat K_j \phi \lor K_j \neg \phi$;
thus, $j$-locality can be characterized syntactically.
(See~\cite{EMM98} for an 
introduction to the logic of local propositions.) 
The notion of $j$-locality has another useful semantic characterization:

\pro\label{pro:localsemantic}
A formula $\phi$ is $j$-local in an interpreted system $\I = (\R,\pi)$
iff there exists a set $\Omega$ of 
$j$-information sets such that $(\I,r,m) \sat \phi$
exactly when
$(r,m) \in \bigcup_{\K \in \Omega} \K$.
\epro

\commentout{
The second definition
describes when a formula is trivial, in the sense that it would be
impossible to
forbid an agent from knowing it. A formula $\phi$ is {\em nontrivial} in
$\I$
if there exists a point $(r,m)$ such that $(\I,r,m) \not\sat \phi$.
Clearly if a formula is trivial (i.e., not nontrivial), then every agent
will always know it.  We cannot expect to prevent agents from knowing
trivial facts.  
More generally, if $\A$ is a set of points, $\phi$ is {\em $\A$-nontrivial
in
$\I$\/} if there is some point $(r,m) \in \A$ such that $(\I,r,m) \sat
\neg \phi$.
We are interested in formulas that are $C(r,m)$-nontrivial.
To see why, note that in a synchronous
system, a formula such as ``the current time is 5'' is known by every
agent at time 5. Thus if $S$ is the synchronous allowability
function defined in Section \ref{sec:secrecy}, such a formula
is not $S(r,5)$-nontrivial for any run $r$.
However, as the following theorem shows, if $\phi$ is
$C(r,m)$-nontrivial and $f$-local, then $i$ will not know $\phi$ at the
point $(r,m)$ if $j$ maintains $C$-secrecy with respect to $i$.
}

The following theorem shows that the semantic characterizations of 
secrecy given in Section~\ref{sec:secrecy} correspond closely
to our intuitions of what secrecy should mean:~agent $j$ maintains
secrecy with respect to $i$ precisely if $i$ cannot rule out
any satisfiable facts that depend only on the local state of $j$.

\thm\label{pro:f-c-sec1}
Suppose that $C$ is an $i$-allowability function.
Agent $j$ maintains $C$-secrecy with respect to agent
$i$ in system $\R$ iff, for every interpretation $\pi$ and point $(r,m)$,
if $\phi$ is $j$-local and $(\I,r',m') \sat \phi$ for some $(r',m') \in
C(r,m)$, 
then $(\I,r,m) \sat P_i \phi$.
\ethm

\commentout{
Recall that possibility and knowledge are dual operators. The dual version
of Theorem~\ref{pro:f-c-sec1} states that 
$j$ maintains secrecy with respect to $i$ precisely if $i$ never knows
any nontrivial facts that depend only on the local state of $j$:

\cor
Suppose $\R$ is a system, and $C$ is an $i$-allowability function.
Agent $j$ maintains $C$-secrecy with respect to agent
$i$ in $\R$ iff  for every interpretation $\pi$, 
if $\phi$ is $j$-local and $C(r,m)$-nonvalid in $\I = (\R,\pi)$, then 
$(\I,r,m) \sat \neg K_i \phi$.
\ecor
}

Since total secrecy is just $C$-secrecy for the allowability function
$C$ such that $C(r,m)$ consists of all point in $\R$, the following
corollary, which gives an elegant syntactic characterization of total
secrecy, is immediate.

\cor\label{cor:totsecchar}
Agent $j$ maintains total secrecy with respect to agent $i$ in system $\R$
iff, for every interpretation $\pi$, if the formula $\phi$ is $j$-local 
and satisfiable in $\I = (\R,\pi)$, then 
$\I \sat P_i \phi$.
\commentout{
The following are equivalent:
\begin{enumerate}
\item[(a)] agent $j$ maintains total secrecy with respect to agent $i$
in system $\R$;
\item[(b)] for every interpretation $\pi$, if the formula $\phi$ is $j$-local 
and satisfiable in $\I = (\R,\pi)$, then 
$\I \sat P_i \phi$.
\end{enumerate}
}
\ecor

Corollary~\ref{cor:totsecchar} says that total secrecy
requires $i$ not to know any $j$-local fact that is 
not valid in $\I$.
A similar result holds for synchronous secrecy. For brevity, and
because we prove more general results in later sections, we 
ignore the details here.
\commentout{
A similar result holds for synchronous secrecy. In synchronous
systems, however, we must exclude formulas that are unsatisfiable at
any time step, since by synchrony $i$ would know any such formula.
Formally, we say that a formula $\phi$ is 
{\em satisfiable at every time step} in an interpreted system $\I$ 
if for every time $m$ there exists 
a run $r$ such that $(\I,r,m) \sat \phi$.
This may seem like a stiff requirement, but since we are only 
interested in comparing formulas at any particular time $m$, we can
work with formulas that are defined to be true at all points not
occurring at time $m$.

\cor
In any system $\R$, the following are equivalent:
\begin{enumerate}
\item[(a)] agent $j$ maintains $S$-secrecy with respect to agent $i$; 
\item[(b)] for every interpretation $\pi$, if the formula $\phi$ is
$j$-local
and satisfiable at each time step in $\I = (\R,\pi)$, then 
$\I \sat P_i \phi$.
\end{enumerate}
\ecor

\prf
Consider the synchronous allowability function $S$, defined in
Section \ref{sec:secrecy}. A formula $\phi$ is satisfiable at
every time step iff it is $S(r,m)$-satisfiable at every
point $(r,m)$. The corollary then follows from Theorem
\ref{pro:f-c-sec1}.
\eprf
} 
We can also give a similar syntactic characterization of run-based secrecy.
For $j$ to maintain total secrecy with respect to $i$, if $\phi$ is
$j$-local, it is always necessary for $i$ to think that $\phi$ was possible.
For run-based secrecy, we require only that $i$ think that $\phi$
is possible sometime in the current run.
Recall that the formula $\di \phi$ means
``$\phi$ is true at some point in the current run''.

\thm\label{pro:syntax_weaktotal}
Agent $j$ maintains run-based secrecy with
respect to agent $i$ in system $\R$ iff, for every interpretation $\pi$,
if $\phi$ is $j$-local and satisfiable in $\I = (\R,\pi)$, 
then $\I \sat P_i \di \phi$.
\ethm

\commentout{
The results in this section
suggest that rather than defining secrecy in semantic
terms---a requirement on all points that an agent considers
possible---it
may be better to define it in syntactic terms, by specifying
what formulas an agent is allowed to know at any point.
The theorems presented demonstrate
that we can recover the semantic
definition syntactically.  Moreover, 
the syntactic definition has a number of advantages. 
For one thing, it allows model checking for secrecy (at least, for
finite-state systems). 
Second, by specifying a noninterference property as a set of formulas
that must be kept secret, it becomes easier to specify exactly what
matters.  For example, 
a system designer may not care that $i$ not know all the formulas that
$f$-$C$-secrecy requires $i$ not to know.
It may be enough that $i$ does not know a small subset of them.  
}
The results of this section show that secrecy has a syntactic
characterization that is equivalent to the semantic characterization.
There are several significant advantages to having such a syntactic
characterization. 
For one thing, it suggests that secrecy can be checked by applying 
model-checking techniques (although techniques would have to be developed to
allow checking $P_i \phi$ for all formulas $\phi$).  Perhaps more
importantly, it suggests some natural generalizations of secrecy that
may be of practical interest.  
For example, it may not be relevant
that $i$ not know {\em all\/} satisfiable formulas.  It may be
enough for a system designer that $i$ does not know only certain
formulas.
This may be particularly relevant for declassification or
downgrading:~if a noninterference property corresponds to a set of 
formulas that must be kept secret from the low-level agent, formulas can be 
declassified by removing them the set.  
Another significant generalization involves replacing knowledge by a
more computational notion, such as {\em algorithmic knowledge}
\cite{FHMV,HP02a}. 
Recall that
the definition of knowledge described in Section \ref{sec:review} 
suffers from the {\em logical omniscience\/} problem:~agents know all
tautologies and know all logical consequences of their 
knowledge \cite{FHMV}.  In the context of security, we are more
interested in what 
{\em resource-bounded\/} agents know.  It does not matter that an agent
with unbounded computational resources can factor and decrypt a message
as long as a resource-bounded agent cannot decrypt the message.
By requiring only that an agent does not {\em algorithmically know\/}
various facts, we can capture secrecy with respect to resource-bounded
agents.

\section{Secrecy in Probabilistic Systems}\label{sec:prob}

The definitions of secrecy that we have considered up to 
are {\em possibilistic}; they consider
only
whether or not an event is possible.
They thus cannot capture what seem like rather serious leakages of
information.  
As a motivating example, consider a 
system $\R$ with two agents Alice and Bob, who we think of as sitting at
separate computer terminals.
Suppose that $\L$ is a language with $n$ words.  
At time 1, Bob inputs a string $x \in \L$ chosen uniformly at
random. At time 2, with probability
$1-\epsilon$, the  
system outputs $x$ directly to Alice's terminal. However, with probability
$\epsilon$, the 
system is struck by a cosmic ray as Bob's input is transmitted to 
Alice, and in this case the system
outputs a random string from $\L$. 
(Bob receives no information about what Alice sees.)
Thus, there are $n(n+1)$ possible runs in this system: $n$ runs where no
cosmic ray hits, and $n^2$ runs where the cosmic ray hits.  Moreover, it is
immediate that  
Bob maintains (possibilistic) synchronous secrecy with respect
to Alice even though, with very high probability, Alice sees exactly what
Bob's input was. 

To reason about the unwanted information flow in this example,
we need to add
probability to the framework.  We can do that in this example by putting
the obvious probability measure on the runs in $\R$:
\ul
\item for each $x \in \L$, the run where Bob inputs $x$ and no cosmic
ray hits (so that Alice sees $x$) gets probability $(1-\epsilon)/n$.
\item for each pair $(x,y) \in \L \times \L$, the run where the cosmic
ray hits, Bob inputs $x$, and Alice sees $y$ gets probability $\epsilon/n^2$.
\eul
If Alice sees $x$, her posterior probability that Bob's input was $x$ is
$${\Pr}_{Alice}(\mbox{Bob typed~} x \mid \mbox{Alice sees $x$}) = \frac{\epsilon
+ n - n\epsilon}{n} = 1 - \frac{n-1}{n}\epsilon.$$ 
If Alice sees $x$, her posterior probability that Bob's input was $y \not= x$
is
$${\Pr}_{Alice}(\mbox{Bob typed~} x \mid \mbox{Alice sees $y$}) = \frac{\epsilon}{n}.$$
Thus, if $\epsilon > 0$, even though Alice never learns {\em with
certainty} that Bob's input was $x$, her probability that Bob input
$x$ rises from $1/n$ to almost 1 
as soon as
she sees an $x$.

In this section 
we introduce definitions of probabilistic secrecy.
The  definitions and the technical results we obtain
closely resemble the definitions and results of the previous two
sections. This is no coincidence. 
As we show in Section~\ref{sec:plaus}, probabilistic and possibilistic
secrecy are instances of a definition of {\em plausibilistic\/}
secrecy for which similar results can be proved in more generality.

\subsection{Defining Probabilistic Secrecy}\label{sec:define_prob_sec}

\commentout{
To reason about probability in systems, we use the
general approach described by Halpern \citeyear{Hal31}. In this section, we 
introduce two 
ways to reason about probability in systems. One emphasizes the {\em subjective} probability 
distributions that agents have on sets of points; the other emphasizes
an {\em objective} 
probability measure on sets of runs. These two approaches are closely
related, as we  
show
in the sections that follow.
} %
To reason about probabilistic security, we need a way to introduce
probability into the multiagent systems framework.  There are actually
two ways of doing this: we can either put a probability on points 
or a probability on runs.  We consider putting a probability on points
first, using a general approach described by Halpern~\citeyear{Hal31}.

Given a system $\R$, 
define a {\em probability assignment\/} $\PR$ to be a function that
assigns to each agent $i$ and point $(r,m)$ 
a probability space $\PR(r,m,i) = (W_{r,m,i}, \F_{r,m,i}, \mu_{r,m,i})$,
where $W_{r,m,i} \subseteq \PT(\R)$ is $i$'s sample space at $(r,m)$ and
$\mu_{r,m,i}$ is a probability measure defined on the subsets of
$W_{r,m,i}$ in $\F_{r,m,i}$.  (That is, $\F_{r,m,i}$ 
is a $\sigma$-algebra that
defines the
measurable subsets of $W_{r,m,i}$.)  We 
call a pair $(\R,\PR)$ a {\em probability system}.  
Given a probability system, we can give relatively straightforward definitions
of probabilistic total secrecy and synchronous secrecy. Rather than requiring
that an agent $i$ think that all states of another $j$ are {\em possible},
we require that all of those states be measurable and equally likely
according to $i$'s subjective probability measure.

\dfn\label{dfn:totprobsec}
Agent $j$ maintains {\em probabilistic total secrecy\/} with respect to
agent $i$ in 
$(\R,\PR)$ if, for all points $(r,m)$, $(r',m')$, and
$(r'',m'')$ in $\PT(\R)$, 
we have
$\K_j(r'',m'') \inter \K_i(r,m) \in
\F_{r,m,i}$, $\K_j(r'',m'') \inter \K_i(r',m') \in \F_{r',m',i}$, and 
$$\mu_{r,m,i}(\K_j(r'',m'') \inter \K_i(r,m)) = 
\mu_{r',m',i}(\K_j(r'',m'') \inter \K_i(r',m').$$
\edfn

Probabilistic total secrecy is a straightforward extension of total
secrecy. Indeed, if for all points $(r,m)$ we have 
$\mu_{r,m,i}(\{(r,m)\}) > 0$, then probabilistic
total secrecy implies total secrecy (as in Definition \ref{dfn:totsecrecy}).

\pro\label{pro:prob_total_sec}
If $(\R,\PR)$ is a probability system such that  
$\mu_{r,m,i}(\{(r,m)\}) > 0$ for all points $(r,m)$
and
$j$ maintains probabilistic total secrecy with respect to $i$ in
$(\R,\PR)$, then $j$ also maintains total secrecy with respect to $i$
in $\R$.
\epro

Like total secrecy,
probabilistic 
total
secrecy is an unrealistic
requirement in practice, and cannot be achieved in synchronous systems. 
To make matters worse, the sets 
$\K_j(r'',m'') \inter \K_i(r,m)$ are typically not measurable according to
what is
perhaps the most common approach for defining $\PR$, as we show in
the next section. 
Thus, even in completely asynchronous systems, total secrecy is usually
impossible 
to achieve
for measurability reasons alone.
\commentout{
While this is a straightforward extension of total secrecy and, as we
shall see, under reasonable assumptions has properties that are
straightforward analogues of total secrecy, it is a somewhat problematic
notion.  For one thing, as we already pointed out in the possibilistic
case, it makes unrealistically strong requirements in practice.  In
addition, as we shall see, according to the standard approach for defining
$\PR$, the set $\K_j(r'',m'') \inter \K_i(r,m)$ is typically not
measurable, so for measurability reasons alone, probabilistic total
secrecy does not hold.
}
Fortunately, the obvious probabilistic analogue of synchronous secrecy
is a more reasonable condition.

\dfn\label{dfn:syncprobsec}
Agent $j$ maintains {\em probabilistic synchronous secrecy\/} with respect to
agent $i$ in 
$(\R,\PR)$ if, for all 
runs $r,r',r''$ and all times $m$, we have
$\K_j(r'',m) \inter \K_i(r,m) \in
\F_{r,m,i}$, $\K_j(r'',m) \inter \K_i(r',m) \in \F_{r',m,i}$, and 
$$\mu_{r,m,i}(\K_j(r'',m) \inter \K_i(r,m)) = 
\mu_{r',m,i}(\K_j(r'',m) \inter \K_i(r',m)).$$
\edfn
Note that if we set up the cosmic ray system of the previous section as 
a probability system in such a way that Alice's probability
on points reflects the posterior probabilities we described for the system, 
it is immediate that Bob does {\em not} maintain probabilistic
synchronous secrecy with respect to Alice.

We now consider definitions of probabilistic secrecy where we start with
a probability on runs.
Define a {\em run-based probability system\/} to be a triple
$(\R,\F,\mu)$, where $\R$ is a system, $\F$ is a $\sigma$-algebra of
subsets of $\R$, and $\mu$ is a probability measure defined on $\F$.
Note that a run-based probability system requires only one probability
measure, rather than a probability measure at each point for each agent.
In practice, 
such a measure is often relatively easy to come by. 
In the same way that a
set of runs can be generated by a protocol, a runs-based probability system can
be generated by a probabilistic protocol: the probability of a set of
runs sharing 
a common prefix can be derived by multiplying the probabilities of the
protocol transitions  necessary to generate the prefix
(see  \cite{Hal31,HT} for further discussion).

Here and throughout the paper, we assume for simplicity that in a
run-based probability system $(\R,\F,\mu)$, $\F$ 
contains all sets of the form $\R(\K_i(r,m))$, for all
points $(r,m)$ and all agents $i$. That is, if a set of runs
is generated by an agent's local state, it is measurable. We also assume
that $\mu( \R(\K_i(r,m)) ) > 0$, 
so that we can condition on information sets.

Recall from Section \ref{sec:secrecy} that run-based total secrecy
requires that, 
for all points $(r,m)$ and $(r',m')$, we have
$\R(\K_i(r,m)) \cap \R(\K_j(r',m')) \not= \emptyset.$ In other words, run-based total
secrecy is a property based on what can happen during runs, rather than points.
In a run-based probability system where all information sets have
positive measure, 
it is easy to see that
this is equivalent to the requirement that
$\mu(\R(\K_j(r',m')) \mid \R(\K_i(r,m))) > 0.$ We strengthen run-based
total secrecy by 
requiring that these probabilities be {\em equal}, for all $i$-information sets.

\dfn
Agent $j$ maintains {\em run-based probabilistic secrecy\/} with respect
to $i$ in $(\R,\F,\mu)$
if for any three points $(r,m), (r',m'), (r'',m'') \in \PT(\R)$, 
\commentout{
the sets
$\R(\K_i(r,m)), \R(\K_i(r',m')),$ and $\R(\K_j(r'',m''))$ are 
measurable, and
}
$$\mu( \R(\K_j(r'',m'')) \mid \R(\K_i(r,m)) ) = \mu( \R(\K_j(r'',m''))
\mid \R(\K_i(r',m')) ). $$ 
\edfn

The probabilities for the cosmic-ray system were defined on sets of runs.
Moreover, facts such as ``Alice sees $x$'' and ``Bob typed $y$'' 
correspond to information sets, exactly as in the definition of run-based
probabilistic secrecy. It is easy to check that Bob does not maintain
run-based probabilistic secrecy with respect to Alice.

In Section~\ref{sec:add_prob}, we consider the connection between
probability measures on points and on runs, and the corresponding connection
between probabilistic secrecy and run-based probabilistic secrecy.
For the remainder of this
section, we consider symmetry in the context of probabilistic secrecy.
In Section \ref{sec:secrecy}, we mentioned that our definitions of secrecy were
symmetric in terms of the agents $i$ and $j$. Perhaps surprisingly,
probabilistic 
secrecy is also a symmetric condition, at least in most cases of interest. 
This follows from a deeper fact: 
under reasonable assumptions,
secrecy (of $j$ with respect to $i$)
implies the probabilistic independence of $i$-information sets and
$j$-information sets.
(See Lemma~\ref{lem:independence} in the appendix for more details.)
Consider probabilities on points:
if there is no
connection whatsoever between $\PR(r,m,i)$ and $\PR(r,m,j)$ in a
probability system $(\R,\PR)$, 
there is 
obviously
no reason to expect secrecy to be symmetric.
However, if 
we assume that the probabilities of $i$ and $j$
at $(r,m)$ are derived 
from
a single common probability measure by conditioning, then 
symmetry follows.
This assumption, which 
holds for the probability systems we will consider here
(and is standard in the economics literature \cite{Morris95}),
is formalized in the next definition.

\dfn
A probability system $(\R,\PR)$ satisfies the {\em common prior
assumption} if there exists a probability space 
$(\PT(\R),\F_{\mathit{cp}},\mu_{\mathit{cp}})$ such 
that for all agents $i$ and points $(r,m) \in \PT(\R)$, 
we have $\K_i(r,m) \in \F_W$,
$\mu_{\mathit{cp}}(\K_i(r,m)) > 0$, and 
$$\PR_i(r,m) = (\K_i(r,m),
\{ U \cap K_i(r,m) \mid U \in \F_W \}, \mu_{\mathit{cp}}
\mid \K_i(r,m)).%
\footnote{Actually, it is more standard in the economics literature not
to require that $\mu_{\mathit{cp}}(\K_i(r,m)) > 0$.  No requirements are
placed on $\mu_{r,m,i}$ if $\mu_{\mathit{cp}}(\K_i(r,m)) = 0$.  See
\cite{Hal21} for a discussion of this issue.}
$$
\edfn
In probability systems that satisfy the common prior assumption, 
probabilistic secrecy is symmetric. 

\pro\label{pro:sync_independence}
If $(\R,\PR)$ is a probability system (resp., synchronous probability system) 
that satisfies the 
common prior assumption with prior probability $\mu_{\mathit{cp}}$, the
following are 
equivalent: 
\begin{itemize}
\item[(a)] Agent $j$ maintains probabilistic total (resp., synchronous) secrecy with respect to $i$.
\item[(b)] Agent $i$ maintains probabilistic total (resp., synchronous) secrecy with respect to $j$.
\commentout{
\item[(c)] 
$\mu_{\mathit{cp}}(\K_j(r',m) \mid \K_i(r,m)) =
\mu_{\mathit{cp}}(\K_j(r',m))$
(resp., $\mu_{\mathit{cp}}(\K_j(r',m) \mid \K_i(r,m)) 
= \mu_{\mathit{cp}}(\K_j(r',m) \mid \PT(m))$, where
$\PT(m)$ is the set of points occurring at time $m$;
that is, the events
$\K_i(r,m)$ and $\K_j(r',m)$ are conditionally independent with respect
to $\mu_{\mathit{cp}}$, given that the time is $m$).
}
\item[(c)] For all points $(r,m)$ and $(r',m')$, 
$\mu_{\mathit{cp}}(\K_j(r',m') \mid \K_i(r,m)) =
\mu_{\mathit{cp}}(\K_j(r',m'))$ (resp., for all points
$(r,m)$ and $(r',m)$, $\mu_{\mathit{cp}}(\K_j(r',m) \mid \K_i(r,m)) 
= \mu_{\mathit{cp}}(\K_j(r',m) \mid \PT(m))$, where
$\PT(m)$ is the set of points occurring at time $m$;
that is, the events
$\K_i(r,m)$ and $\K_j(r',m)$ are conditionally independent with respect
to $\mu_{\mathit{cp}}$, given that the time is $m$).
\end{itemize}
\epro

In run-based probability systems 
there is a single measure $\mu$ that 
is independent of the agents, and we have symmetry provided that 
the system is synchronous or both agents have perfect recall.
(If neither condition holds, secrecy may not be symmetric, as illustrated by
Example \ref{xam:no_ind}.)

\pro\label{pro:run_based_independence}
If $(\R,\F,\mu)$ is a run-based probability system that is either
synchronous or one where agents $i$ and $j$ both have perfect recall, then
the following are 
equivalent:
\begin{itemize}
\item[(a)] Agent $j$ maintains run-based probabilistic secrecy with respect
to $i$.
\item[(b)] Agent $i$ maintains run-based probabilistic secrecy with respect
to $j$.
\item[(c)] For all points $(r,m), (r',m') \in \PT(\R)$,
$\R(\K_i(r,m))$ and $\R(\K_j(r',m'))$ are probabilistically independent
with respect to $\mu$.
\end{itemize}
\epro

\subsection{From Probability on Runs to Probability on Points}\label{sec:add_prob}
\commentout{
Where are the probability assignments in probability systems coming
from?  
Halpern and Tuttle \citeyear{HT} give a general approach for
defining probability assignments in systems that captures some
reasonable intuitions.  
We review the main ideas here.

Just as a (joint) protocol generates a set of runs, a
probabilistic joint protocol generates a set of runs with a probability
measure on the runs.  
}
In the last section we described two ways of adding probability to
systems: putting a probability on points and putting a probability on runs.
In this section, we discuss an approach due to Halpern and Tuttle
\citeyear{HT} for connecting the two approaches.

\commentout{
Let $(\R,\F,\mu)$ be a run-based probability system. 
Here, and throughout the rest of the paper, we
assume 
for simplicity
that $\F$ contains all sets of the form $\R(\K_i(r,m))$, for all
points $(r,m)$ and all agents $i$. That is, if a set of runs
is generated by an agent's local state, it is measurable. We also assume
that $\mu( \R(\K_i(r,m)) ) > 0$, so that all of these events have positive probability.
(We sometimes omit the set $\F$ and write simply $(\R,\mu)$ if 
$\F = 2^{\R}$ or if $\F$ is irrelevant to the discussion.)
} %

Given an agent $i$ and a point $(r,m)$, we would like to derive the
probability measure $\mu_{r,m,i}$ from $\mu$ by conditioning $\mu$ on
$\K_i(r,m)$, the information that  
$i$ has at the point $(r,m)$.  The problem is that $\K_i(r,m)$ is a set of
{\em points}, not a set of {\em runs}, so 
straightforward conditioning does not work.
To solve this problem,
we condition $\mu$ on $\R(\K_i(r,m))$,
the set of runs going through $\K_i(r,m)$, 
rather than on 
$\K_i(r,m)$. Conditioning is always well-defined, 
given our assumption that $\R(\K_i(r,m))$ has positive measure.

We can now define a measure
$\mu_{r,m,i}$ on the points in $\K_i(r,m)$ as follows.  If $\S \subseteq \R$
and $A \subseteq \PT(\R)$, let
$A(\S)$ be the set of points in $A$ that lie on runs in $\S$; that is,
$$A(\S) = \{(r',m') \in A: r' \in \S\}.$$
In particular, $\K_i(r,m)(\S)$ consists of the points in $\K_i(r,m)$ that
lie on runs in $\S$.
Let $\F_{r,m,i}$
consist of all sets of the
form $\K_i(r,m)(\S)$, where 
$\S \in \F$.  
Then define 
$$\mu_{r,m,i}(\K_i(r,m)(\S)) = \mu(\S \mid \R(\K_i(r,m)).$$
It is easy to check that if $U \subseteq \K_i(r,m)$ is measurable with
respect with respect to $\mu_{r,m,i}$, then $\mu_{r,m,i}(U) =
\mu(\R(U) \mid \R(\K_i(r,m)))$.  
We say that the resulting probability system $(\R,\PR)$ is {\em
determined\/} by the run-based probability system  $(\R,\F,\mu)$, 
and call $\mu$ the {\em underlying measure}.
We call a probability system 
{\em standard\/} if it is determined by a run-based probability system.

Note that {\em synchronous} standard probability
systems satisfy the common 
prior assumption, as defined in the previous section. 
If we assume that all runs are measurable, then we can simply take
$\mu_{\mathit{cp}}(r,m) = \mu(r)/2^{m+1}$.  This ensures that the time $m$
points have the same relative probability as the runs, which is exactly
what is needed.   More generally, if $\PT(m)$ is the set of time $m$
points and $\S$ is a measurable subset of $\R$, we take
$\mu_{\mathit{cp}}(\PT(m)(\S)) = \mu(\S)/2^{m+1}$.  
It follows from Proposition~\ref{pro:sync_independence} that
probabilistic synchronous secrecy is symmetric in synchronous standard
systems.  

\commentout{
In a synchronous standard system, 
the sets $\K_j(r'',m) \inter
\K_i(r,m)$ that arise in the definition of probabilistic synchronous
secrecy are in $\F_{r,m,i}$, since they are the set of points in
$\K_i(r,m)$ on the runs in $\R(\K_j(r'',m) \inter \K_i(r,m))$. 
Because of
this, we can characterize synchronous secrecy entirely in terms of the
underlying run-based probability system.

\lem\label{lem:run_based_sync_sec}
Suppose that $(\R,\PR)$ is 
the standard system determined by the synchronous
run-based probability system $(\R,\F,\mu)$. Agent $j$ maintains
probabilistic secrecy with respect to agent $i$ in $(\R,\PR)$ iff for
all runs $r,r',r''$ and all times $m$,
$$ \mu ( \R(\K_j(r'',m) ) \mid \R(\K_i(r,m)) ) = \mu ( \R(\K_j(r'',m))
\mid \R(\K_i(r',m)) ). $$ 
\elem

\prf
For $\K_j(r'',m)$ to be measurable with respect to $\mu_{t,m,i}$,
$\K_j(r'',m) \cap \K_i(t,m)$  must take the form 
$\K_i(t,m)(\S)$, for some $\S \in \F$. In synchronous systems, the set
$\S = \R(\K_j(r'',m))$ 
satisfies this requirement,
because the only runs where $j$ has the local state $r''_j(m)$ are runs
where $j$ has this local state at time $m$.
Because $\mu_{t,m,i}(\K_j(r'',m)) = \mu( S \mid \R(\K_i(t,m)) )$, 
we have
$$\mu_{r,m,i} (\K_j(r'',m)) = \mu ( \R(\K_j(r'',m))  \mid \R(\K_i(r,m)) ) $$
and
$$\mu_{r',m,i} (\K_j(r'',m)) = \mu ( \R(\K_j(r'',m))  \mid
\R(\K_i(r',m)) ). $$
\eprf

The analogue of Lemma~\ref{lem:run_based_sync_sec} does not hold for
probabilistic total secrecy, because 
the sets $\K_j(r'',m) \inter \K_i(r,m)$ that arise in the definition
are not, in general, measurable in standard
systems.  
To see this, consider a system 
$\R$ that consists of two runs $r_1$ and $r_2$, 
each with probability $1/2$.
Agent 1 is in local state
$l_1$ throughout $r_1$ and in local state $l_2$ throughout $r_2$.
(He only knows what run he is in.)
Agent 2 has
local states $l_0', l_1', l_2', \ldots$, and is in local state $l_j'$ at the
points $(r_1,j)$, $(r_2, 2j)$, and $(r_2,j+1)$.  Note that 
$\K_1(r_1,0) \inter \K_2(r_1,0) = \{(r_1,0)\}$ and $\K_1(r_2,0) \inter
\K_2(r_2,0) = \{(r_2,0), (r_2,1)\}$.  However, $(r_1,0)$ is not
measurable with respect to $\mu_{r_1,0,1}$ and $\{(r_2,0), (r_2,1)\}$ is
not measurable with respect to $\mu_{r_2,0,1}$.  In fact, the only sets
measurable with respect to $\mu_{r_1,0,1}$ is the set consisting of all
the points on $r_1$ (which has probability 1) and the empty set (which
has probability 0); the same is true for $\mu_{r_2,0,1}$.  In the standard
probability system determined by $(R,\mu)$, there is no way for agent 1
to determine the likelihood of being at any particular time in $r_1$ or
$r_2$.  Thus, while the notion of total secrecy 
is perfectly well-defined for probability systems, it is generally 
not very useful unless we are willing to assign probabilities to whether or not
it is a particular time.

Lemma \ref{lem:run_based_sync_sec} shows us that we can characterize
probabilistic synchronous 
secrecy in terms of an underlying probability measure on runs.
It turns out that a deeper result is also true, in synchronous standard systems
with perfect recall: probabilistic synchronous secrecy is equivalent to
run-based secrecy 
in the underlying run-based probability system. This can be viewed as a slightly weaker
probabilistic analogue of Proposition \ref{pro:sync<->weak}. (Perfect recall is
necessary, as illustrated by Example \ref{xam:need_recall}.)
}

In synchronous standard systems with perfect recall, probabilistic
secrecy and run-based probabilistic secrecy coincide.
(We remark that Example \ref{xam:need_recall} shows that
the requirement of perfect recall is necessary.)
This provides further evidence that our notions of probabilistic secrecy 
are appropriate in synchronous systems.

\pro\label{pro:sync_prob_sec}
If $(\R,\PR)$ is the standard system determined by the
synchronous run-based 
probability system $(\R,\F,\mu)$, and agents $i$ and $j$ have perfect
recall in $\R$, then
agent
$j$ maintains run-based probabilistic secrecy with respect to $i$ in
$(\R,\F,\mu)$ iff  
$j$ maintains probabilistic synchronous secrecy with respect to $i$ in $(\R,\PR)$.
\epro
\commentout{
In asynchronous standard systems, it does not hold in
general because of issues of nonmeasurability.  
The sets $\K_j(r'',m'') \inter \K_i(r,m)$ that arise in the definition
of secrecy are guaranteed to be measurable (i.e., in $\F_{r,m,i}$) in
the synchronous case, where $m'' = m$, since $\K_j(r'',m'') \inter
\K_i(r,m)$ is the set of points in
$\K_i(r,m)$ on the runs in $\R(\K_j(r'',m) \inter \K_i(r,m))$. 
On the other hand, in the asynchronous case, $\K_j(r'',m'') \inter
\K_i(r,m)$ is not, in general, measurable.  This again shows that total
secrecy is somewhat problematic in the asynchronous case.
}

\subsection{Characterizing Probabilistic Secrecy}\label{sec:probsecchar}

We now demonstrate that we can characterize probabilistic secrecy syntactically, 
as in the nonprobabilistic case. To do so, we must first explain how to 
reason about probabilistic formulas.
Define an {\em interpreted probability system\/} $\I$ to be a tuple
$(\R,\PR,\pi)$, where $(\R,\PR)$ is a probability system.
In an interpreted probability system we can give semantics to
syntactic statements of probability. 
We 
are
most interested in formulas of the form
$\Pr_i (\phi ) = \alpha$ 
(or similar formulas with $\leq$, $>$, etc.,
instead of $=$). 
Such formulas were given semantics by Fagin, Halpern, and Megiddo
\citeyear{FHM}; we follow their approach here. 
Intuitively, a 
formula such as $\Pr_i (\phi) = \alpha$ is true 
at a point $(r,m)$ if, according to $\mu_{r,m,i}$, the probability that
$\phi$ is true is given by $\alpha$.  More formally,
$  (\I,r,m) \models {\Pr}_i(\phi) = \alpha $ if
$$  \mu_{r,m,i}(\{(r',m') \in \K_i(r,m) : (\I,r',m') \models \phi \}) = \alpha. $$ 
Similarly, we can give semantics to $\Pr_i(\phi) \leq \alpha$ and
$\Pr(\phi) > \alpha$, etc., as well as conditional formulas such as
$\Pr(\phi \, | \, \psi) = \alpha$.
Note that although these formulas talk
about probability, they are either true or false at a given state.

The semantics for a formula such as $\Pr_i(\phi)$ implicitly assumes
that the set of points in $\K_i(r,m)$ where $\phi$ is true is
measurable.  While there are ways of dealing with nonmeasurable sets
(see \cite{FHM}), here we assume that all 
relevant sets are
measurable.  This is certainly 
true
in synchronous standard systems
determined by a a run-based system where all sets of runs are measurable.
More generally, it is true in a probability system $(\R,\PR)$ where, for
all $r$, $m$, $i$, all the sets in the probability space $\PR(r,m,i)$
are measurable.  

The first result shows that we can characterize probabilistic total 
and synchronous secrecy.

\thm\label{thm:sync_prob_syntax}
\begin{itemize}
\item[(a)] If $(\R,\PR)$ is a probabilistic system, then 
agent $j$ maintains probabilistic total secrecy with respect to
agent $i$ iff,
for every interpretation $\pi$
and formula $\phi$ that is $j$-local in $\I = (\R,\PR,\pi)$, 
there exists a constant $\sigma$ such that 
$\I \sat \Pr_i(\phi) = \sigma$.
\item[(b)] 
If $(\R,\PR)$ is a synchronous probabilistic system, then 
agent $j$ maintains probabilistic synchronous secrecy with respect to
agent $i$ iff, 
for every interpretation $\pi$, time $m$, 
and formula $\phi$ that is $j$-local in $\I = (\R,\PR,\pi)$, 
there exists a constant $\sigma_m$ such that 
$(\I,r,m) \sat \Pr_i(\phi) = \sigma_m$ for all runs $r \in \R$.
\end{itemize}
\ethm

We can also characterize run-based secrecy in standard systems using the 
$\di$ operator. 
For this characterization,
we need the additional assumption of
perfect recall.  

\thm\label{thm:runbased_prob_syntax}
If $(\R,\PR)$ is a standard probability system where agent $j$ has
perfect recall, then agent $j$ maintains 
run-based probabilistic secrecy with
respect to agent $i$ iff, for every interpretation $\pi$
and every formula $\phi$ that is $j$-local
in $\I = (\R, \PR, \pi)$, there exists a constant $\sigma$ such that
$\I \sat \Pr_i(\di \phi) = \sigma$. 
\ethm
Example~\ref{xam:recall_for_syntax} demonstrates that the assumption of
perfect recall is necessary in Theorem~\ref{thm:runbased_prob_syntax} and that
synchrony alone does not suffice.

\commentout{
Note that if we assume perfect recall, run-based probabilistic
secrecy has the same properties as its nonprobabilistic analogue: it is
symmetric, has an elegant syntactic characterization, and follows from
probabilistic synchronous secrecy in synchronous systems.
These results again emphasize that
``sensible'' definitions of secrecy rely on the assumption of 
perfect recall---especially in asynchronous systems.
}
\commentout{
The requirement that $i$ and $j$ have perfect recall is necessary in
Proposition~\ref{pro:async-symmetric} and~\ref{pro:sync_prob_sec};
see Examples~\ref{xam:need_recall} and~\ref{xam:no_ind}.  Similarly,
the requirement that either $j$ have perfect recall or that $\R$ is
synchronous is necessary in Theorem~\ref{thm:probsec}; see
Example~\ref{xam:recall_for_syntax}.  These results again emphasize that
secrecy in the absence of perfect recall is rather problematic.
}

\commentout{
Proposition~\ref{pro:sync_prob_sec}, together with~\ref{pro:sync<->weak},
demonstrate the suitability of our definitions for synchronous systems where
agents have perfect recall. In such systems we can define secrecy in a way
that protects an agent's current state, as well as information about her
past or future states, in a way that does not place nonsensical restrictions
on the points that agents consider possible. In asynchronous systems, or
in systems where perfect recall is not present, it seems to be
inherently difficult to give appropriate definitions of secrecy. We hope
that the results of this paper 
will help to clarify these difficulties, and that our definitions of secrecy 
will be applicable to a wide range of situations.
}

\subsection{Secrecy in Adversarial Systems}\label{sec:nondeterm}

It is easy to capture our motivating ``cosmic-ray system'' example using
a synchronous standard system
because we assumed
a probability on the set
of runs. 
Furthermore, 
it is not hard to show
that Bob does not maintain synchronous secrecy with respect to Alice in this
system.  
However, there is an important and arguably
inappropriate assumption that was made when we modeled the cosmic-ray
system, namely, that we were given 
the probability 
with which Bob inputs various strings.
While we took that probability to be uniform, that was not a 
necessary 
assumption: any
other probability distribution would have served to make our
point.  The critical assumption was that there is a well-defined
distribution that is known to the modeler.  However, in many cases the
probability distribution is 
{\em not}
known.
In the ``cosmic ray'' example, if we think of the strings as words in
natural language, it may not be reasonable to view all strings as
equally likely.  Moreover, the probability of a string may depend on the
speaker: it is unlikely that a teenager would have the same distribution as an
adult, or that people having a technical discussion would have the 
same distribution as people discussing a movie.   

This type of situation, where there is an initial nondeterministic step
followed by a sequence of deterministic or probabilistic steps, is quite
common. 
The nondeterministic step could determine the choice of speaker, the
adversary's protocol, 
or the input to a 
probabilistic
protocol.
Indeed, it has been argued \cite{Rabin82,Vardi85} that any setting where
there is a mix of nondeterministic, probabilistic, and deterministic
moves
can be reduced to one where there is an initial
nondeterministic move followed by probabilistic or deterministic moves.
\commentout{
Typically, we want to prove that
an agent maintains probabilistic secrecy without making 
assumptions about the user's input distribution. 
More precisely, we want to be able to model systems in such a way as to
allow users to make nondeterministic choices to which we are not willing
to assign a probability measure, and to reason about secrecy in such systems.
A common technique for dealing with this uncertainty
is to view the adversary as following some (possibly probabilistic)
{\em protocol} (cf. \cite{Rabin82,Vardi85}), coming from some set of possible
protocols.  
Halpern and Tuttle \citeyear{HT} argue that doing this allows us
to view the
system as a {\em collection} of probability spaces, one corresponding 
to each selection of the nondeterministic choices that different 
agents could make.
For each fixed protocol, we have a purely probabilistic system,
with a well-defined probability on runs. (Note that we can assume that
the nondeterministic choices are made at the {\em start} of a run,
because, as Halpern and Tuttle demonstrate, we can ``compress'' or
``factor out'' the nondeterministic choices made throughout the run 
into a single choice which is determined at the beginning of a run.)
}
In such a setting, we do not have one probability distribution over the
runs in a system. Rather, we can partition the set of runs according to
the nondeterministic initial step, and then use a separate probability
distribution for the set of runs corresponding to each initial step.  
For example, consider a setting with a single agent and an
adversary.  The agent uses a 
protocol $p$, and the adversary uses one of a set 
$\{q_1, \ldots, q_n\}$ of protocols.  
The system $\R$ consists of all the runs generated by running 
$(p,q_k)$ for $k = 1, \ldots, n$.
$\R$ can then be partitioned into $n$ subsets 
$D_1, \ldots, D_n$, where
$D_j$ consists the runs of the joint protocol $(p,q_j)$.
While we may 
not
want to assume a probability on how likely the
adversary is to use $q_j$, 
typically there is
a natural probability
distribution on each set $D_j$. 
Note that we can capture uncertainty about a speaker's distribution over
natural language strings in the same way; each protocol corresponds to a
different speaker's ``string-production algorithm''.

Formally, situations where there is a nondeterministic choice
followed by a sequence of probabilistic or deterministic choices can be
characterized by an {\em adversarial probability system\/}, which is a tuple
$(\R,\D,\Delta)$, 
where $\R$ is a system, 
$\D$ is a 
countable
partition of $\R$, 
and 
$\Delta = \{(D,\F_D,\mu_D): D \in \D\}$ is a set 
of probability spaces, where $\mu_D$ is a probability
measure on the $\sigma$-algebra $\F_D$ (on $D \in \D$)
such that, for all agents $i$, points $(r,m)$, and cells $D$, 
$\R(\K_i(r,m)) \cap D \in \F_D$ and, 
if $\R(\K_i(r,m)) \inter D \ne \emptyset$, then 
$\mu_D( \R(\K_i(r,m)) ) > 0$.
\footnote{We actually should have written $\mu_D( \R(\K_i(r,m)) \inter
D)$ rather than $\mu_D( \R(\K_i(r,m)))$ here, since $\R(\K_i(r,m))$ is
not necessarily in $\F_D$ (and is certainly not in $\F_D$ if
$\R(\K_i(r,m))$ is not a subset of $D$).  
For brevity we 
shall
continue to abuse notation and 
write $\mu_D(U)$ as shorthand for $\mu_D(U \inter D)$.} 
There are several ways of viewing the cosmic-ray example as an
adversarial probability system.  If we view the input as a
nondeterministic choice, then we can
take $D(x)$ to consist of all runs where the
input is $x$, and let $\D = \{D(x): x \in \L\}$.  
The measure $\mu_x$ on $D(x)$ is obvious:
the one run in $D(x)$ where the cosmic ray does not strike gets
probability $1-\epsilon$; the remaining $n$ runs each get probability
$\epsilon/n$.  Note that we can assign a probability on $D(x)$ without
assuming anything about Bob's input distribution.  
Alternatively, we can assume there are $k$ ``types'' of agents (child,
teenager, adult, etc.), each with their own distribution over inputs.
Then the initial nondeterministic choice is the type of agent.  
Thus, the set of runs is partitioned into sets $D_j$, $j=1, \ldots, k$.  
We assume that agents of type $j$ generate inputs according to
probability $\Pr_j$.
In each set $D_j$, there is one run where Bob inputs $x$ and the cosmic ray
does not strike; it has probability $\Pr_j(x) (1-\epsilon)$.  There are
$n$ runs where Bob inputs $x$ and the cosmic ray strikes; each gets
probability $\Pr_j(x)\epsilon/n$.

We can identify an adversarial probability system with a set of
run-based probability systems, by viewing
the measures in $\Delta$ as constraints on a single measure on $\R$.
Let $\F_\D = \sigma( \bigcup_{D \in \D} \F_D )$, the $\sigma$-algebra 
generated by the measurable sets of the 
probability spaces of $\Delta$. (It is straightforward
to check that $U \in \F_\D$ iff $U = \bigcup_{D \in \D} U_D$,
where $U_D \in \F_D$.)
Let $\M(\Delta)$ consist of all measures $\mu$ on $\F$ such that
(1) for all $D \in \D$,  if $\mu(D) > 0$ then $\mu\mid D = \mu_D$
(i.e., $\mu$ conditioned on $D$ is $\mu_D$)
and (2) for all agents $i$ and points $(r,m)$, there exists some cell $D$
such that $\R(\K_i(r,m)) \inter D \ne \emptyset$ and $\mu(D) > 0$.
It follows from these requirements and our assumption that 
that if 
$\R(\K_i(r,m)) \inter D \ne \emptyset$ 
then $\mu_D(\R(\K_i(r,m) \inter D) > 0$ that $\mu(\R(\K_i(r,m)) > 0$ for
all agents 
$i$ and points $(r,m)$.
We can thus associate $(\R,\D,\Delta)$ with
the set of run-based probability systems $(\R,\F_\D,\mu)$, for
$\mu \in \M(\Delta)$.  

Rather than defining secrecy in adversarial systems directly, we give a
slightly more general definition.  Define a {\em generalized run-based
probability system\/} to be a tuple $(\R,\F,\M)$, where $\M$ is a set of
probability measures on the $\sigma$-algebra $\F$.  Similarly, define a
{\em generalized probability system\/} to be a tuple $(\R,{\bf PR})$, where
${\bf PR}$ is a set of probability assignments.  We can define secrecy
in generalized (run-based) probability systems by considering secrecy
with respect to each probability measure/probability assignment.

\dfn
Agent $j$ maintains {\em probabilistic total (resp. synchronous)
secrecy\/} with respect to agent $i$ in the generalized probabilistic
system $(\R,{\bf PR})$ if, for all $\PR \in {\bf PR}$, 
$j$ maintains probabilistic total (resp. synchronous) secrecy with respect to
$i$ in $(\R,\PR)$.  Agent $j$ maintains {\em run-based
secrecy\/} with respect to agent $i$ in the generalized probabilistic
run-based system $(\R,\F,\M)$ if, for all $\mu \in \M$, 
$j$ maintains run-based probabilistic secrecy with respect to
$i$ in $(\R,\F,\mu)$.  
\edfn

It is now straightforward to define secrecy in an adversarial systems 
by reducing it to a generalized probabilistic system.
Agent $j$ maintains run-based probabilistic secrecy with respect to $i$
in $(\R,\D,\Delta)$ if $j$ maintains run-based probabilistic secrecy
with respect to $i$ in $(\R,\F_\D,\M(\Delta))$.  Similarly, agent $j$
maintains total (resp. synchronous) secrecy with respect to $i$ in
$(\R,\D,\Delta)$ if $j$ maintains total (resp. synchronous) secrecy with
respect to $i$ in $(\R,{\bf PR})$, where ${\bf PR}$ consists of all the
probability assignments determined by the run-based probability systems
$(\R,\F_\D,\mu)$ for $\mu \in \M(\Delta)$.  
A straightforward analogue of Proposition
\ref{pro:run_based_independence} holds for adversarial systems;
again, secrecy is symmetric in the presence of assumptions such as
perfect recall or synchrony.

\subsection{Secrecy and Evidence}\label{sec:evidential}

Secrecy in adversarial probability systems turns out to be closely
related to the notion of {\em evidence\/} in hypothesis testing
(see \cite{Kyburg83} for a good overview of the literature).
Consider this simple example: someone gives you a coin, which may be fair
or may
be double-headed. You have no idea what the probability is that the coin
is fair,
and it may be exceedingly unlikely that the coin is double-headed.
But suppose you then observe that the lands heads on each of 1,000
consecutive tosses.  Clearly this observation
provides strong {\em evidence} in favor of the coin being double headed.

In this example there are two hypotheses: that the coin is fair and
that it is double-headed.  Each hypothesis places a probability on the
space of observations.  In particular, the probability of seeing 1000
heads if the coin is fair 
is $1/2^{1000}$, and the 
probability of seeing
1000 heads if the coin is double-headed is 1.  While we can talk of an
observation being more or less likely with respect to each hypothesis,
making an observation does {\em not\/} tell us how likely an hypothesis
is.  No matter how many heads we see, we do not know the  probability
that the coin is double-headed unless we have the prior probability of
the coin being double headed.  In fact, a straightforward computation
using Bayes' Rule shows that if the prior probability of the coin being
double-headed is $\alpha$, then the probability of the coin being
double-headed after seeing 1000 heads is 
$\frac{\alpha}{\alpha + ((1-\alpha)/2^{1000}}$.

In an adversarial probability system $(\R,\D,\Delta)$, the initial
nondeterministic choice plays the role of an hypothesis.  For each
$D \in \D$, $\mu_D$ can be thought of as placing a probability on
observations, given that choice $D$ is made.  These observations then
give evidence about the choice made.  Agent $i$ does not obtain evidence
about the choices made if the probability of any sequence of observations
is the same for all choices.  

\commentout{
The same notion extends to the idea of secrecy. Even if a user has no idea
about the underlying probability space, some of his observations may provide
evidence in favor of a particular protocol choice for some other
user. It is precisely
this situation that secrecy aims to avoid. 
To make this precise, we must first
characterize the systems where these ideas are relevant.

Suppose that $(\R,\D,\Delta)$ is an adversarial probability system such
that
$\D = P \times Q$ is a countable partition of $\R$.
Intuitively, $P$ and $Q$ are the sets of protocols that
two agents $i$ and $j$ will use.
As before, $\Delta$ is a set of probability spaces, one for each partition.
Suppose also that the local states of
$i$ and
$j$ encode which cell in $\D$ they are currently in, i.e., which protocol
they have chosen to use. Formally, if $(r,m) \in \K_i(r',m')$, then there
exists $p \in P$ and $q,q' \in Q$ such that $r \in (p,q)$ and $r' \in (p,q')$.
(The requirement for $j$ is symmetric.)  To make the notation simpler, let
$\R(p)$ be the set of runs where $i$ uses protocol $p$, $\R(q)$ be the set of runs
where $j$ uses protocol $q$, and $\R(p,q) = \R(p) \inter \R(q)$.

We are interested in the case where $j$'s
protocol is kept secret from $i$. To make this precise, let $f$ be
a $j$-information function such that $f(r,m) = q$, where $r \in (p,q)$ for
some $p \in P$.
                                                                                                               
Given such a system, we can take advantage of the fact that there is a
single probability space $(\mu_{(p,q)},\F_{(p,q)}) \in \Delta$ to define secrecy
in terms of how $i$'s observations provide evidence for $j$'s use of
some particular protocol. The requirement is that any observation made
by $i$ is equally likely regardless of which protocol $j$ is using.
\dfn\label{dfn:evidential}
Agent $j$ maintains {\em evidential secrecy} with respect to $i$ 
in the adversarial probability system $(\R, P \times Q,\Delta)$ if, for
all $p \in P$, all $q, q' \in Q$, and all points $(r,m)$,
$$\mu_{(p,q)}(\R(\K_i(r,m)) \inter \R(p,q)) = \mu_{(p,q')}(\R(\K_i(r,m))
\inter \R(p,q')).$$
\edfn
                                                                                                               
\dfn\label{dfn:evidential}
Agent $j$ maintains {\em evidential secrecy} with respect to $i$ 
in the adversarial probability system $(\R, P \times Q,\Delta)$ if, for
all $p \in P$, all $q, q' \in Q$, and all points $(r,m)$,
$$\mu_{(p,q)}(\R(\K_i(r,m)) \inter \R(p,q)) = \mu_{(p,q')}(\R(\K_i(r,m))
\inter \R(p,q')).$$
\edfn

For a given observation $o$, the quantities $\mu_h(o)$, for different
hypotheses 
$h$, are often referred to as ``likelihood measures''. In cases where we may not
have an {\em a priori} distribution on which hypotheses are likely, likelihood
measures may be the best way to compare different hypotheses. Ideally,
of course, 
we'd like to be able to quantify the likelihood of a give hypotheses, given the
observation, but in the absence of prior probabilities we are often forced to
invert the probabilities to reason about the likelihood of the observation, 
given different hypotheses. In a formal sense, evidential secrecy requires that
these likelihood measures should provide no new information, regardless of the
observation made by an observer.

\begin{itemize}
\item $\mu( e \mid \R(p,q)) = \mu_{(p,q)}(e)$ for all measurable subsets
$e \subseteq \R(p,q)$.
\item $\mu( \R(p,q) ) > 0$ for all $p \in P, q, \in Q$.
\item $\mu( \R(q) | \R(p)$ = $\mu( \R(q) )$ for all $p \in P, q \in Q$.
\end{itemize}
                                                                                                               
As before, we want to consider the set of all possible measures consistent
with the
measures on individual cells of $\R$, but with the added restriction that
the choice
of protocols by $i$ and $j$ must be probabilistically independent. (It is
not the system's
fault if $i$ and $j$ meet up before using the system to decide which
protocols to use!)
                                                                                                               
\thm\label{thm:commented_out_evidential}
Let $(\R, P \times Q, \Delta)$ be an adversarial probability system that
is synchronous
or where both $i$ and $j$ have perfect recall. Agent $j$ maintains
evidential $f$-secrecy with respect to $i$ iff $j$ maintains
run-based
probabilistic $f$-secrecy with respect to $i$ for each measure $\mu \in
\M(\Delta)$.
\ethm

Evidential secrecy allows us to simplify the analysis and definition of secrecy when
nondeterministic choices matter.
Furthermore, it is a sensible definition even in cases where the
number of
protocols may be uncountable, because we need to deal only with
probability measures
within individual cells. In this sense, evidential secrecy may be seen as a generalization
of secrecy.
Finally, the connection between secrecy and evidential secrecy allows us
relate our work
to that of Gray and Syverson~\cite{gray98}, as discussed in the next
section.
}
\dfn\label{dfn:evidential}
Agent $i$ {\em obtains no evidence for the initial choice\/} in 
the adversarial probability system $(\R, \D,\Delta)$ if, for
all $D, D' \in \D$ and all points $(r,m)$ such that 
$\R(\K_i(r,m)) \inter D \ne \emptyset$ and $\R(\K_i(r,m)) \inter D' \ne
\emptyset$, we have 
$$\mu_{D}(\R(\K_i(r,m))) = \mu_{D'}(\R(\K_i(r,m))).$$
\edfn

\noindent Roughly speaking, $i$ obtains no evidence for initial choices
if the initial choices (other than $i$'s own choice) are all secret.  
The restriction to cells such
that $\R(\K_i(r,m)) \inter D \ne \emptyset$ and $\R(\K_i(r,m)) \inter D'
\ne \emptyset$  
ensures
that $D$ and $D'$ are both compatible with $i$'s
initial choice.  

To relate this notion to secrecy, 
we consider adversarial probability systems with a little more structure.
Suppose that for
each agent $i = 1, \ldots, n$, there is a set $\INIT_i$ of
possible initial choices.  (For example, $\INIT_i$ could consist
of a set of possible protocols or a set of possible initial inputs.)
Let $\INIT = \INIT_1 \times \cdots \times \INIT_n$ consist of all tuples
of initial choices.  For $y_i \in \INIT_i$, let $D_{y_i}$ consist of all
runs in $\R$ where agent $i$'s initial choice is $y_i$; if $y = (y_1,
\ldots, y_n) \in \INIT$, then $D_y = \inter_{i=1}^n D_{y_i}$ consists of 
all runs where the initial choices are characterized by $y$.
Let $\D = \{D_y: y \in \INIT\}$.  
To model the fact that $i$ is aware of his initial choice, we require that
for all points $(r,m)$ and agents $i$, there exists $y$ such that 
$\R(\K_i(r,m)) \subseteq D_y$.
If $\D$ has this form and each agent $i$ is aware of his initial choice,
we call $(\R,\D,\Delta)$ the adversarial system {\em determined by $\INIT$}.

If $i$ obtains no evidence for the initial choice, she cannot learn anything 
about the initial choices of other agents.  To make this precise in our
framework, 
let $\M^{\INIT}_i(\Delta)$ consist of the measures $\mu \in \M(\Delta)$
such that 
for all cells $D_{(y_1, \ldots, y_n)}$, we have
$\mu(D_{(y_1, \ldots, y_n)}) = \mu(D_{y_i}) \cdot
\mu(\inter_{j \ne i} D_{y_j}),$ i.e., such that the initial choices made
by agent $i$ are independent of the choices made by other agents.
Intuitively, if the choices of $i$ and the other agents are
correlated, $i$ learns something about the other agents' choices simply
by making 
his own choice. We want to rule out such situations. 
Note that because all the information sets have positive probability
(with respect to all $\mu \in \M(\Delta)$)
and, for all $i$, there exists an information set $\K_i(r,m)$ such that
$D_{y_i} \supseteq \R(\K_i(r,m))$, 
the sets $D_{y_i}$ must
also have positive probability. 
It follows that $\INIT$ and $\D$ must be countable.

Given $i$, let $i^-$ denote the ``group agent'' consisting of all agents
other than $i$.  (In particular, if the system consists of only two
agents, then $i^-$ is the agent other than $i$.)
The local state of $i^-$ is just the tuple of local
states of all the agents other than $i$.  
Let $f_{i^-}$ be the $i^-$-information function that maps a global state to
the tuple of $(i^-)$'s initial choice.  As we observed in
Section~\ref{sec:secrecy}, our definitions apply without change to new
agents that we ``create'' by identifying them with functions on global
states.  In particular, our definitions apply to $i^-$.

\thm\label{thm:evidential}
Let $(\R, \D, \Delta)$ be the adversarial probability system determined
by $\INIT$ and suppose that $\R$ 
is either synchronous or a system where 
$i$ has perfect recall.
Agent $i$ obtains no
evidence for the initial choice in $(\R,\D,\Delta)$ iff agent $i^-$ maintains
generalized run-based probabilistic $f_{i^-}$-secrecy with respect to $i$ 
in $(\R,\M^{\INIT}_i(\Delta))$.
\ethm

The assumption of either symmetry or perfect recall is necessary because the
proof relies on the symmetry of run-based secrecy (as established by
Proposition~\ref{pro:run_based_independence}). We do not need to assume
perfect recall for agent $i^-$ because the theorem deals with
$f_{i^-}$-secrecy and, on every run, $f_{i^-}$ is constant.
It therefore 
follows that the ``agent'' associated with $f_{i^-}$ (in the sense
described in Section~\ref{sec:secrecy}) has perfect recall even if $i^-$ does
not.

Thinking in terms of evidence is often simpler than thinking in terms of
run-based probabilistic secrecy.  This connection between evidence and
secrecy is particularly relevant when it comes to relating
our work to that of Gray and Syverson~\citeyear{gray98};
see Section~\ref{sec:protocols}.  
\section{Plausibilistic Secrecy}\label{sec:plaus}

So far, we have given definitions of 
secrecy for nonprobabilistic systems,
for
probability systems (where uncertainty is 
represented
by a single
probability measure), 
and for
generalized probability systems
(where uncertainty is represented by a set of probability
measures).  All of these definitions turn out to be special cases of
secrecy with respect to a general representation of uncertainty called a
{\em plausibility measure} \cite{FrH7,FrH5Full}.  
By giving a general
definition, we can cull 
out the essential features of all the definitions, as well as 
point
the way to defining notions of secrecy with respect to other
representations of uncertainty
that may be useful in practice.

Recall that a probability space is a tuple $(W,\F,\mu)$, where
$W$ is a set of worlds, $\F$ is an algebra of measurable subsets of $W$,
and $\mu$ maps sets in $\F$ to elements of $[0,1]$ such that the axioms of
probability are satisfied. A plausibility space is a direct generalization of
a probability space. We simply replace the probability measure $\mu$ with a
plausibility measure $\Pl$, which maps from sets in $\F$ to elements of an
arbitrary partially ordered set. If $\Pl(A) \leq \Pl(B)$, then $B$ is
at least as plausible as $A$. Formally, a plausibility space is a tuple
$(W,\F,\mb{D},\Pl)$, where $\mb{D}$ is a domain of plausibility values
partially ordered 
by a relation $\leq_\mb{D}$, and where $\Pl$ maps from sets in $\F$ 
to elements  of $\mb{D}$ 
in such a way that if $U \subseteq V$, then $\Pl(U) \le_\mb{D} \Pl(V)$.
We assume that $\mb{D}$ contains two special elements denoted 
$\top_\mb{D}$ and $\bot_\mb{D}$, 
such that $\Pl(W) = \top_\mb{D}$ and $\Pl(\emptyset) = \bot_\mb{D}$. 

As shown in \cite{FrH7,Hal31}, 
all standard representations of uncertainty can be viewed as instances of
plausibility measures.  We consider a few examples here that will be
relevant to our discussion:
\begin{itemize}
\item It is straightforward to see that 
a probability measure is a plausibility measure.
\item We can capture the notion of ``possibility'' using the trivial
plausibility measure
$\Pltriv$ that assigns the empty set plausibility 0
and all 
other sets 
plausibility 1.  That is, $\mb{D} = \{0,1\}$, $\Pltriv(\emptyset) = 0$, and
$\Pltriv(U) = 1$ if $U \ne \emptyset$. 
\item 
A set $\M$ of probability
measures on a space $W$ can be viewed as a single plausibility measure.
In the special case where $\M$ is a finite set, say $\M = \{\mu_1,
\ldots, \mu_n\}$, we can take $\mb{D}_{\M}$ to consist of $n$-tuples in
$[0,1]^n$, with the pointwise ordering, and define $\Pl_{\M}(U) = (\mu_1(U),
\ldots, \mu_n(U))$.  Clearly $\Pl_{\M}(\emptyset) = (0,\ldots, 0)$ and
$\Pl_{\M}(W) = (1,\ldots, 1)$, so $\bot_{\mb{D}_{\M}} = (0,\ldots,0)$ and 
$\top_{\mb{D}_{\M}} = (1,\ldots,1)$.  If $\M$ is infinite, we consider a
generalization of this approach.  Let $\mb{D}_{\M}$ 
consist of all functions from $\M$ to $[0,1]$. 
The pointwise order on functions
gives a partial order on $\mb{D}_{\M}$; thus, $\bot_{\mb{D}_{\M}}$ is the 
constant function 0, and $\top_{\mb{D}_{\M}}$ is the constant function 1.
Define the plausibility measure $\Pl_{\M}$ by taking $\Pl_{\M}(U)$ to be the
function $f_U$ such that $f_U(\mu) = \mu(U)$, for all $\mu \in \M$.
\end{itemize}
\commentout{
Clearly this definition generalizes probabilistic secrecy in both
probability systems.  It also generalizes secrecy in generalized
probability systems (since we can replace the set ${\bf PR}$ of
probability assignment by a single plausibility assignment such that
associates with each point $(r,m)$ and agent $i$ the plausibility
measure corresponding to the set of plausibility measures $\PR(r,m,i)$
for $\PR \in {\bf PR}$.  It also generalizes possibilistic secrecy,
since we can take $\PL(r,m,i)$ to be the trivial plausibility measure
with domain $\K_i(r,m)$ (so that a set gets plausibility 1 iff its
intersection with $\K_i(r,m)$ is nonempty, and otherwise gets
plausiblity 0).
}

We can define secrecy using plausiblity measures
by direct analogy with the probabilistic case.
Given a system $\R$, define a {\em plausibility assignment\/} $\PL$ on
$\R$ to be a 
function that assigns to each agent $i$ and point $(r,m)$  a
plausibility space $(\W_{r,m,i},\F_{r,m,i},\Pl_{r,m,i})$; define a
{\em plausiblity system\/} to be a pair $(\R,\PL)$, where $\PL$ is a
plausibility assignment on $\R$.  
We obtain definitions of total
plausibilistic secrecy and synchronous plausibilistic secrecy
by simply replacing ``probability''
by ``plausibility'' 
in Definitions \ref{dfn:totprobsec} and \ref{dfn:syncprobsec}.

Given a plausibility measure $\Pl$ on a system $\R$, 
we would like to define run-based plausibilistic secrecy and
repeat the Halpern-Tuttle construction to generate standard plausibilistic
systems. 
To do this, we need a notion of conditional plausibility. 
To motivate the definitions to come, we start by describing conditional
probability spaces. The essential idea behind conditional probability
spaces,
which go back to Popper~\citeyear{Popper68} and de Finetti~\citeyear{Finetti36},
is to treat conditional probability, rather than unconditional probability,
as the primitive notion. 
A {\em conditional probability measure\/} $\mu$ takes two arguments
$V$ and $U$; $\mu(V,U)$ is generally written $\mu(V \mid U)$. 
Formally, a {\em conditional probability space\/} is a tuple $(W,\F,\F',\mu)$
such that $\F$ is a $\sigma$-algebra over $W$, $\F'$ is a nonempty
subset of $\F$ that is closed under supersets 
in $\F$ (so that if $U \in \F'$, $U \subseteq V$, and $V \in \F$, then
$V \in \F$),
the domain of $\mu$ is
$\F \times \F'$, and the following conditions are satisfied:
\ul
\item $\mu( U \mid U) = 1$ if $U \in \F'$.
\item if $U \in \F'$ and $V_1, V_2, V_3, \ldots$ are pairwise disjoint elements of
$\F$, then $\mu(\union_{i=1}^{\infty} V_i \mid U ) = \sum_{i=1}^{\infty} \mu( V_i \mid U)$;
 \item $\mu( U_1 \inter U_2 \mid U_3 ) = \mu( U_1 \mid U_2 \inter U_3)
\cdot \mu(U_2 \mid U_3)$ 
if $U_1 \in \F$ and $U_2 \inter U_3 \in \F'$.
\eul
The first two requirements guarantee that, for each fixed $U \in \F'$,
$\mu(\cdot \mid U)$ is an unconditional probability measure.  
The last requirement guarantees that the various conditional probability
measures ``fit together''.  As is standard, we identify unconditional
probability with conditioning on the whole space, and write $\Pr(U)$ as
an abbreviation for $\Pr(U \mid W)$.

Given an unconditional probability space $(W,\F,\mu)$, we immediately
obtain a conditional probability 
space by taking $\F'$ to consist of all sets $U$ such that $\mu(U) \not= 0$
and defining conditional probability in the standard way.
 However, starting with 
conditional probability is more general in the sense that it is possible to extend an unconditional 
probability space to a conditional probability space where $\F'$ contains sets $U$ such that $\mu(U) = 0$.
In other words, there exist conditional probability spaces $(W,\F,\F',\mu)$ such that
$\mu( U \mid W ) = 0$ for some $U \in \F'$.  This generality is useful for reasoning about
secrecy, because (as we shall see) it is sometimes useful to be able to
 condition on sets that have 
a probability of 0. 
 up needing to assume 
To generalize conditional probability to the plausibilistic setting, we need
to define operators $\oplus$ and $\otimes$ that act as analogues of $+$ and $\times$
for probability; 
these operators 
add
useful algebraic structure to the plausibility spaces
we consider. We extend the notion of an {\em  algebraic plausibility
spaces} \cite{FrH7,Hal25,Hal31} 
to allow an analogue of countable additivity.
We briefly sketch the relevant details here.

A {\em countably-additive algebraic conditional plausibility space\/}
(cacps)
is a tuple 
$(W,\F,\F',\Pl)$ such that
\begin{itemize}
\item $\F$ is a $\sigma$-algebra of subsets of $W$;
\item $\F'$ is a nonempty subset of $\F$ that is
closed under supersets
in $\F$;
\item there is a partially-ordered domain $\mb{D}$ such that, for each $V \in
\F'$, $\Pl(\cdot \mid V)$ is a 
plausibility measure on $(W,\F)$ with range $\mb{D}$  (so, intuitively, the
events in $\F'$ are the ones for which conditioning is defined); and
\item there are functions  $\oplus: \mb{D}^{\infty} \rightarrow \mb{D}$ and
$\otimes: \mb{D} \times \mb{D} \rightarrow \mb{D}$ such that: 
\ul
\item if $U \in \F'$,  $V_1, V_2, \ldots$, 
are pairwise disjoint elements of $\F$, 
and $J$ is some subset of $\{1, 2, 3, \ldots$ such that $\Pl(V_i) =
\bot$ for $i \in J$,   
then
$$\Pl( \union_{i=1}^{\infty} (V_i \mid U ) =
\oplus_{i=1}^\infty \Pl(V_i \mid U)
\oplus_{{i \notin J}} \Pl(V_i \mid U).$$ 
\item if $U_1, U_2, U_3 \in \F$ and $U_2 \cap U_3 \in \F'$, then
$$\Pl( U_1 \cap U_2 \mid U_3) = \Pl(U_1 \mid U_2 \cap U_3) \otimes \Pl(
U_2 \mid U_3).$$ 
\item $\otimes$ distributes over $\oplus$, more precisely, 
$a \otimes (\oplus_{i=1}^\infty b_i) = \oplus_{i=1}^\infty (a \otimes
b_i)$ if $(a,b_i), (a,\oplus_{i=1}^\infty b_i) \in
\Dom(\otimes)$ and $(b_1,b_2, \ldots ), (a \otimes b_1, a \otimes b_2 ,
\ldots ) \in \Dom(\oplus),$ where $\Dom(\oplus) = \{(\Pl(V_1\mid U),
\Pl(V_2 \mid U), \ldots): 
V_1, V_2,  \ldots \in \F$ are pairwise disjoint and $U \in \F'\}$
and $\Dom(\otimes) = \{(\Pl(U_1 \mid U_2 \inter U_3),\Pl(U_2 \mid U_3)):
U_2 \inter U_3
\in F', \, U_1, U_2, U_3 \in \F\}$.  (The reason that this property is
required only for 
tuples in $\Dom(\oplus)$ and $\Dom(\otimes)$ is discussed shortly.)
\item if $(a,c), \, (b,c) \in \Dom(\otimes)$, 
$a \otimes c \leq b \otimes c$, and $c \not= \bot$, then $a \leq b$.
\commentout{
\item if $(d_1, d_2, d_3, \ldots), (d_1', d_2', d_3', \ldots)  \in
\Dom(\oplus)$, $\oplus_{i=1}^n d_i \le d'' \le (\oplus_{i=1}^n d_i) \oplus
(\oplus_{i=n+1}^\infty d_i'$ for all $n$, then $d'' =
\oplus_{i=1}^\infty d_i$.
\item if  $(d_1, d_2, \ldots), \, (d_1', d_2', \ldots) \in \Dom(\oplus)$
and $d_i \le d_i'$ for $i = 1, 2, \ldots$, then  
$\oplus_{i=1}^\infty d_i \le \oplus_{i=1}^\infty d_i'$.
}
\eul
\end{itemize}
To understand the reason for the restriction to $\Dom(\oplus)$ and
$\Dom(\otimes)$, consider probability.  In that case, $D$ is $[0,1]$,
and we take $\oplus_{i=1}^\infty b_i $ to be $\max(\sum_{i=1}^\infty b_1,1)$.  It is not too hard to
show that the distributive property does not hold in general if
$\sum_{i=1}^\infty b_i > 1$ (consider, for example $a = 1/2$, $b_1 = b_2
= 2/3$, and $b_i = 0$ for $i \ge 3$); however, it does hold if
$\sum_{i=1}^\infty b_i \le 1$, a property 
that
is guaranteed to hold
if there exist 
sets $V_1, V_2, \ldots$ that are pairwise disjoint and a set $U$ such
that $b_i = \mu(V_i \mid U)$ for some probability measure $\mu$.

It can be shown (see \cite{Hal25,Hal31}) that
the constraints on cacps's imply that $\bot$ acts as an identity element for
$\oplus$ and that $\top$ acts as an identity element for $\otimes$, just
as
$0$ and $1$ do for addition and multiplication, 
as long as we
restrict to tuples in $\Dom(\oplus)$ and $\Dom(\otimes)$, respectively,
which is all we care about in our proofs. 
The constraints also imply
that $\Pl(U \mid U) = \top$ for $U \in \F$.
All the plausibility measures we considered earlier can be viewed as examples
of cacps's. 
First, the trivial plausibility
measure $\Pltriv$ is a cacps if we take $\oplus$ to be $\max$ and $\otimes$ 
to be $\min$.
A conditional probability space (as just defined) is a cacps simply 
by defining $\oplus$ as above, so that $\oplus_{i=1}^\infty b_i =
\max(\sum_{i=1}^\infty b_i,1)$, and taking $\otimes$ to be
multiplication. 
If we have a set $\M$ of probability measures on a space $W$, we can 
construct a conditional plausibility measure $\Pl_{\M}$ in essentially the same way that we constructed
an unconditional plausibility measure from the set $\M$, so that $\Pl_{\M}( V \mid U)$
is the function $f_{V \mid U}$ from measures in $\M$ to $[0,1]$ such
that 
$f_{V \mid U}(\mu) = \mu( V \mid U)$ if $\mu(U) \ne 0$, and
$f_{V \mid U}(\mu) = *$, where $*$ is a special ``undefined'' value,
if $\mu(U) = 0$.
To get a cacps, we simply define $\oplus$ and $\otimes$ pointwise (so
that, for example, $f \oplus g$ is that function such that 
$(f \oplus g)(\mu) = f(\mu) \oplus g(\mu)$).
There are subtleties involved in defining the set $\F'$ on which
conditioning is defined---in particular, 
care is required when dealing with sets $U$ such that $\mu(U) > 0$ for
some, but not all, of the 
measures in $\M$. These issues do not affect the results of this paper
because we assume that the 
information sets on which we condition have positive probability, so we
ignore them here. 
See Halpern~\citeyear{Hal31} for more details.

\commentout{

Given a 
domain
$D$, suppose that we have a notion of a limit, such
if a sequence $d_1, d_2, \ldots$ is convergent with limit $d$, we write 
$d = \lim_{n \rightarrow \infty} d_n$. (Convergence may arise in a metric space, or, for example,
if we consider pointwise convergence of a sequence of functions whose range is a metric space.)
Now, suppose that the domain $D$ of a plausibility measure $\Pl$ is such a space, and that
$\otimes$ distributes over limits, i.e., 
$$d \otimes \lim_{n \rightarrow \infty} d_n = \lim_{n \rightarrow \infty} d \otimes d_n.$$
If a sequence $d_1, d_2, \ldots$ is convergent, with limit $d$, we write 
$d = \lim_{n \rightarrow \infty} d_n$. Given a sequence $d_1, d_2, \ldots$, if the
sequence of partial sums $\bigoplus_{i=1}^{n} d_i$ is convergent, with limit $d$,
we write $d = \bigoplus_{i=1}^{\infty} d_i$. (Note that $\otimes$ distributes over the
infinite sum if it distributes over finite sums.) These definitions are useful
when dealing with an infinite plausibility space. 
We say that a plausibility space is {\em countably algebraic} if it is algebraic,
if its domain $D$ admits limits over which $\otimes$ distributes, and if whenever
$\Pl(U) \not= \bot$ and $V_1, V_2, \ldots$ is countable sequence
of pairwise disjoint elements of $\F$, we have
$$ \Pl( \bigcup_{i=1}^{\infty} V_i \mid U) = \bigoplus_{i=1}^{\infty} \Pl( V_i \mid U). $$ 

\commentout{
A {\em conditional plausibility space (cps)\/} is a tuple $(W, \F,\F',\Pl)$,
where $\Pl(U \mid V) $ is defined if $U \in \F$ and $V \in \F'$.  ($\F$
and $\F'$ must satisfy certain properties, but for the purposes of this
discussion, we can take $\F$ to be an arbitrary algebra and $\F'$ to
consist of all $U \in \F$ such that $\Pl(U) \ne bot$.)  For each fixed
$V \in \F'$, $\Pl(\cdot \mid V)$ is a plausiblity measure.  A cps is
{\em algebraic\/} if there are function $\oplus$ and $\otimes$ such
that
\begin{itemize}
\item $\Pl(V \union V' \mid U) = \Pl(V \mid U) \oplus \Pl(V' \mid U)$ if
$V \inter V' = \emptyset$ and 
\item $\Pl(U_1 \inter U_2 \mid U_3) = \Pl(U_1 \mid U_2 \inter U_3) 
\otimes \Pl(U_2 \mid U_3)$ if $U_2 \inter U_3 \in \F',$ $U_1, U_2, U_3
\in \F$;
\item $\otimes$ distributes over $\oplus$
\item if $a \otimes c \le b \otimes c$, and $c \ne \bot$, then $a \le b$.
\end{itemize}
}

It is essentially trivial to define a countably algebraic plausibility
space (CAPS) 
corresponding to the trivial plausiblity measure: $\Pltriv(V|U)$ is 1 if
$V \inter U \ne \emptyset$ and 0 otherwise.  This also gives an algebraic
plausibility measure, with $\oplus = \max$ and $\otimes = \times$.
The extension to a countably algebraic plausibility measure follows easily by taking
limits as if 0 and 1 were embedded in the standard real metric space.

We can also define a CAPS
corresponding to a set $\M$ of probability measures: $\Pl_{\M}(\cdot \mid U)$, 
is just the plausibility measure corresponding to the set of
conditional probability measures.
(Some
care must be taken in
the case that $U$ has probability 0 with
respect to some---but not all---of the probability measures in $\M$; again,
see \cite{Hal31} for details.)  In this case, 
$\oplus$, $\otimes$, and convergence are all defined pointwise.
}
Define a {\em run-based plausibility system\/} to be a cacps
$(\R,\F,\F',\Pl)$.  
Instead of requiring that $\mu(\R(\K_i(r,m))) > 0$ as in the probabilistic
case, we now require that $\R(\K_i(r,m)) \in \F'$ for all agents $i$ and
points $(r,m)$.  
This
requirement guarantees that conditioning on 
$\R(\K_i(r,m))$ is defined, but is 
easier to work with than the
requirement that $\mu(\R(\K_i(r,m))) > 0$.  We can now
repeat the Halpern-Tuttle construction to generate standard 
plausibilistic systems. 
With this construction, we can explain how the results of 
Sections~\ref{sec:define_prob_sec},~\ref{sec:add_prob}, 
and~\ref{sec:probsecchar} carry over to the 
more general plausibilistic setting. In general, the results extend 
by replacing $+$ and $\times$ consistently in the proofs by $\oplus$
and $\otimes$, but some care is required. We summarize the details here without
stating them as formal results; a technical discussion appears in the appendix.
\ul
\item Proposition \ref{pro:sync_prob_sec} generalizes to
run-based plausibility systems. 
\item Theorems \ref{thm:sync_prob_syntax} and \ref{thm:runbased_prob_syntax} 
carry over to the plausibilistic setting (with essentially the same
proofs) once we define a language for reasoning about plausibility
analogous to the language for reasoning about probability, 
with formulas of the form $\Pl_i(\phi) = \alpha$.
\item Proposition \ref{pro:sync_independence} generalizes, given a
common prior  
$\Pl_{\mathit{cp}}$, provided that $\otimes$ is commutative. 
For total secrecy we require that for all points $(r,m)$ we have
$\Pl_{\mathit{cp}}(\K_i(r,m) \mid \PT(\R)) \ne \bot$ and 
$\Pl_{\mathit{cp}}(\K_j(r,m) \mid \PT(\R)) \ne \bot$; 
similarly, for synchronous secrecy we
require that for all points we have
$\Pl_{\mathit{cp}}(\K_i(r,m) \mid \PT(m)) \ne \bot$ and 
$\Pl_{\mathit{cp}}(\K_j(r,m) \mid \PT(m)) \ne \bot$.
\item Proposition \ref{pro:run_based_independence} generalizes provided
that $\otimes$ 
is commutative and that for all points $(r,m)$ we have
$\Pl(\R(\K_i(r,m)) \mid \R) \ne \bot$ and $\Pl(\R(\K_j(r,m)) \mid \R) \ne \bot$.
\item Theorem \ref{thm:evidential} can be extended once define
adversarial plausibility systems appropriately.
\eul

\commentout{
\noindent {\bf \large Note:} In order for the proof to work, there are several subtle details that
need to be worked out. I'm putting them here so that we can keep the details around until
the paper's finished. 
\ul
\item In order to use Proposition \ref{pro:plaus_independence} and Corollary \ref{cor:plaus_symmetry},
we require all the cells to be in $\F'$. This is {\em not} a strict generalization of the probabilistic case,
where cells may have zero probability. There are two options for dealing with this. We could just assume
that cells are in $\F'$ as part of the definition of $\M_i$ above -- I prefer this option and
I've instituted it above.
The other option is to rewrite Proposition \ref{pro:plaus_independence} to make it more general. To do that we need to
assume that the plausibility measure is ``acceptable'' so that cells not in $\F'$ have plausibility of $\bot$.
We also need to carefully check that we can pull out $\bot$-terms from a countable $\oplus$ summation, and 
rewrite the proof so that all these details are taken care of. (Information sets are in $\F'$ by assumption;
this properly mimics the probabilistic case.)
\item For the forward direction of the proof of 4.13 we also need the assumption that 
if $\R(K_i(r,m)) \subseteq D_{y_i}$, then
$\Pl( \R(\K_i(r,m)) \mid D_{y_i} ) \not= \bot$. (This is necessary for the symmetry lemmas.)
We could also assume this directly, but I think we get more mileage out of following the probabilistic case,
so that's what I've done for now.

Note that in the probabilistic case we require that there exists a cell $D$ such that
$\mu(D) > 0$ and $\R(\K_i(r,m)) \inter D \not= \emptyset$, and that if the latter condition holds
$\mu_D(\R(\K_i(r,m)) \inter D) > 0$. If we follow that approach here and require that
$\Pl_D( \R(\K_i(r,m)) \inter D ) \not= \bot$ when 
that set is nonempty for some cell $D$ such that $\Pl(D) \not= \bot$, then we get the result.
(See appendix for details.)
\eul
{\bf \large (End of Note)}
} %

\commentout{
Then we have the following generalization of Theorem~\ref{thm:evidential}.

\thm\label{thm:plaus_evidential}
Let $(\R, \D, \Delta)$ be the adversarial plausibility system determined
by $\INIT$ and suppose that $\R$ 
is either synchronous or a system where $i$ has perfect recall.
Agent $i$ obtains no (plausibilistic)
evidence for initial choice in $(\R,\D,\Delta)$ iff agent 
$i^-$ maintains generalized run-based plausibilistic $f_{i^-}$-secrecy with
respect to $i$  in $(\R,\M^{\INIT}_i(\Delta))$.
\ethm
} %

These results demonstrate
the essential unity of our definitions and theorems
in the probabilistic and 
nonprobabilistic 
cases, and suggest further
generalizations.  In particular,
it may 
be worthwhile to
consider definitions of secrecy that use representations of uncertainty
that are based on representations of uncertainty that are  more
qualitative than probability.
For example, in the cosmic-ray example, we might consider a measure with
three degrees of likelihood: ``impossible'', ``possible but very
unlikely'', ``possible and 
likely''. 
This would let us handle examples where strange, unlikely things might happen,
while maintaining the simplicity of the nonprobabilistic definitions presented
in Section~\ref{sec:secrecy}.

\section{Related Work}\label{sec:rw}

We are certainly not the first to discuss formal definitions of
\commentout{
secrecy. Indeed, a multitude of related definitions have been proposed
over the last two two or three decades. 
There are several reasons for the abundance of definitions. Two in
particular are worth pointing out here. One reason is that it has been
secrecy; many definitions have been proposed over the last two decades. 
particular are worth noting. One reason is that it has been difficult to
capture a suitable definition of secrecy that adequately restricts information
flow without being unreasonably strong. (Our definitions of total secrecy, for
example, are very ``secure'', but totally unreasonable for many systems that
exhibit some degree of synchrony.) Another reason is that early definitions of 
security were expanded and modified to facilitate verification (e.g., through unwinding 
conditions), and to ensure that properties are composable in the sense that two secure
systems, when composed, form a secure system.

Some of these definitions have been important technical milestones in the search 
for properties that are verifiable and composable. However, we believe that some
of the definitions of noninterference-style properties that have
been proposed are problematic. In particular, we believe
that such definitions are sometimes insufficient {\em as a basic notion of secrecy}.
While a property such as forward correctability is undoubtedly useful if
a system designer is interested in a trace-based security property that is maintained
through system composition, we do not believe that it captures a natural intuition of
what secrecy means. Indeed, in the rush to come up with an ``ideal'' security property
that is composable, that facilitates verification, and that restricts information flow in a
reasonable way, the computer security community seems to have overlooked the task of
deciding what secrecy means in the abstract. It has been difficult for the security community
to agree on an ideal definition of secrecy even in the very restricted case where
a system designer is not worried about issues of timing and
probability. When those issues do come  
into play, we believe that it is crucial to start with formal definitions that obviously capture
intuitions of knowledge and secrecy, 
and we hope that our work makes a serious contribution to this effort.

That said, we have absolutely no argument with technical definitions that satisfy properties such
as composability. It may be useful to draw analogy with inductive proofs.
When one is trying to prove a proposition inductively, it is
often helpful to start with a stronger inductive hypothesis that allows the inductive
step to work. The stronger inductive hypothesis is useful because it facilitates the
proof and because it {\em implies} the proposition we want to prove. In the same way,
we believe that a property like forward correctability is useful because it implies
secrecy and because two forward-correctable systems, when composed, 
also satisfy forward-correctability
(and thus secrecy), {\em not} because we believe that the property is a particularly elegant
definition of secrecy. To quote McCullough \citeyear{87}: ``we take [deducibility security]
to be our base notion of security; whatever other definitions of security we
come up with, it must imply deducibility security.'' 
} %
secrecy: many definitions have been proposed over the last two decades. 
One reason for this is that researchers have sought an ``ideal'' definition
of security that is, for example, easy to verify and composable (in the sense
that if two systems maintain secrecy then their composition does too).
While we certainly agree that composability and verifiability are 
important properties, we believe that
the intuition behind secrecy should be isolated from stronger 
properties that 
happen to
imply secrecy--especially when we have to worry about subtle issues such as
probability and nondeterminism.
In this section we consider how our definitions relate to other
attempts to define information-flow conditions.
We show in particular how it can capture
work that has been done in the synchronous setting, the asynchronous
setting, and the probabilistic setting.  
Because there are literally 
dozens of papers that have, in one way or
another, defined notions of secrecy or privacy, this section is in no
way meant to be comprehensive or representative. 
Rather, we have chosen
examples that inspired our definitions, or examples for which our
definitions give some insight. 
In light of our earlier comments, we also focus on definitions that have
tried to capture the essence of secrecy, rather than notions that have
been more concerned with 
issues like composability and verification.

One important strand of literature to which we do not compare our work
directly here is the work on defining information flow and noninterference
using process algebras related to CCS and CSP; see, for example,
\cite{focardi94,focardi01,ryan99,ryan00}.  
\commentout{
Much recent work dealing with information flow and noninterference
has focused on systems specified using process algebras related to
CCS~\cite{focardi94} and 
CSP~\cite{ryan99}, including papers by Focardi and Gorrieri~\citeyear{focardi01} and Ryan,
Schneider, Goldsmith, Lowe, and Roscoe~\citeyear{ryan00}. Process-algebraic definitions have
several important advantages. They provide an elegant way of describing distributed systems
compactly and compositionally. Furthermore, a wide variety of process equivalences have been
proposed to characterize differences in system behavior
(see
van Glabeek~\citeyear{vanglabbeek01} 
for a survey) and verification tools based on
process equivalence can 
be used to decide whether a system meets its specification.
Unfortunately, space precludes a formal discussion of process-algebraic
approaches to secrecy 
definitions.}
We believe that these definitions too are often best understood in terms of
how they capture our notions of secrecy.  However, a careful discussion
of this issue would take us too far afield.
In future work we plan to consider the issue in detail, by
describing how 
processes can be translated to the runs and systems framework 
in a way that captures their semantics and then
showing how some of the 
process-algebraic definitions can be recast as examples of secrecy. 
In \cite{HalOn03} we give one instance of such a translation: we
show how definitions of anonymity given using CSP by
Schneider and Sidiropoulos \citeyear{schneider96} can be captured in the
runs and systems framework.  

\subsection{Secrecy in Trace Systems}\label{sec:sts}

Many papers in computer security define notions of secrecy 
(often referred to as ``noninterference'')
using using trace-based models. Traces are usually defined as 
sequences of input and output events, where each event
is associated with some agent
(either as an input that she provides or an output that she sees). 
However, there have been some subtle differences among
the trace-based models. In some cases, infinite
traces are used; in others, only finite traces.  In addition,
some models assume that the underlying systems are synchronous,
while others implicitly assume asynchrony.
Although ``asynchronous'' system models have been more common, we first
consider synchronous trace-based systems. 

Both McLean~\citeyear{mclean94} and 
Wittbold and Johnson~\citeyear{Wittbold&Johnson} present 
their definitions of security in the context of synchronous
input/output traces. These
traces are essentially restricted versions of the runs introduced
in this paper. Here we consider a slightly simplified version of 
McLean's framework and describe two
well-known noninterference properties within the framework.

Let $LI$ and $HI$ be finite sets of high-level and low-level input variables,
and let $LO$ and $HO$ be finite sets of high-level and low-level
output variables. We assume that these sets are pairwise disjoint.
A tuple $t = \langle l_i,h_i,l_o,h_o\rangle$ 
(with $l_i \in LI, h_i \in HI$, $l_o \in LO$, and $h_o \in HO$) 
represents a snapshot of a system at a given
point in time; it describes the input provided
to the system by a low-level agent $L$ and a high-level agent $H$, 
and the output sent by the system to $L$ and $H$.
A {\em synchronous trace\/} $\tau = \langle t_1,
t_2,\ldots\rangle$ is a sequence of such tuples.  
It represents an infinite
execution sequence
of the entire system by describing the input/output behavior of the
system at 
any given point in time.%
\footnote{The traces are said to be synchronous because the input and
output values 
are specified for each agent at each time step, and both agents
can infer the time simply by looking at the number of system outputs
they have seen.}
A {\em synchronous trace system\/}
is a set $\Sigma$ of synchronous traces, representing
the possible execution sequences of the system.

In a synchronous trace system, the local state 
of an agent can be defined using a
{\em trace projection function}. 
For example, let 
$|_L$
be the
function projecting $\tau$ onto the low-level input/output
behavior of $\tau$, so that if 
$$
\tau = \langle \langle l_i^{(1)}, h_i^{(1)}, l_o^{(1)}, h_o^{(1)}\rangle,
     \langle l_i^{(2)}, h_i^{(2)}, l_o^{(2)}, h_o^{(2)}\rangle, \ldots \rangle,
$$
then
\[ \tau |_L = \langle \langle l_i^{(1)}, l_o^{(1)}\rangle, 
	          \langle l_i^{(2)}, l_o^{(2)}\rangle, \ldots\rangle. \]
Similarly, we can define a function $|_H$ that extracts 
high-level input/output traces and a function $|_{HI}$ that extracts
just high-level {\em input} traces.

Given a trace $\tau  = \langle t_1, t_2, \ldots \rangle$, 
the {\em length $k$ prefix\/} of $\tau$ is 
$\tau_k = \langle t_1, t_2,
\ldots, t_k \rangle$,
i.e., the finite sequence containing the first $k$ state tuples of the
trace $\tau$. Trace projection functions apply to trace prefixes in the
obvious way.

It is easy to see that synchronous trace systems can be viewed as
systems in the multiagent systems framework.
Given a trace $\tau$, we can define the run $r^{\tau}$ such that 
$r^{\tau}(m) = (\tau_m |_L, \tau_m |_H )$.  
(For simplicity, we have omitted the environment state from the global
state in this construction, since it plays no role.)
Given a synchronous trace system $\Sigma$, let 
$\R(\Sigma) = \{r^{\tau}: \tau \in \Sigma\}$. It is easy to check that 
$R(\Sigma)$ is synchronous, and that
both agents $L$ and $H$ have 
perfect recall.

McLean defines a number of notions of secrecy in his framework. 
We consider two of the best known here: {\em separability}
\cite{mclean94}  and {\em
generalized noninterference}~\cite{mccullough87}.
Separability, as its
name suggests, ensures
secrecy between the low-level and high-level agents, whereas
generalized noninterference
ensures that the low-level agent is unable to know anything about
high-level input
behavior.

\dfn
\label{dfn:sep}
A synchronous trace system $\Sigma$ satisfies {\em separability} if, for
every pair of traces  
$\tau, \tau' \in \Sigma$, there exists a trace $\tau'' \in \Sigma$ such
that
$\tau'' |_L = \tau |_L$ and $\tau'' |_H = \tau' |_H$.
\edfn

\dfn
A synchronous trace system $\Sigma$ satisfies {\em generalized
noninterference} if, for every pair of traces  
$\tau, \tau' \in \Sigma$, there exists a trace $\tau'' \in \Sigma$ such
that
$\tau'' |_L = \tau |_L$ and $\tau'' |_{HI} = \tau' |_{HI}$.
\edfn

These definitions are both special cases of nondeducibility, 
as
discussed in
Section \ref{sec:secrecy}: take the set of worlds $W$ to be $\Sigma$, 
the information function $g$ to be $|_L$, and the information
function $h$ to be 
$|_H$ (for separability) and $|_{HI}$ (for generalized 
noninterference).%
\footnote{Actually, it is not difficult to see that if
the information functions $g$ and $h$ 
are restricted to trace projection functions, then nondeducibility is
essentially equivalent in expressive power to {\em selective interleaving
functions}, the 
mechanism for defining security properties introduced by McLean 
\citeyear{mclean94}.
}
In our framework,
separability essentially corresponds to synchronous secrecy, whereas
generalized noninterference corresponds to 
synchronous $|_{HI}$-secrecy.
The following proposition makes this precise.
Let $f_{hi}$ be the information function that
extracts a high-level input trace prefix from a point in exactly the
same way that $|_{HI}$ extracts it from the infinite trace.

\pro
\label{pro:sep->sync-sep}
If a synchronous trace system $\Sigma$ satisfies separability
(resp., generalized noninterference), 
then 
$H$ maintains synchronous secrecy 
(resp., synchronous $f_{hi}$-secrecy)
with respect to $L$  in $\R(\Sigma)$.
\epro

\prf
We prove the result for separability.  The proof for generalized
noninterference is similar and left to the reader.
Suppose that 
$\Sigma$ satisfies separability.
Let $r^{\tau}$ and $r^{\tau'}$ be runs in $\R(\Sigma)$.  We want to show
that, for all times $m$, we have that $\K_L(r^{\tau},m) \inter
\K_H(r^{\tau'},m) \ne \emptyset$.
Since $\sigma$ satisfies separability, there exists a trace $\tau'' \in
\Sigma$ such that  
$\tau'' |_L = \tau |_L$ and $\tau'' |_H = \tau' |_H$. It
follows immediately 
that $\tau''_m |_L = \tau_m |_L$ and $\tau''_m |_H = \tau'_m |_H$. 
Thus, $(r^{\tau''},m) \in \K_L(r^{\tau},m) \inter \K_H(r^{\tau'},m)$.
\eprf

The converse to Proposition~\ref{pro:sep->sync-sep} is not quite true.
There is a subtle but significant difference between McLean's framework
and ours. McLean works with {\em infinite\/} traces;
separability and generalized noninterference are defined 
with respect to {\em traces} rather than sets of {\em points} (i.e., trace
prefixes). 
To see the impact of this,
consider a system $\Sigma$ where the high-level agent inputs either
infinitely
many 0's or infinitely many 1's. The output to the low-level agent is
always finitely many 0's followed by infinitely 1's, except for a single
trace where the high-level agent inputs infinitely many 0's and the
low-level agent inputs infinitely many 0's. 
Thus, the system consists of the following traces, where we have omitted
the low-level 
inputs since they do not matter,
and the high-level outputs, which can taken to be constant:
$$\begin{array}{ll}
(0^k 1^{\infty}, 0^{\infty}), \ \ k=0, 1, 2, 3, \ldots\\ 
(0^k 1^{\infty}, 1^{\infty}), \ \ k=0, 1, 2, 3, \ldots\\ 
(0^{\infty}, 0^{\infty}).
\end{array}$$
In the system $\R(\Sigma)$, $H$ maintains synchronous secrecy and thus
synchronous $f_{hi}$-secrecy with respect to $L$, 
because by looking at any finite trace
prefix, 
$L$
cannot tell whether the high-level
inputs have been 0's or 1's. 
However, 
$\Sigma$ does not satisfy separability or generalized interference.
If 
$L$
``sees'' infinitely many 0's, he immediately knows that the 
high-level inputs have been 0's. 
This seems unreasonable.
After all, agents only makes observations at finite points in time.  

Note that if $\tau$ is a trace where the low-level outputs
are all 0's and the high-level inputs are all 1's, each finite
prefix of the trace $\tau$ is a prefix of a trace in $\Sigma$,
even though $\tau$ is not in $\Sigma$.
This turns out to be the key reason that the system
satisfies synchronous secrecy but not separability.

\dfn A synchronous trace system $\Sigma$ is {\em limit closed\/}
\cite{Emerson} if, for all synchronous traces $\tau$, we have 
$\tau \in \Sigma$ 
iff for every time $k$ there exists a trace $\tau' \in
\Sigma$ such that $\tau'_k = \tau_k$. 
\edfn

Under the assumption of limit closure, we do get the converse to
Proposition~\ref{pro:sep->sync-sep}.

\pro
\label{pro:sync-sep->sep}
A limit-closed synchronous trace system $\Sigma$ satisfies separability
(resp. generalized noninterference)
iff
$H$ maintains synchronous secrecy 
(resp., synchronous $f_{hi}$-secrecy)
with respect to $L$ in $\R(\Sigma)$.
\epro

While we believe that it is unreasonable in general to assume that an agent's
view includes the entire run (as McLean's definitions implicitly do),
these results nonetheless demonstrate the close connection between our
definition of  
synchronous $f$-secrecy and security properties such as separability and
generalized noninterference.
Up to now we have considered a synchronous trace model, where the input
and output events of high and low users occur in lockstep. 
However, many
trace-based definitions of security are given in an asynchronous setting. 
We consider a number of definitions of secrecy in this setting.
For uniformity we use the terminology 
of Mantel~\citeyear{mantelthesis03}, who has carefully compiled a variety
of well-known trace-based properties into a single framework.

In Mantel's framework, traces are not infinite sequences of 
input/output event {\em tuples}, but finite sequences 
of input/output events. For example, 
if $l$ and $l'$ are low-level events while  
$h$ and $h'$ are high-level events, a possible system trace could
be 
\[ \tau = \langle l, h, l, h', h', l', l', l, h \rangle. \]
As with synchronous trace systems, we denote a projection function
for a set $A$ by $|_A$.
Thus, if $\tau$ is defined as above, we have
\[ \tau |_L = \langle l, l, l', l', l \rangle, \]
where $|_L$ is the low-level projection function.
Note that because asynchronous traces are sequences of events, rather than
tuples, the low-level projection function ignores high-level events altogether.
This means that a low-level view of the system may remain completely unchanged even as 
many, many high-level input events occur.

An {\em asynchronous trace system\/} is a set of traces that is closed
under trace prefixes.
There is a straightforward way of associating with each system a set of
runs.  A set $T$ of traces is {\em run-like\/} if, for all traces
$\tau_1$ and $\tau_2$ in $T$, either $\tau_1$ is a prefix of $\tau_2$ or
$\tau_2$ is a prefix of $\tau_1$.  Intuitively, a run corresponds to a
maximal 
run-like set of traces.
More formally, let $T$ be a maximal set
of run-like traces.  Note that if $T$ is infinite, then for all $n \ge 0$
there exists exactly one trace in $T$ of length $n$ (where the length of
$\<t_0, \ldots, t_{n-1}\>$ is $n$); if $T$ is finite, then there is some $N
\ge 0$ such that $T$ has exactly one trace of length $n$ for all $n \le N$.
If $T$ is infinite, let the run $r^T$ be such that 
$r^T(m) = \<\tau^m|_L, \tau^m|_H\>$, where $\tau^m$ is the unique
trace in $T$ of length $m$.
If $T$ is finite, let $r^T$ be such that
$r^T(m) = \<\tau^m|_L, \tau^m|_H\>$ if $m \le N$, where
$N$ is the length of the longest trace in $T$, and $r^T(m) = r^T(N)$ if
$m \ge N$; that is, the final state repeats forever.  Given an
asynchronous trace 
system
$\Sigma$, let $\R(\Sigma)$ denote the set of
all runs of the form $r^T$, where $T$ is a maximal set of run-like
traces in $\Sigma$.

Trace-based security properties are usually expressed as closure
properties on sets of traces, much like our possibilistic definitions of
secrecy;  
see \cite{Mantel00} for more details.
We focus here on definitions of asynchronous separability and
generalized noninterference, given by Zakinthinos and Lee \citeyear{ZL97}.
\dfn
An asynchronous trace system $\Sigma$ satisfies {\em asynchronous separability} if, for
all traces $\tau, \tau' \in \Sigma$, if $\tau''$ is a trace that results
from an arbitrary interleaving of the traces $\tau |_L$ and $\tau' |_H$,
then $\tau'' \in \Sigma$.
\edfn
The definition of generalized noninterference is slightly more complicated, because the
trace that results from interleaving does not include high inputs:
\dfn
An asynchronous trace system $\Sigma$ satisfies {\em asynchronous generalized noninterference} \
if, for all traces $\tau, \tau' \in \Sigma$, if $\tau''$ is a trace that results
from an arbitrary interleaving of the traces $\tau |_L$ and $\tau' |_{HI}$, there exists
a trace $\tau'''$ such that $\tau''' |_{L \cup HI} = \tau'' |_{L \cup HI}$.
\edfn

It is straightforward to relate these definitions to secrecy. Exactly as in
the synchronous case, let $f_{hi}$ be an information function that
extracts a high-level input trace prefix from a point: if
$r^T(m) = \<\tau|_L, \tau|_H\>$, let 
$f_{hi}(r^T,m) = \tau|_{HI}$.

\pro\label{pro:zl->totsec}
If $\Sigma$ is an asynchronous trace system that satisfies asynchronous
separability (resp. asynchronous generalized noninterference), then 
$H$ maintains total secrecy (resp. total $f_{hi}$-secrecy) with
respect to $L$ in $\R(\Sigma)$.
\epro

The converse of Proposition~\ref{pro:zl->totsec} does not necessarily hold.
We demonstrate this by providing a counterexample that works for both
separability and generalized noninterference. 
Suppose that there are no high output events, only one low
output event $l_o$, and arbitrary sets $LI$ and $HI$ of low and high
input events, respectively.
Consider the system consisting of all traces $\tau$ involving these
events such that $l_o$ occurs at most once in $\tau$, and when it
occurs, it does not follow any high input events.
In $\R(\Sigma)$, $H$ maintains total secrecy and $f_{hi}$-secrecy with
respect to $L$, 
because any local state for $L$ is compatible with any local state for $H$. (Because the
system is asynchronous, $L$ learns nothing by seeing $l_o$: 
when $L$ sees $l_o$, he thinks it possible that arbitrarily many high input
events could have occurred {\em after} $l_o$.
Furthermore, $L$ learns nothing about $H$
when he does {\em not\/} see $l_o$: it is always possible that no high
input events have 
occurred and that $l_o$ may yet occur.) However, $\Sigma$ does not satisfy asynchronous
separability or asynchronous generalized noninterference, because
interleavings where a high input event precedes
$l_o$ are ruled out by construction. 

\commentout{
It seems intuitively clear, in this example, that $L$ obtains no
information about the 
actions of $H$ even though the system satisfies neither separability nor generalized 
noninterference. But there do exist systems where secrecy is not as
obvious. Consider 
an input-total system $\Sigma'$ such that the set of output events consists of a single
low-level output event $l_o$ and a single high-level output event $h_o$. Suppose that
$\Sigma'$ contains all traces such that $l_o$ and $h_o$ may each occur at most once
in a given trace, and such that if both $h_o$ and $l_o$ occur in the same trace, $l_o$
must come {\em after} $h_o$.
The system $\R(\Sigma')$ satisfies total secrecy, because a given low-level trace is
consistent with any high-level trace (which may contain an arbitrary number of high
input events, and at most one occurrence of $h_o$). However, $\Sigma'$ does not satisfy
asynchronous separability because given a trace containing both output events, any
interleaving where $l_o$ comes after $h_o$ is not a valid trace.
}

This example illustrates a potential weakness of our approach to secrecy. 
Although $H$ maintains total secrecy with respect to $L$ in
$\R(\Sigma)$, there is a sense in which $L$ learns something about
$H$. 
Consider a point $(r,m)$ in $\R(\Sigma)$ at which $L$ has not seen
$l_o$.  
At that point, $L$ knows that if a high event has occurred, he will
never see $l_o$.
This knowledge does not violate secrecy, because it does not
depend on 
the local state of $H$; it is not an $H$-local fact. But there is a
sense in which 
this fact can be said to be ``about'' $H$. 
It is information about a correlation between high events and a
particular low event.
Is such information leakage a problem?  We have not been able to
construct an example where it is.  But it is worth pointing out 
that all of our definitions of
secrecy aim to protect the local state of some particular user. Any
``secret information'' that cannot be characterized as a local
proposition is not protected. 

In any case,
we {\em can} show that total secrecy and separability are equivalent if we assume
a particularly strong form of asynchrony that rules out a temporal dependence 
between high and low events.
Formally, $\Sigma$ is {\em closed under interleavings\/} if for all
asynchronous traces $\tau$ and $\tau'$, if $\tau \in \Sigma$,
$\tau'|_L = \tau|_L$ and $\tau'|_H = \tau'|_H$, then $\tau' \in \Sigma$.
Though this requirement allows $L$ to learn about high events that may occur in
the future (or that have possibly occurred in the past), it rules out any knowledge
of the ordering of high and low events in a given run.
With this requirement, total secrecy and asynchronous separability coincide.

\pro\label{pro:zl_iff_totsec}
If $\Sigma$ is an asynchronous trace system that is closed under
interleavings, then $\Sigma$ satisfies asynchronous separability iff
$H$ maintains total secrecy with respect to $L$ in $\R(\Sigma)$.
\epro

A similar result is true for generalized noninterference and
$f_{hi}$-secrecy if we 
modify the definition of closure under interleavings to allow $L$ to learn something
about the ordering of high output events; we omit the details here.

\commentout{
The definitions of asynchronous secrecy and generalized nonessentiallyinterference
capture secrecy (as we've seen), but they also require a particularly
strong form of asynchrony, such that all interleavings are possible.

Formally, $\Sigma$ is {\em closed under $(A,B)$-interleavings} for sets of events
$A$ and $B$ if for every $\tau$ in $\Sigma$, if $\tau'$ is an interleaving of
$\tau \mid_A$ and $\tau \mid_B$, then there exists $\tau'' \in \Sigma$ such that
$\tau'' \mid_{A \cup B} = \tau'$.

\pro\label{pro:zl_iff_totsec}
If $\Sigma$ is an asynchronous trace system that is closed under
$(L,H)$-interleavings (resp. $(L,HI)$-interleavings), then $\Sigma$
satisfies asynchronous separability
(resp. asynchronous generalized noninterference) iff
$H$ maintains total secrecy (resp. total $f_{hi}$-secrecy) with
respect to $L$ in $\R(\Sigma)$.
\epro
} %

\commentout{
One of our main points is that the intuitions behind secrecy are very simple,
and based on ideas about an agent's lack of knowledge, or perhaps probabilistic
independence. When applied to different {\em system frameworks}, secrecy may mean
very different things. But we claim that it's important to formalize intuitions
of secrecy in a general way that applies to a wide variety of system
frameworks. 
In our view, trace-based approaches do not facilitate this generality. 
}
\commentout{
It is worthwhile to note that Zakinthinos and Lee assume ``input totality'', that is,
that for any trace $\tau$, any any input event $e$, there exists a trace $\tau'$
that results from concatenating $e$ onto $\tau$. The system above is not a valid system,
then, because it contains only output events.
Even with input totality, however, if we consider the smallest input total system generated by
$\langle a, x \rangle$ and its prefixes,
we don't get equivalence because $x$ always comes after $a$, while
asynchronous separability requires that there be interleavings where the $a$ comes first.

Zakinthinos and Lee also define generalized noninterference.
\dfn
An asynchronous trace system $\Sigma$ satisfies asynchronous generalized noninterference
if, for all traces $\tau \in \Sigma$, for all traces $\tau' \in \lles(\tau)$, 
for all traces $t \in \interleave(HI^*,\mc{L}(\tau'))$, there exists $s \in \lles(\tau)$
such that $t = \mc{LHI}(s)$, where $\mc{LHI}$ extracts low-level events and high-level 
input events. 
\edfn

Zakinthinos and Lee do not define $HI^*$, or define what $\interleave$ might mean if it takes
a set as an argument, but it seems reasonable to believe that
the term $\interleave(HI^*,\mc{L}(\tau))$ refers to 
the set of traces that can be derived by interleaving the low-level events of $\tau$,
in any arbitrary way, with any syntactically possible trace 
consisting only of high-level input events. Note that if $HI^*$ 
consists of all syntactically possible traces
consisting only of high-level input events, that $HI^* \subseteq \Sigma$, due to 
input totality.

The following restatement emphasizes the similarity to
separability. (Note that  
we cannot simply require that an arbitrary interleaving be in the system,
because that  
leaves out high events. Note also that input totality is crucial for this restatement.)
\pro
An asynchronous trace system $\Sigma$ satisfies asynchronous generalized noninterference
iff for all traces $\tau, \tau' \in \Sigma$, 
if $\tau''$ is a trace that results from an arbitrary interleaving of $\mc{L}(\tau)$
and $\mc{HI}(\tau')$, then there exists $\tau''' \in \Sigma$ such that
$\mc{LHI}(\tau''') = \tau''.$
\epro

Once again, it would be nice to prove a theorem showing that this is equivalent
to total secrecy (where the high information function extracts only input events),
but it is not true, again because arbitrary interleavings are required. 
For example, let $a$ be a low-level output event, and let $\Sigma$ be 
the smallest input total system that includes the trace $\langle a \rangle$.
The system clearly satisfies secrecy; the $a$ does not tell the low-level
user 
anything at all. Yet if $y$ is a high-level input event, $\langle a, y \rangle$ is
a trace of the system, but $\langle y, a \rangle$ is not.

It is not clear why Zakinthinos and Lee want these arbitrary interleavings, given that they
do not matter to what would seem to be the intuitive ``view'' of the
system held by low and 
high-level users. They give no justification for the use of interleavings as the 
basis of security. 
[[ There might be some use of interleaving functions in ~\cite{GM82}, but I couldn't find
my copy of the paper. I could look at it post-Ireland... ]]
}

\commentout{
For example, they define separability as the 
following requirement: for every two traces $\tau$ and $\tau'$, every
possible interleaving 
$\tau''$ of the subtraces $\mc{L}(\tau)$ and $\mc{H}(\tau')$ is a valid
trace in 
the system. 
Here, separability says that the low-level agent is unable to rule out 
any possible high-level input/output sequences, as in Definition
\ref{dfn:sep}. 
Notice that
since traces can be of arbitrary length, and since arbitrary
interleavings of low-level and high-level traces must be valid traces of
the system, the systems 
considered by Zakinthinos and Lee must be completely asynchronous, so that
agents have 
absolutely no idea of what the time is. 
As we show in the full paper, this asynchronous notion of separability
corresponds to our notion of total secrecy; similarly Zakinthinos and
Lee's notion of generalized 
noninterference 
corresponds to $f_{hi}$-secrecy.
}

\subsection{Secrecy and User Protocols}\label{sec:protocols}

Generalized noninterference, within the context of McLean's synchronous
input/output traces, 
captures the intuition that the low-level agent cannot rule out any
high-level 
input traces. But is protecting the input of the high-level agent enough to 
guarantee secrecy? In more than one way, it is not. The first problem, of course,
is the use of possibility as an uncertainty measure. The second problem
is captured by the following example, which illustrates the
fact that the string input by a high-level user may be completely divorced from 
the message she wants to send. Consider the following 
synchronous trace
system, a simplified version of one described
by Wittbold and Johnson~\citeyear{Wittbold&Johnson} and Gray and
Syverson~\citeyear{gray98}.  

\begin{itemize}

\item
All input/output values are restricted to be either 0 or 1.

\item 
At each time step $k$, the high-level output value $h_o^{(k)}$ is 
nondeterministically chosen to be either 0 or 1.

\item
At each time step $i$, the low-level output value $l_o^{(k)}$ is
$h_o^{(k-1)} \oplus h_i^{(k)}$, where $\oplus$ is the exclusive-or
operator and $h_i^{(k)}$ is the high-level input at time $i$. 

\item For completeness, suppose that the low-level output at time 1
is 0, since $h_o^{(k-1)}$ is not defined at the first time step. (What
happens at the first time step is unimportant.)
\end{itemize}

The set of traces that represents this system satisfies noninterference.
At any time $i$, $l_o^{(k)}$ depends only on $h_i^{(k)}$ and
$h_o^{(k-1)}$.
But as the following table shows, any value of $l_o^{(k)}$ is consistent
with any value of $h_i^{(k)}$:

\begin{center}
\begin{tabular}{|c|c|c|}	\hline
$h_i^{(k)}$ & $h_o^{(k-1)}$ & $l_o^{(k)}$ \\ \hline
0 & 0 & 0 \\
0 & 1 & 1 \\
1 & 0 & 1 \\
1 & 1 & 0 \\ 	\hline
\end{tabular}
\end{center}

Thus,
given traces $\tau, \tau'$, we can construct a new trace
$\tau''$ by taking the
low-level input/output of $\tau$ and the high-level input of
$\tau'$. For the
high-level output of the resulting trace, we take 
$h_o^{(k)} = l_o^{(k+1)} \oplus h_i^{(k+1)}$.
Because
$\tau''$ is a valid trace of the system, the
system satisfies generalized noninterference. 

The problem with this system is that a malicious high-level agent (for
example,
a ``trojan horse'' program) who knows how the system works can transmit
arbitrary 
strings of data
directly to the low-level agent. If the high-level agent wants
to
transmit the bit $x$ at time $k$ and sees the high-level output bit $y$
at time $k-1$, then she can ensure that the low-level output is $x$ at
time $k$
by inputting the bit $x \oplus y$ at time 
$k$.

Wittbold and Johnson~\citeyear{Wittbold&Johnson}
point out that examples such as this show that generalized noninterference
does not guarantee security.
We claim
that the fundamental problem is not with generalized noninterference
{\em
per se}, but rather with 
an underlying system model that assumes that everything relevant to the
state of 
the high-level agent can be captured using input/output traces. 
\commentout{
If part of the high-level agent's state 
includes some string, that 
string should be part of the state of the agent.
Similarly, if the agent has a protocol that involves sending a string,
the protocol should be represented in the agent's local state.
In other words, we believe that it isn't the {\em definition} of secrecy
that is 
problematic, but rather the {\em system model} for which secrecy is defined.
Indeed, this example reinforces the importance of having an abstract
definition of secrecy, where the system includes everything that might
be relevant. If we are worried about protecting the input of a high-level
user when trojan horses might be present, it is clear that the system model
must account for the behavior of the high-level user somehow;
if we are only worried about the actual strings input by the user, this may not
be necessary.
\footnote{In a deterministic system, this kind of attack isn't even possible. Should we
formalize this?}
}
If we model an agent's local state so that it includes a protocol for
transmitting a specific string to another agent, generalized noninterference
does not ensure secrecy.

To deal with the problem they noted,
Wittbold and Johnson introduced {\em nondeducibility on strategies}
(NOS).
We modify their definition 
slightly so that it 
is compatible with
McLean's framework of synchronous traces.
A {\em protocol} $\mb{H}$ is a function from a high-level input/output
trace prefix $\tau_k |_H$ to a high-level input value $h_i \in
HI$. Intuitively,  
a protocol tells the agent $H$ what to do at each step, given what he
has already 
seen and done.
A trace $\tau$ is {\em consistent} with a protocol $\mb{H}$ 
if, for all $k$, $\mb{H}( \tau_{k-1} |_H ) = h_i^{(k)}$, where
$h_i^{(k)}$ is the high-level input value of the $k$th tuple in $\tau$.
A synchronous trace system $\Sigma$ {\em satisfies nondeducibility on
strategies\/} if, for  
all traces $\tau \in \Sigma$ and every high-level strategy
$\mb{H}$ consistent with 
some trace in $\Sigma$,
there exists a trace $\tau'$ that is consistent with $\mb{H}$ 
such that
$\tau' |_L = \tau |_L$. 
If the protocol of the high-level agent is included as part of
her local state, and $f_{\strat}$ is an $H$-information function
that extracts the protocol of the high-level agent from 
the local state, then it is straightforward to show that NOS is just 
synchronous $f_{\strat}$-secrecy.
\commentout{
\footnote{In fact, 
if we make the reasonable assumption that $H$'s
protocol
is the same at every point of a
run, so that it 
does not depend on time, NOS is also a special case of total
$f_{\strat}$-secrecy.} 
Because an agent's local state is intended to include all the relevant parts
of an agent's knowledge, is seems natural to include the agent's protocol in
its state. After all, if the agent does not know her protocol,  
or if the traces of the system are inconsistent with the protocols being used,
nondeducibility
on strategies is 
an unhelpful definition.
}
\commentout{
They define probabilistic noninterference for a much more
specific system representation than we do, because they are concerned with
specifying multilevel security properties for systems that can be
described entirely in terms of traces of input and output events, and 
conditional probability distributions that characterize the data that a user will
choose to input into the system at any point.
The ultimate goal is to provide end-to-end
security guarantees that apply even if a ``trojan horse'' uses knowledge of 
how the system is constructed in order to covertly transmit information 
to low-level users. In some respects, their work can be seen as a generalization
of {\em nondeducibility on strategies}, as introduced by Wittbold and
Johnson~\citeyear{Wittbold&Johnson}. The definitions given by Gray and Syverson
were an inspiration for the work we present in this paper, and they also serve as
particularly relevant examples of security properties that are special
cases of the kind we propose.
}

Gray and Syverson \citeyear{gray98} extend 
NOS to probabilistic systems
using the Halpern-Tuttle framework.
In Gray and Syverson's terminology,
low-level and high-level agents use 
probabilistic protocols $\mb{L}$ and $\mb{H}$, respectively.
Again, the protocols $(\mb{H}$ and $\mb{L})$ determine what the agents
$H$ and $L$ will input next, given 
what they have 
seen and done so far. 
The system is assumed to have a fixed probability distribution $\mb{O}$  
that determines its output behavior, given the inputs and outputs seen
so far. 
Formally, for each trace prefix $\tau$ of length $k$, 
$\mb{H}(\cdot \mid (\tau |_H))$ is a probability measure on high-level
input events that occur at time
$k+1$, given the projection of $\tau$ onto the
high-level input/output; similarly, $\mb{L}(\cdot \mid (\tau |_L))$ is
a probability measure on low-level 
input events that occur at time $k+1$
and $\mb{O}(\cdot \mid \tau )$ is a probability measure on output events
that occur at time $k+1$, given $\tau$.  
Gray and Syverson require that the choices made by $H$, $L$, and the
system at each time step be probabilistically independent.  With this
assumption, $\mb{H}$, $\mb{L}$, and $\mb{O}$ determine a conditional
distribution that we denote $\mu_{\mb{H},\mb{L},\mb{O}}$, where
$$\mu_{\mb{L},\mb{H},\mb{O}} (\langle l_i, h_i, l_o, h_o \rangle \mid \tau )
   = \mb{L}( l_i \mid (\tau |_L)) \cdot \mb{H}( h_i \mid (\tau |_H))
\cdot \mb{O}(l_o,h_o \mid \tau).$$      
Let $\Lambda$ and $\Gamma$ be 
countable
sets of protocols for the low-level
and high-level agents, respectively.%
\footnote{Gray and Syverson take $\Lambda$
and $\Gamma$ to consist of 
{\em all possible}
probabilistic protocols for the low-level and
high-level agent, respectively, but their approach 
still makes 
sense if
$\Lambda$ and $\Gamma$ are 
arbitrary 
sets of protocols,  
and it certainly seems reasonable to assume that there are only
countably many protocols that $H$ and $L$ can be using.}
\commentout{
In order to deal with their definitions in our framework, we will assume that only a 
{\em countable} number of protocols are possible. In this sense, our setup is 
strictly more restrictive than theirs; on the other hand, a countable
number of viable protocols does not seem like a huge restriction,
especially given Gray and Syverson's assumption that the protocol is
actually encoded in the local state of the agents.
}
Given $\Lambda$, $\Gamma$, and $\mb{O}$ (and, implicitly, sets of low
and high input and output values),  
we can define an adversarial probability system
$\R^*(\Lambda,\Gamma,\mb{O})$ in a straightforward way.  Let $\Sigma$
consist of all synchronous traces over the input and out values.
For each joint protocol $(\mb{L},\mb{H}) \in \Lambda \times \Gamma$, 
let $\R(\Sigma,\mb{L},\mb{H})$ consist of all runs defined as in our earlier
mapping from synchronous traces to runs, except that now we include
$\mb{L}$ in the low-level agent's local state and $\mb{H}$ in the
high-level agent's local state.  Let $\R(\Sigma,\Lambda \times \Gamma) =
\union_{(\mb{L},\mb{H}) \in \Lambda \times \Gamma}
\R(\Sigma,\mb{L},\mb{H})$.  
We can partition $\R(\Sigma,\Lambda \times \Gamma)$ according
to the joint protocol used; let $\D(\Lambda,\Gamma)$ denote this
partition.  Given the 
independence assumptions, the joint protocol $(\mb{L},\mb{H})$ also 
determines a probability $\mu_{\mb{L},\mb{H},\mb{O}}$ on
$\R(\Sigma,\mb{L},\mb{H})$.   Let $\Delta(\Lambda,\Gamma,\mb{O}) =
\{\mu_{\mb{L},\mb{H},\mb{O}}: \mb{L} \in \Lambda, \mb{H} \in \Gamma\}$.
Let
$\R^*(\Lambda,\Gamma,\mb{O}) = (\R(\Sigma,\Lambda\times \Gamma), \D,
\Delta\}$.
We can now define Gray and
Syverson's notion of secrecy in the context of these adversarial systems.

\dfn
An adversarial system $\R^*(\Lambda,\Gamma,\mb{O})$ satisfies
{\em probabilistic noninterference\/} if, 
for all low-level protocols
$\mb{L} \in \Lambda$, points $(r,m)$ 
where $L$'s protocol is $\mb{L}$,
and high-level protocols $\mb{H}, \mb{H'} \in \Gamma$, 
we have
$\mu_{(\mb{L},\mb{H},\mb{O})}(\K_L(r,m)) = \mu_{(\mb{L},\mb{H'},\mb{O})}(\K_L(r,m)).$
\edfn

\thm The following are equivalent:
\begin{itemize}
\item[(a)] $\R^*(\Lambda,\Gamma,\mb{O})$ satisfies probabilistic
noninterference;
\item[(b)] $L$ obtains no evidence about $H$'s protocol (in the
sense of Definition~\ref{dfn:evidential}) in $\R^*(\Lambda,\Gamma,\mb{O})$;
\item[(c)] $H$ maintains generalized run-based probabilistic
$f_{\strat}$-secrecy 
with respect to $L$ in $(\R(\Sigma,\Lambda\times \Gamma),\M^{\INIT}(\Delta(\Lambda,\Gamma,\mb{O})))$,
where $f_{\strat}$ 
is the information function that maps from $H$'s local state to $H$'s protocol;
\item[(d)] $H$ maintains generalized probabilistic synchronous
$f_{\strat}$-secrecy with respect to $L$ in 
the standard generalized probability system determined by
$(\R(\Sigma,\Lambda\times
\Gamma),\M^{\INIT}(\Delta(\Lambda,\Gamma,\mb{O})))$. 
\end{itemize}
\ethm
\prf The fact that (a) implies (b) is immediate from the definitions,
since $H$'s initial choice is
the protocol $\mb{H}$.  
The equivalence of (b) and (c) follows from Theorem~\ref{thm:evidential}.
Finally, since the traces in $\Sigma$ are synchronous,
the equivalence of (c) and (d) follows from
Proposition~\ref{pro:sync_prob_sec}. \eprf

\commentout{
We want to show that probabilistic noninterference is equivalent to
synchronous plausibilistic $f_{\strat}$-secrecy, for some plausibility measure $\Pl$,
of high-level agents with respect to low-level agents. 
To do so we use the
techniques of Section \ref{sec:nondeterm}, as well as an extra assumption that reflects
the assumption that high and low protocols are chosen independently.

Consider the $\sigma$-algebra $\S$ on $\R$ generated by points of $\R$.
That is, let $\S$ is the smallest $\sigma$-algebra generated 
by
$$\S' = \{ \{ r' \in \R \mid r'(m) = r(m) \} \mid (r,m) \in \P(\R) \}.$$
Note that a probability measure defined on $\S'$ extends uniquely to $\S$,
because $\S'$ is closed under finite intersections~\cite{Billingsley}.
(Within the set of runs generated by a single adversary, probabilities on
$\S'$ can be determined unambiguously by the adversary and by the system's probability
distribution on outputs.)

Now, consider the set $\M = \M(\Delta)$ of probability measures
on $\S$ such that $\mu \in \M$ if all of the following conditions hold:
\ul
\item $\mu( e \mid \R(\mb{A}) ) = \mu_{\mb{A}}( e )$ for
all points $(r,m) \in \P(\R)$ and all adversaries $\mb{A}$, where
$e = \R(f_L^{-1}(f_L(r,m)))$.
\item $\mu(\R(\mb{A})) > 0$ for all adversaries $\mb{A} = (\mb{L},\mb{H})$
\item $\mu( \R(\mb{H}) \mid \R(\mb{L}) ) = \mu( \R(\mb{H}) )$ for all adversaries
$\mb{A} = (\mb{L},\mb{H})$.
\eul
In other words,
we want to consider all measures $\mu$ that give all protocols positive
probability
such that (a) the conditional probability of the low-level traces matches the
probability assigned by the protocol-specific probability measures described
by Syverson and Gray and (b) the high-level and low-level
protocols are selected independently.
The set $\M$ is nonempty.
For an example of a measure $\mu \in \M$, let
$\mb{L}_1, \mb{L}_2, \ldots$ and 
$\mb{H}_1, \mb{H}_2, \ldots$
be countable enumerations of $\Lambda$ and $\Gamma$, and define $\mu$ on $\S$ such 
that whenever $s \in \S'$
(so that $s \subseteq \R(\mb{A})$, for some $\mb{A} = (\mb{L}_j,\mb{H}_k)$), we have
$$ \mu( s ) = \frac{ \mu_{\mb{A}}(s) }{ 2^{j+k} },$$
so that $\mu(\R(\mb{L}_j)) = 1/2^j$, $\mu(\R(\mb{H}_k)) = 1/2^k$, and
$\mu(\R(\mb{L}_j,\mb{H}_k)) = 1/2^{j+k}$. This ensures that all the conditions are
satisfied to have $\mu \in \M$.

[[ My probabilistic definitions in the previous version, where we had a definition
of secrecy with respect to a set of probability measures, made this easier. Here
we no longer have a nice
definition for ``secrecy with respect to a set of probability
measures''. We can't use 
the adversarial definitions directly because they don't deal with the independence of
high and low strategies. If the definition of adversarial secrecy included this
extra requirement, this result would be more direct. ]]

\thm\label{thm:probNI}
A probabilistic trace system $(\R,\D,\Delta)$ satisfies probabilistic noninterference
iff, for each measure $\mu \in \M(\Delta)$, the high-level agent
maintains run-based probabilistic $f_{\strat}$-secrecy with respect to the low-level
agent in $(\R,\M(\Delta))$.
\ethm

This theorem (whose proof, in the appendix, is somewhat involved) 
illustrates the fact that even in complex systems involving probability and nondeterminism,
the notion of noninterference---emphasizing the inability of a high-level agent to affect
low-level behavior---can be coincident with the notion of secrecy or nondeducibility,
which emphasizes the inability of a low-level agent to deduce 
anything about high-level behavior.

} %

\commentout{
Probabilistic noninterference is thus equivalent to a very natural definition of secrecy, 
assuming that we choose the correct information extraction functions (i.e., those that extract the 
low-level trace prefixes and the high-level protocols) correctly. Note that in our setup, it is
in fact the low-level traces that maintain secrecy with respect to the high-level protocols. This
seems backwards, but recall that Gray and Syverson are concerned with whether or not the high-level
agent is able to {\em interfere} with the low-level agent. It therefore follows that low-level events
should not depend on the protocol of the high agent. It turns out that this is equivalent to a 
requirement that the high-level agent's protocol maintains secrecy with respect to low-level events.

\pro\label{pro:sec_probNI}
Let $(R,\Delta)$ be a probabilistic trace system. In the corresponding probabilistic system $(\R,\M)$,
for each $\mu \in \M$, the low-level agent $L$ maintains synchronous probabilistic $f_L$-secrecy with
respect to the protocol of the high-level agent iff the high
level agent's protocol  
maintains secrecy with respect to the low-level events given by $f_L$.
\epro
}

\commentout{

\subsection{Private Information Retrieval}                                                                                                             
Although we have discussed secrecy in the context of noninterference and
security, essentially the same definitions apply in other contexts.
Chor, Goldreich, Kushilevitz, and Sudan \citeyear{CGKS98},
for example, consider ``private information retrieval''. Suppose that a
database user wants to query a database, but wants some guarantee that
no one will be able to figure out from his queries exactly what
information  he is after.  To make this precise, suppose that a database
is characterized by a string $x$ of $n$ bits, and that a user is
interested in finding out bit $x_i$.
Further assume that the database information is replicated at $k$ servers,
and that there is no communication among the different servers.
The user has a (possibly randomized) strategy for querying each server
about the database.  For simplicity, we assume that the user queries
each server $s$ exactly once.
We want it to be the case that the user can
reconstruct the desired bit from the $k$ replies it receives, but that
no individual server can deduce anything about which bit the user is 
interested in.
Since the user's strategy may be randomized, the actual query that the
user poses to server $s$ may depend on some random input.  For
simplicity, assume that the random inputs come from some finite space
$RI$ of inputs, and $\Pr$ is the uniform distribution over $RI$.
Let $Q_s(i,r)$ be the query that the user poses to
server $s$ if he is interested in bit $x_i$ and gets random input $r \in RI$.
The server responds with an answer $A_s(x,q)$, based on the database $x$
and the query $q$, and the user uses a reconstruction function $Re$ to
reconstruct the index $x_i$ from the answers retrieved. 
A {\em private information retrieval scheme}, then, is a tuple $(\Q,\A,Re)$,
where $\Q$ is the set of query-generation functions and $\A$ is the set of answer 
functions used by each server.
If $Q$ is the set of possible queries,
privacy for such a scheme requires that
for every query function $Q_s \in \Q$, for each $i, j \in \{1, \ldots, n\}$,
and for each $q \in Q$,
$$ \Pr(\{ r \mid Q_s(i,r) = q \}) = \Pr(\{ r \mid Q_s(j,r) = q \}). $$
That is, we require the probability that the server $s$ is asked $q$ when the user is
interested in $x_i$ 
to be
the same as the probability that $s$ is asked $q$
when the user is interested in $x_j$.

Now, with each database index $i$, set of $k$ random inputs from $RI$, and
resulting set of $k$ queries, we can associate a run $r$. At time $0$,
suppose that the user picks an index $i$ and generates the queries $q_s$
to be sent to each individual server $s$. [[ Do we want to deal with time
here? We could model the replies sent by the server, even though they're 
irrelevant to privacy. We do need to specify the local states of the user
and the different servers.]]

Let $\R$ be the set of possible runs.  We can partition this set
according to the indices to get an adversarial probability system
$(\R,\D,\Delta)$. (Within a partition, there is a uniform probability on
runs given by the uniform probabilities on random inputs in $RI$.)
Taking the user to be $u$,
let $f_u$ be a $u$-information function that extracts the index from
the user's state.
                                                                                                             
\pro
The private information retrieval scheme $(\Q,\A,Re)$ satisfies the
privacy
condition iff the user $u$ maintains run-based probabilistic
$f_u$-secrecy with respect to each server $s$ in the adversarial system
$(\R,\D,\Delta)$.
\epro
                                                                                                             
\prf
Wait until we agree on the details.
                                                                                                             
Sketch: suppose that privacy holds, and let $\mu \in \M(\Delta)$ be a
measure on $\R$
consistent with $\Delta$. We'll show that the database maintains secrecy
with respect
to the ``index choice'', and $f_u$-secrecy follows by symmetry. We have
$ \mu_i(q) = \mu_j(q) $, so $\mu( q | i ) = \mu( q | j )$, and this is
secrecy.
                                                                                                             
The converse works the same way.
\eprf
                                                                                                             
} %

\commentout{
\subsection{Process Algebras and Multiagent Systems}

Recent work in information flow and noninterference
has focused on systems that are specified using process algebras
related to CCS~\cite{focardi94} and
CSP~\cite{ryan99}. 
Milner~\citeyear{milner89} gives an excellent introduction to CCS,
while Focardi and Gorrieri~\citeyear{focardi01} provide an overview of
recent work on noninterference properties expressed using CCS.
Ryan, Schneider, Goldsmith, Lowe, and Roscoe~\citeyear{ryan00} give a
thorough introduction to CSP and describe how it can be used to
describe a number of security properties.

Definitions of noninterference properties 
that 
are based on process
algebras have several important advantages. 
With process algebras it is easy to describe distributed systems
compactly even if the resulting systems have trace
sets that are large or infinite.
In addition, the process-algebra approach provides an elegant way of
constructing systems compositionally,
as well as tools for verifiation (based on proving process equivalence).
Finally, the process algebra approach can express features of systems that
cannot be expressed using input/output traces;
these features can be quite significant in the context of
security (see \cite{focardi01,ryan99}).  

With regard to the last point, it may seem that our approach suffers
from the limitations of other trace-based approaches, since runs are
essentially traces. This is not the case. By using local states in a
natural way, 
the runs-and-systems approach can in fact capture some of the standard
examples that cannot be captured using input-output trace-based approaches.
Consider the following 
simplistic 
example (discussed by both Ryan
and Schneider \citeyear{ryan99} and Focardi and Gorrieri
\citeyear{focardi01}).  
Let $p$ and $q$ be the following two processes:
\[p = a.b.0 + a.b.0\]   
and
\[q = a.(b.0 + c.0),\]
represented graphically in Figure~\ref{fig:bisim}.
\begin{figure}[hbt] 
\begin{center}
\centerline{\epsfig{figure=proc_alg.eps, height=4cm}}
\caption{Processes $p$ and $q$.\label{fig:bisim}}
\end{center}
\end{figure}
(We assume that the reader is familiar with the basic syntax and
semantics of process algebra; see \cite{milner89}.)
These two systems have the same set of input/output traces,
namely the set $\{ \langle a \rangle, \langle a,b \rangle, \langle a,c
\rangle \}$.
However, the behavior of the two systems is different. If $b$ and $c$
represent input events, the two systems have different
{\em refusal sets} after executing the trace $\langle a \rangle$.
After $q$ executes $a$, it moves to a state where it can execute the
event $b$ {\em or} the event $c$, while $p$, after executing $a$, moves
to a state where it has no choice about the next event it executes
(either $b$ or $c$). These systems are 
thus not equivalent
according to 
most of the standard notions of equivalence considered in the literature 
(including bisimulation equivalence and testing
equivalence~\cite{milner89}).

Although the two processes have the same set of traces, they do not have
to be equivalent according to the runs-and-systems approach either, as
long as local states are defined appropriately.
There are several ways to capture the processes $p$ and $q$ as systems
in our framework.  
One possibility is to 
assume that there is one
agent who performs the actions $a$, $b$, and $c$ according to 
either process $p$ or process $q$.
\commentout{
The agent decides (nondeterministically) what protocol she is
following.  In both cases, there are two runs, one corresponding to the
trace 
$\langle a, b \rangle$
and the other corresponding to the trace 
$\langle a, c \rangle$.  (There
may also be a run or runs corresponding to the trace $a$ if we assume
that the system is deadlocked for some reason after the agent performs
$a$.  We start by considering the case where there are two traces.)  
For simplicity, assume that the environment state just records the
actions that were taken.  (Thus, the environment records the
input-output relation.)  
The agent's nondeterministic choice is reflected in her local state.  In
the case of $p$, the agent's initial local state must reflect her choice
of whether 
to run protocol $\langle a, b \rangle$ or protocol 
$\langle a, c \rangle$.  This choice continues to
be reflected in her state at time 1.  
With process $q$, the agent does not have to make a choice at time 0;
her initial local state simply
reflects the fact that her next action is $a$. At time 1, she must
choose between $b$ and $c$.  
Thus, in the system corresponding to $p$,
the agent can distinguish the initial points of the two runs (since she
has made different choices); on the other hand, in the system
corresponding to $q$, she cannot distinguish the initial states of the
runs.  
If we also want to consider the trace $a$, the system
corresponding to $p$ has two more runs (depending on the initial choice
of the agent); the system corresponding to $q$ has just one more run.  
} %
In the systems corresponding to both $p$ and $q$, there are two runs,
one corresponding to the 
trace $\langle a, b \rangle$ and another corresponding to the trace
$\langle a, c \rangle$. In the case of process $p$, the agent's state
at time 0 should reflect that she has to make a nondeterministic choice between 
the two runs, and at time 1 should reflect that she has already performed $a$
and has no choice. 
as to whether she will now perform $b$ or $c$. 
In the case of process $q$, the agent does not have to make a choice at time 0,
and can instead postpone the choice until time 1, when either $b$ or $c$ 
may be performed. In both cases the agent's local state should encode the choices
that have been made and the choices that are available, as well as the
sequence of events that she has performed up to the given time.
Thus, although both systems have two runs, the local states in the runs
are different, and reflect the differences between processes $p$ and $q$.

The model just discussed is synchronous.  However, we can also take an
asynchronous version of the model.  In this case, there would be
infinitely many runs, corresponding to the different times the actions
were performed.  We can, of course, also capture other constraints on
timing (such as upper bounds on how long it takes to perform actions),
which seem to be less easy to model in the process algebra framework.
In any case, the key point is that there are quite natural ways to
capture $p$ and $q$ as {\em different} systems.

In general, 
the most appropriate way of describing processes as systems depends on
the notion of process equivalence being used.  There are many different
notions of process equivalence discussed in the literature.
(For a survey of a wide variety of process equivalences, 
see van Glabeek~\citeyear{vanglabbeek01}.)
When process algebras are used to model security properties, the choice of 
process equivalence is particularly critical,
because the notion of equvalence used implicitly encodes the types of
attacks that the adversary can perofrm.
We want to translate processes to systems so as to preserve
equivalences, in the sense that the systems corresponding to two processes
$p$ and $q$ are identical if and only if $p$ and $q$ are equivalent.
We conjecture that this can be done for all reasonable notions of
equivalence.  While stating such a conjecture precisely and proving it
is well beyond the scope of this paper, we illustrate the intuitions by
considering on standard notion of equivalence, {\em ready-trace equivalence}.

There are several ways to  
define ready-trace semantics; we use
the terminology of van Glabbeek
\citeyear{vanglabbeek01}.
The trace
$$ \tau = \langle (a_1, A_1), \ldots, (a_n, A_n) \rangle $$
is a ``ready trace'' of process $p$ if
$$ p \goesto{a_1} p_1 \goesto{a_2} \ldots \goesto{a_n} p_n, $$
and for each $1 \leq i \leq n$, $A_i$ is the set of events that 
can be performed at the next step.
For example, one ready trace of the process $p$ is
$\langle (a,\{b\}), (b,\emptyset) \rangle$ and one ready trace
of the process $q$ is $\langle (a,\{b,c\}),(c,\emptyset) \rangle.$
We can capture deadlock using ready traces, by taking
$A_i = \emptyset$.  
Let $\RT(p)$ be the set of ready traces of a process $p$. 
We say that two processes $p$ and $q$ are {\em ready-trace equivalent} if $\RT(p) = \RT(q)$.
Ready-trace equivalence is strictly finer than testing equivalence and
strictly  
weaker than bisimulation~\cite{vanglabbeek01}.

To construct a system corresponding to a process $p$, we construct runs
from sets of ready traces in exactly the same way that we did in 
Section~\ref{sec:sts}. Given $\RT(p)$, let $\R_{RT}(p)$ be the set of runs
$r^T$ such that $T$ is a run-like subset of $\RT(p)$. As before, if $T$
is infinite then we have $r^T(m) = \tau^m$ (the unique trace of length
$m$ in $T$); otherwise, if $N$ is the length of the longest ready trace
in $T$, then  
$r^T(m) = \tau_m$ when $m \leq N$ and $r^T(m) = \tau_N$ when $m > N$.

\pro\label{pro:ready_traces}
If $p$ and $q$ are processes, then $\RT(p) = \RT(q)$ if and only if 
$\R_{RT}(p) = \R_{RT}(q)$.
\epro

As Proposition~\ref{pro:ready_traces} suggests, it may be possible to
embed the process algebraic approach in the runs and systems framework.  
Nevertheless, as we said earlier, there are certainly some advantages in
working directly with process algebras.
We would like to explore further connections between
our definitions and those provided by the process algebra
approaches. 
One obvious place to start would be to demonstrate that definitions
of noninterference given for process algebras are in fact instances
of our definitions of secrecy, for particular translations from processes
to systems. Other issues to consider include the following:
\begin{itemize}
\item Ryan and Schneider~\citeyear{ryan99} give
a general definition of noninterference in terms of an
equivalence relation $\approx$ and 
a {\em Constrain} operator. This definition seems
to be similar to our definition of $f$-$C$-secrecy, 
but we have not yet made the connection precise.
\commentout{
However, they
identify ``nondeducibility'' with the instance of
nondeducibility discussed by Sutherland at the end of his paper, which
is essentially generalized noninterference.  
Thus, the way they capture
nondeducibility does {\em not\/} capture our notion of $f$-total
secrecy. Their ``most liberal generalisation''
\cite[Section~4.2]{ryan99}, presented in terms of an abstraction
operator and a {\em Constrain\/} operator, does seem to correspond to
$f$-total secrecy, although we have not formalized the relationship.
However, it does not seem that their paper (or any of the other papers on the
process algebra approach) has an analogue to $f$-$C$-secrecy, nor is
there an analogue to the notion of syntactic security.
}
\item 
Can notions like semisynchronous
secrecy and resource-bounded secrecy could be 
handled cleanly in a process algebra approach? 
\item 
How do our definitions of probabilistic secrecy relate
to recent work adding probability to definitions of noninterference based
on the process algebra approach~\cite{aldini01,smith03}?
\end{itemize}

} %

\section{Conclusion}
\label{sec:conclusion}

We have defined general notions of secrecy for systems where multiple
agents interact over time, and have given syntactic characterizations
of our definitions that connect them to logics of knowledge and
probability. We have applied our definitions to the problem of
characterizing the absence of information flow, and have shown how our
definitions can be viewed as a generalization of a variety of
information-flow definitions that have been proposed in the past.

We are not the first to attempt to provide a general framework for
analyzing secrecy; see, for example,
\cite{focardi01,Mantel00,mclean94,ryan99,ZL97}
for some other attempts.
However, we believe that our definitions are more
closely related to the intuitions that people in the field have had,
because those definitions have often been expressed in terms of the
knowledge of the agents who interact with a system.

Our definitions of probabilistic secrecy, and their plausibilistic
generalizations, demonstrate the underlying simplicity and unity
of our definitions. Likewise, our results on the symmetry of secrecy
illustrate the close connection between notions of secrecy and
independence. The definitions and results that we have presented,
and their underlying intuitions of knowledge and independence, do
not depend on the particular system representation that we describe
here, so they should be broadly applicable. 

Indeed, although we have discussed secrecy largely with respect
to the kinds of input and output systems that have been popular
with the theoretical security community, our definitions of
secrecy apply in other contexts, such as protocol analysis,
semantics for imperative programming languages, and database theory.
Chor, Goldreich, Kushilevitz, and Sudan \citeyear{CGKS98},
for example, consider the situation where a user wants to 
query a replicated database for some specific database item, but wants 
a guarantee that no one will be able to determine, based on his
query, which item he wants. 
It is not hard to show that the
definition of privacy given by Chor et al.~is a special case of secrecy
in an adversarial system with a cell corresponding to each possible item
choice. 

There are several possible directions for future work. 
One is the verification of secrecy properties. 
Because we have 
provided syntactic characterizations of several secrecy properties
in terms of knowledge and local propositions, it would seem that 
model-checking techniques could be applied directly. 
(Van der Meyden \citeyear{vandermeyden98} gives some recent results on
model-checking in the runs and systems framework.) However,
verifying a secrecy property requires verifying an infinite set of
formulas, and developing techniques to do this efficiently would 
seem to require some nontrivial advances to the state of the art 
in model checking. Of course, to the extent that we are interested 
in more limited forms of secrecy, where an agent is restricted from knowing
a small set of formulas, knowledge-based model-checking techniques 
may be immediately applicable. At any rate, we emphasize that 
it is our goal in this paper to provide general techniques for the
specification, rather than the verification, of secrecy. 

Another direction for future work is a careful consideration of how
secrecy definitions can be weakened to make them more useful in practice.
Here we briefly consider some of the issues involved:

\ul

\item {\em Declassification:} Not all facts can be kept secret in a real-world
computer system. The canonical example is password checking, where a system is
forced to release information when it tells an attacker that a password is invalid.
Declassification for information-flow properties has been addressed by,
for example, Myers, Sabelfeld, and Zdancewic~\citeyear{MSZ04}. It would be interesting
to compare their approach to our syntactic approach to secrecy, keeping in mind 
that our syntactic definitions can be easily weakened simply by removing facts from
the set of facts that an agent is required to think are possible.

\item {\em Computational secrecy:} Our definitions of secrecy are 
most appropriate for attackers with unlimited
computational power, since agents ``know'' any fact that follows logically
from their local state, 
given the 
constraints of the system. Such an assumption is unreasonable for most
cryptographic 
systems, where secrecy depends on the inability of attackers to solve difficult
computational problems. The process-algebraic approach advocated by
Mitchell, Ramanathan, Scedrov, and Teague~\citeyear{MRST04} and the work
on probabilistic algorithm knowledge of Halpern and
Pucella~\citeyear{HalPuc02:tark} 
may help to shed light on how definitions of secrecy can be weakened to
account for 
agents with computational limitations.

\item {\em Quantitative secrecy:} Our definitions of probabilistic
secrecy require independence: an agent's posterior probability distribution
on the possible local states of a secret agent must be exactly the same as his
prior distribution. This requirement can be weakened using the information-theoretic
notions of entropy and mutual information.  Rather than requiring that 
no information 
flow from one user to another, we can quantitatively bound the mutual information 
between their respective local states.
Information-theoretic approaches to secrecy have been discussed by Wittbold and
Johnson~\citeyear{Wittbold&Johnson}, and more recently by Clark, Hunt, and 
Malacaria~\citeyear{CHM02}, Lowe~\citeyear{Lowe02}, and Di Pierro, Hankin, and Wiklikcy~\citeyear{DHW02}.

\item {\em Statistical privacy:} In some systems, such as databases that release 
aggregate statistical information about individuals, our definitions of secrecy are much too
strong because they rule out the release of any useful information. Formal definitions of secrecy
and privacy for such systems have recently been proposed by Evfimievski, Gehrke, and
Srikant~\citeyear{EGS03} and by Chawla, Dwork, McSherry, Smith and Wee~\citeyear{CDMSW05}.
These definitions seek to limit the information that an attacker can learn about a user
whose personal information is stored in the database.  It would be
interesting to cast those 
definitions as weakenings of secrecy.

\eul

These weakenings of secrecy are all conceptually different, but it seems highly likely that
there are relations and connections among them. We hope that our work will help to clarify 
some of the issues involved. 
\commentout{
We have defined general notions of secrecy in multiagent systems, and
shown the connection between our definitions and standard epistemic
notions.  We then applied the definition to the problem of
characterizing absence of information flow.  

Of course, we are not the first to attempt to provide a general
framework for analyzing security properties.  (See, for example,
\cite{focardi01,Mantel00,mclean94,ryan99,ZL97}
for some other attempts.)  However, we believe that our
definitions come closer to the intuitions that people in the field have
had, since those intuitions
are often expressed in terms of (a lack of) knowledge.
Our approach has a number of other advantages over previous approaches:
\begin{itemize}
\item As we have shown, it 
can be easily extended to deal with probabilistic secrecy.
\item It can deal with the whole spectrum from synchrony to total
asynchrony, without assuming a specific underlying system model.
\item It can be generalized 
to deal with more specific security concerns, 
so that we can talk about the secrecy of a particular formula as well
as total secrecy.
\item The characterization in terms of knowledge allows extensions to
resource-bounded knowledge, which seems appropriate for dealing with
security in the presence of resource-bounded adversaries.
\end{itemize}

There are a number of possibilities for future work; we highlight
a few of them here.
\begin{itemize}
\item 
There has been some recent work on model checking epistemic
specifications of anonymity properties \cite{vandermeyden02}.
Since we have defined syntactic analogues for various properties using
epistemic logic, it would seem that model-checking techniques can be
applied to the various secrecy properties we have considered.  However,
verifying a secrecy property requires verifying an infinite set of
formulas.  Developing techniques to do this would seem to require some
nontrivial advances to the state of the art in model checking.  Of
course, to the extent that we are interested in more limited secrecy,
where one agent does not know only certain formulas, that model-checking
techniques should be immediately applicable.
\item  Compositionality has been an important concern in 
security that
we have not addressed at all. (See \cite{mccullough87,mclean94} for
some early work on compositionality.)
It would be interesting to understand the extent to which
our notions are preserved under composition.
\item As we said at the end of Section~\ref{sec:CSP}, it would be
worthwhile to understand better the exact relationship between our
approach and the process-algebraic approach.
\item As we have said, using syntactic definitions of secrecy opens the
door to considering notions of secrecy more appropriate for
resource-bounded users.   This would then allow considering secrecy
combined with imperfect cryptography.
\end{itemize}
} %

\paragraph{Acknowledgments:}  We thank Riccardo Pucella, Andrei
Sabelfeld, and Ron van der Meyden 
for useful discussions.
We also thank 
the CSFW reviewers, who provided
helpful feedback and criticism, and Niranjan Nagarajan, who 
suggested a probability measure that led to Example~\ref{xam:no_ind}.

\appendix

\section{Examples of Systems}\label{app:examples}

In this section, we give examples of simple systems that show the
limitations of various theorems.  All the systems involve only two
agents, and we ignore the environment state.
We describe each run using the notation
$$\langle (X_{i,1},X_{j,1}), (X_{i,2},X_{j,2}), (X_{i,3},X_{j,3}),  \ldots \rangle,$$
where $X_{i,k}$ is the local state of agent $i$ at time $k$. 
For asynchronous systems, we assume that the final global state---$(X_{i,3},X_{j,3})$, 
in the example above---is repeated infinitely. 
For synchronous systems we need different
states at each time step, so we assume that
global states not explicitly listed 
encode the time in some way, so change at each time step.
For notational simplicity, we use the same symbol for a local state and
its corresponding information set.

\xam\label{xam:need_recall}
Suppose that the 
synchronous system
$\R$ consists of the following two runs:
\ul
\item $r_1 = \langle (X,A), (Y_1,B_1), (Y_2,B_2), \ldots \rangle$
\item $r_2 = \langle (Z,A), (Y_1,C_1), (Y_2,C_2), \ldots \rangle$
\eul
Note that
agent 2 has perfect recall in $\R$, but
agent 1 does not (since at time 0 agent 1 knows the run, but at all
later times, he does not).  
It is easy to check that 
agent $2$ maintains synchronous secrecy with respect to $1$,
but not run-based 
secrecy, since $\R(B_1) \inter \R(Z) = \emptyset$.

For the same reasons, if we take the probability measure $\mu$ on $\R$
with $\mu(r_1) = \mu(r_2) = 1/2$, probabilistic synchronous secrecy and
run-based probabilistic secrecy do not coincide.
This shows that the perfect recall requirement is necessary in both 
Propositions~\ref{pro:sync<->weak} and~\ref{pro:sync_prob_sec}.
\exam

\xam\label{xam:no_ind}
Suppose that the $\R$ consists of the following three runs (where, in
each case, the last state repeats infinitely often):
\ul
\item $r_1 = \langle (X,A) \ldots \rangle$
\item $r_2 = \langle (X,B), (Y,A) \ldots \rangle$
\item $r_3 = \langle (Y,A) \ldots \rangle,$
\eul

It is easy to see that agent 2 maintains run-based secrecy with respect
to agent 1 in $\R$, 
but not total secrecy or synchronous secrecy (since, for example, $Y
\inter B = \emptyset$).

Now consider a probability measure $\mu$ on $\R$ 
such $\mu(r_1) = \mu(r_3) = 2/5$, and
$\mu(r_2) = 1/5$. Then $\mu(\R(A) \mid \R(X)) = \mu(\R(A) \mid \R(Y)) = 1$ and 
$\mu(\R(B) \mid \R(X)) = \mu(\R(B) \mid \R(Y)) = 1/3$, so 
agent 2 maintains run-based probabilistic secrecy with respect to $1$ in
$\R$.  
$1$ does not maintain probabilistic secrecy with respect to $2$ in
$(\R,\mu)$, 
since $\mu(\R(X)\mid\R(A)) = 3/5$, while $\mu(\R(X)\mid\R(B)) = 1$.
Thus, if the agents do not have perfect recall and the system is not
synchronous, then run-based probabilistic secrecy is not 
necessarily
symmetric.
\exam

\xam\label{xam:recall_for_syntax}
Suppose that the synchronous system $\R$ consists of the following four runs:
\ul
\item $r_1 = \langle(X,A),(Y_1,C_1), (Y_2,C_2), \ldots \rangle$
\item $r_2 = \langle(X,B),(Y_1,D_1), (Y_2,D_2), \ldots \rangle$
\item $r_3 = \langle(Q,A),(R_1,D_1), (R_2,D_2), \ldots \rangle$
\item $r_4 = \langle(Q,B),(R_1,C_2), (R_2,C_2), \ldots \rangle$
\eul
Note that agent 2 does not have perfect recall in $\R$, although agent 1
does. Let $\mu$ give each of these runs equal probability.  It is easy to
check that for all $i \geq 1$,
$\mu(\R(A) \mid \R(X)) = \mu(\R(A) \mid \R(Q)) = 1/2$, 
$\mu(\R(B) \mid \R(X)) = \mu(\R(B) \mid \R(Q)) = 1/2$,
$\mu(\R(C_i) \mid \R(X)) = \mu(\R(C_i) \mid \R(Q)) = 1/2$, and
$\mu(\R(D_i) \mid \R(X)) = \mu(\R(D_i) \mid \R(Q)) = 1/2$.
Because $\R(X) = \R(Y_i)$ and $\R(Q) = \R(R_i)$ for all $i \geq 1$, 
it follows that agent $2$ maintains run-based probabilistic secrecy with
respect to 1 in $(\R,\mu)$.

Now, let $p$ be a primitive proposition and let $\pi$ be an
interpretation such that $p$ is true if 2's local state is either $A$ or
$D_1$.  Thus, $p$ is 2-local in $\I = (\R,\mu,\pi)$.  Since
$\mu(\R(A) \cup \R(D_1) \mid \R(X)) = 1$ while  
$\mu(\R(A) \cup \R(D_1) \mid \R(Q)) = 1/2$, there is no constant $\sigma$
such that $\I \sat \Pr_1(\di p ) = \sigma$.  This shows that the
assumption that agent $j$ has perfect recall is necessary in
Theorem~\ref{thm:runbased_prob_syntax}.
\commentout{
Suppose that the $\R$ consists of the following four runs (where, in
each case, the last state repeats infinitely often):
\ul
\item $r_1 = \langle (X,A) \ldots \rangle$
\item $r_2 = \langle (X,B), (X,C) \ldots \rangle$
\item $r_3 = \langle (Y,A), (Y,B) \ldots \rangle$
\item $r_4 = \langle (Y,C) \ldots \rangle$
\eul
Note that agent 2 does not have perfect recall in $\R$, although agent 1
does. 
Let $\mu$ give each of these runs equal probability.  It is easy to
check that 
$\mu(\R(A) \mid \R(X)) = \mu(\R(A) \mid \R(Y)) = 1/2$, 
$\mu(\R(B) \mid \R(X)) = \mu(\R(B) \mid \R(Y)) = 1/2$, and
$\mu(\R(C) \mid \R(X)) = \mu(\R(C) \mid \R(Y)) = 1/2$. 
It follows agent $2$ maintains run-based probabilistic secrecy with
respect to 1 in $(\R,\mu)$.
Let $p$ be a primitive proposition and let $\pi$ be an
interpretation such that $p$ is true if 2's local state is either $A$ or
$B$.  Thus, $p$ is 2-local in $\I = (\R,\mu,\pi)$.  Since
$\mu(\R(A) \cup \R(B) \mid \R(X)) = 1$, while  
$\mu(\R(A) \cup \R(B) \mid \R(Y)) = 1/2$, there is constant $\sigma$
such that $\I \sat \Pr_1(\di p ) = \sigma$.  This shows that the
assumption that agent $j$ has perfect recall is necessary in
Theorem~\ref{thm:runbased_prob_syntax}.
} %
\exam

\section{Proofs for Section \ref{sec:nonprobsec}}

\opro{pro:timing}
If $\R$ is a system where $i$ and $j$ have perfect recall, 
$C$ depends only on timing, and
$j$ maintains $C$-secrecy with respect to $i$, then
$j$ maintains run-based secrecy with respect to $i$.
\eopro

\prf
Given $(r,m)$ and $(r',m')$, we must find a run
$r''$ and times $m_1$ and $m_2$ such that 
$r_i''(m_1) = r_i(m)$ and $r_j''(m_2) = r_j'(m')$.
Because $C$ depends only on timing,
there exists a point $(r,n)$ such that $(r',m') \in C(r,n)$. 
The proof now splits into two cases:
\ul
\item 
Suppose that $n \geq m$.
By $C$-secrecy,
there exists a point $(r'',m_2)$ such that $r''_i(m_2) = r_i(n)$
and $r_j''(m_2) = r_j'(m')$. Because $i$ has perfect recall, 
there exists 
some $m_1 \leq m_2$ such that $r''_i(m_1) = r_i(m)$. 
\item
Suppose that $m > n$. Because $C$ depends only
on timing, there exists $n' \geq m'$ such that 
$(r',n') \in C(r,m)$. By $C$-secrecy, there exists
a point $(r'',m_2)$ such that $r''_i(m_2) = r_i(m)$
and $r_j''(m_2) = r_j'(n')$. Because $j$ has perfect recall,
there exists some $m_1 \leq m_2$ such that 
$r''_j(m_1) = r'_j(m')$.
\eul
\eprf

\opro{pro:sync<->weak} 
If $\R$ is a synchronous system where both $i$ and $j$ have
perfect recall, then agent $j$ maintains synchronous
secrecy with respect to $i$ iff $j$ maintains run-based
secrecy with respect to $i$.
\eopro

\prf
Suppose that agent $j$ maintains synchronous 
secrecy with respect to $j$ 
in $\R$. Because both $i$ and $j$
have perfect recall, $j$ maintains run-based secrecy with
respect to $i$ by Proposition~\ref{pro:timing}.

Conversely, suppose that $j$ maintains run-based secrecy 
with respect to $i$. Given runs $r,r' \in \R$ and any time $m$,
there exists a run $r''$, and times $n$ and $n'$, such that
$r_i''(n) = r_i(m)$ and $r_j''(n') = r_j'(m)$. By synchrony,
$m = n = n'$, and we have 
$r_i''(m) = r_i(m)$ and $r_j''(m) = r_j'(m)$. Thus $j$ maintains
synchronous secrecy with respect to $i$.
\eprf

\opro{pro:localsemantic}
A formula $\phi$ is  $j$-local in an interpreted system $\I = (\R,\pi)$
iff there exists a set $\Omega$ of 
$j$-information sets such that $(\I,r,m) \sat \phi$
whenever $(r,m) \in \bigcup_{\K \in \Omega} \K$.
\eopro

\prf
Suppose that $\phi$ is $j$-local. Let
$$\Omega = \{ \K_j(r,m) \mid (\I,r,m) \sat \phi \}. $$
If $(\I,r,m) \sat \phi$, then $\K_j(r,m) \in \Omega$ by
definition, so $(r,m) \in \bigcup_{\K \in \Omega} \K$.
Likewise, if $(r,m) \in \bigcup_{\K \in \Omega} \K$, then
$(r,m) \in \K_j(r',m')$ for some $(r',m')$ such that
$(\I,r',m') \sat \phi$. By $j$-locality, $(\I,r,m) \sat \phi$.

Conversely suppose that there exists a set of
$j$-information sets $\Omega$ such that $(\I,r,m) \sat \phi$
whenever $(r,m) \in \bigcup_{\K \in \Omega} \K$. We need to show
that $\phi$ is $j$-local. Suppose that $r_j(m) = r'_j(m')$.
If $(\I,r,m) \sat \phi$, then
$(r,m) \in \K_j(r'',m'')$ for some $\K_j(r'',m'') \in \Omega$,
and clearly $(r',m') \in \K_j(r'',m'') \subseteq \bigcup_{\K \in \Omega} \K$ 
too, so $(\I,r',m') \sat \phi$ by assumption. 
\eprf

\commentout{
\othm{pro:f-c-sec1}
Suppose $\R$ is a system, and $C$ is an $i$-allowability function.
Agent $j$ maintains $C$-secrecy with respect to agent
$i$ in $\R$ iff  for every interpretation $\pi$, 
if $\phi$ is $j$-local and $C(r,m)$-nontrivial in $\I = (\R,\pi)$, then 
$(\I,r,m) \sat \neg K_i \phi$.
\eothm

\prf
Suppose that $j$ maintains $C$-secrecy with respect to $i$
in $\R$. Let $\pi$ be an interpretation, let $(r,m)$ be a point in
$\PT(\R)$, and let $\phi$ be a formula that  
is $j$-local and $C(r,m)$-nontrivial in $\I = (\R,\pi)$.
Because $\phi$ is $C(r,m)$-nontrivial, there exists a point $(r',m') \in C(r,m)$
such that $(\I,r',m') \not\sat \phi$. Because $j$ maintains $C$-secrecy
with respect to $i$, there
exists
a point $(r'',m'') \in C(r,m)$ such that 
$(r'',m'') \in \K_i(r,m) \inter \K_j(r',m'))$.
Because $\phi$ is $j$-local, 
$(\I,r'',m'') \not\sat \phi$. Thus $(\I,r,m) \sat \neg K_i \phi$,
as required.

For the converse,
given $(r,m) \in \PT(\R)$ and $(r',m') \in C(r,m)$, let
$E = \PT(\R) - \K_j(r',m')$.
Let $\pi$ be an interpretation such 
that $\pi(r'',m'')(p) = {\bf true}$ iff $(r'',m'') \in E$.
Let $\I = (\R,\pi)$.
Because $(\I,r',m') \sat \neg p$, it follows that $p$ is
$C(r,m)$-nontrivial.  Moreover, 
it is clear from the definition that $p$ is $j$-local.
By assumption, 
$(\I,r,m) \sat \neg K_i p$.  Thus, there exists some point $(r'',m'')
\notin E$ such 
that $(r'',m'') \in \K_i(r,m)$.  Because $(r'',m'') \in \K_i(r,m)$, it
follows 
that $(r'',m'') \in C(r,m)$, and since $(r'',m'') \notin E$, it must be
the case that $\K_j(r'',m'') = \K_j(r',m')$.  Thus, $j$ maintains
$C$-secrecy with respect to $i$ in $\R$.
\eprf
} %

\othm{pro:f-c-sec1}
Suppose that $C$ is an $i$-allowability function.
Agent $j$ maintains $C$-secrecy with respect to agent
$i$ in system $\R$ iff, for every interpretation $\pi$ and point $(r,m)$,
if $\phi$ is $j$-local and $(\I,r',m') \sat \phi$ for some $(r',m') \in
C(r,m)$, 
then $(\I,r,m) \sat P_i \phi$.
\eothm

\prf
Suppose that $j$ maintains $C$-secrecy with respect to $i$
in $\R$. Let $\pi$ be an interpretation, let $(r,m)$ be a point,
and let $\phi$ be a formula that  
is $j$-local such that $(\I,r',m') \sat \phi$ for some $(r',m') \in C(r,m)$.
By $C$-secrecy, there
exists a point $(r'',m'') \in \K_i(r,m) \inter \K_j(r',m')$.
Because $\phi$ is $j$-local, 
$(\I,r'',m'') \sat \phi$. Thus $(\I,r,m) \sat P_i \phi$,
as required.

For the converse,
given $(r,m) \in \PT(\R)$ and $(r',m') \in C(r,m)$, let
$\pi$ be an interpretation such 
that $\pi(r'',m'')(p) = {\bf true}$ iff $(r'',m'') \in \K_j(r',m')$.
Let $\I = (\R,\pi)$.
Clearly, $p$ is $j$-local.
By assumption, 
$(\I,r,m) \sat P_i p$.  Thus, there exists some 
point $(r'',m'') \in \K_i(r,m)$ such 
that $(\I,r'',m'') \sat p$. By definition, $(r'',m'') \in \K_j(r',m')$.
Because $(r'',m'') \in \K_i(r,m) \cap \K_j(r',m')$, $j$ maintains
$C$-secrecy with respect to $i$ in $\R$.
\eprf

\othm{pro:syntax_weaktotal}
Agent $j$ maintains run-based secrecy with
respect to agent $i$ in system $\R$ iff, for every interpretation $\pi$,
if $\phi$ is $j$-local and satisfiable in $\I = (\R,\pi)$, 
then $\I \sat P_i \di \phi$.
\eothm

\prf
Suppose that $j$ maintains run-based secrecy with respect to $i$.
Let $\pi$ be an interpretation and let $\phi$ be a $j$-local
formula formula that is satisfiable in $\I = (\R,\pi)$.
Choose a point $(r,m)$.
Because $\phi$ is satisfiable, 
there exists a point $(r',m')$ such that $(\I,r',m') \sat \phi.$
Because $j$ maintains run-based secrecy with respect to $i$,
there exist a run $r''$ and times $n$ and $n'$ such
that $r_i''(n) = r_i(m)$ and $r_j''(n') = r'_j(m')$.
By $j$-locality, $(\I,r'',n') \sat \phi$. It follows that
$(\I,r'',n) \sat \di \phi$, 
and that $(\I,r,m) \sat P_i \di \phi$, as desired.

For the converse, given points $(r,m)$ and $(r',m')$, let
$\pi$ be an interpretation such 
that $\pi(r'',m'')(p) = {\bf true}$ iff $(r'',m'') \in \K_j(r',m')$.
We must show that 
$\R(\K_i(r,m)) \cap \R(\K_j(r',m')) \not= \emptyset$.
Clearly $p$ is $j$-local and satisfiable, so
$(\I,r,m) \models P_i \di p$. Thus,
there exists a point $(r'',n) \in \K_i(r,m)$ such that
$(\I,r'',n) \models \di p$. By definition of $p$, 
there exists $n'$ such that $(r'',n') \in \K_j(r',m')$.
It follows that 
$r'' \in \R(\K_i(r,m)) \cap \R(\K_j(r',m'))$.
\eprf

\commentout{
\othm{pro:syntax_weaktotal}
In any system $\R$, agent $j$ maintains run-based secrecy with
respect to agent $i$ iff for every interpretation $\pi$,
if $\phi$ is $j$-local in $\I = (\R,\pi)$, 
then either 
$\I \sat K_i \neg \di \phi$ or $\I \sat \neg K_i \neg \di \phi$.
\eothm

\prf
Suppose that $j$ maintains run-based secrecy with respect to $i$.
Let $\pi$ be an interpretation, let $\phi$ be a $j$-local
formula in $\I = (\R,\pi)$, and let $(r,m) \in \PT(\R)$. 
Suppose that $\I \not\sat \K_i \neg \di \phi$. It follows
that there exists a point $(r',m')$ such that $(\I,r',m') \sat \phi.$
We will show that $\I \sat \neg \K_i \neg \di \phi$, as
required. Let $(r,m)$ be an arbitrary point. By run-based 
secrecy, there is a run $r''$ and times $n$ and $n'$ such
that $r_i''(n) = r_i(m)$ and $r_j''(n') = r'_j(m')$.
By $j$-locality, $(\I,r'',n') \sat \phi$. It follows that
$(\I,r'',n) \sat \di \phi$, that $(\I,r'',n) \not\sat \neg \di \phi$,
and thus that $(\I,r,m) \sat \neg \K_i \neg \di \phi$, as needed.

For the converse, given $(r,m), (r',m') \in \PT(\R)$, let
Let $\pi$ be an interpretation such 
that $\pi(r'',m'')(p) = {\bf true}$ iff $(r'',m'') \in \K_j(r',m')$.
We must show that 
$$\R(\K_i(r,m)) \cap \R(\K_j(r',m')) \not= \emptyset.$$
We have $\I \not\sat \K_i \neg \di p$, because 
$(\I,r',m') \sat \di p$. Therefore, $\I \models \neg \K_i \neg \di p$.
In particular, $(\I,r,m) \models \neg \K_i \neg \di p$. Thus
there exists a point $(r'',n) \in \K_i(r,m)$ such that
$(\I,r'',n) \models \di p$. It follows that 
$r'' \in \R(\K_i(r,m)) \cap \R(\K_j(r',m'))$.
\eprf
}

\section{Proofs for Section \ref{sec:prob}}

\opro{pro:prob_total_sec}
If $(\R,\PR)$ is a probability system such that  
$\mu_{r,m,i}(\{(r,m)\}) > 0$ for all points $(r,m)$
and $j$ maintains probabilistic total secrecy with respect to $i$ in
$(\R,\PR)$, then $j$ also maintains total secrecy with respect to $i$
in $\R$.
\eopro

\prf
Suppose that $j$ maintains probabilistic
total secrecy with respect to $i$ in $(\R,\PR)$, and
$(r,m)$ and  $(r',m')$ are arbitrary points.
Then (taking $(r'',m'') = (r',m')$ in the
definition)
we have
$\mu_{r,m,i}(\K_j(r',m') \inter \K_i(r,m)) =
\mu_{r',m',i}(\K_j(r',m') \inter \K_i(r',m'))$.  But $(r',m') \in \K_j(r',m')
\inter \K_i(r',m')$, so $\mu_{r',m',i}(\K_j(r',m') \inter \K_i(r',m')) \ge
\mu_{r',m',i}(\{(r',m')\}) > 0$, by assumption. 
Thus, $\mu_{r,m,i}(\K_j(r',m') \inter \K_i(r,m)) > 0$, from which it
follows that $\K_j(r',m') \inter \K_i(r,m) \ne \emptyset$.
\eprf

The following result is proved by Gill, van der Laan, and
Robins~\citeyear{GillLR97}; see also Gr\"unwald and
Halpern~\citeyear[Theorem 3.1]{GH02}. 
(A more general version is stated and proved as Proposition
\ref{pro:plaus_independence}.) 

\lem\label{lem:independence}
\commentout{
Let $X$ be a probability space with an associated
measure $\mu$, let $Y \subseteq X$, and let $\Psi$ and $\Omega$ 
be sets of measurable subsets of $Y$ such that there exist countable subsets 
$\Psi' \subseteq \Psi$ and $\Omega' \subseteq \Omega$, both which partition $Y$.
The following are equivalent:
\ul
\item[(a)] Let $\psi \in \Psi$. For any $\omega, \omega' \in \Omega$ with positive probability, 
$\mu( \psi \mid \omega) = \mu( \psi \mid \omega')$. 
\item[(b)] Let $\omega \in \Omega$. For any $\psi, \psi' \in \Psi$ with positive probability,
$\mu( \omega \mid \psi) = \mu( \omega \mid \psi')$.
\item[(c)] For any $\psi \in \Psi$ and
$\omega \in \Omega$, both with positive measure, we have $\mu( \psi \mid \omega ) = \mu( \psi \mid Y)$.
That is, $\Psi$ and $\Omega$ are conditionally independent partitions, given Y.
\eul
\elem

\prf
We will first show that (a) implies (c).
Suppose that (a) holds, and that $\psi \in \Psi$ and $\omega \in \Omega$.
Because $\Omega'$ is a countable partition of $Y$, and $\psi \in Y$, we have
$$\mu(\psi) = \sum_{\omega' \in \Omega'} \mu( \psi \cap \omega' ). $$
By our assumption, we have $\mu( \psi | \omega') = \mu( \psi | \omega)$
for each $\omega' \in \Omega'$, and thus
$$ \mu( \psi \cap \omega' ) = \mu( \psi \mid \omega') \cdot \mu( \omega') = \mu( \psi \mid \omega) \cdot \mu( \omega').$$
Therefore,
\begin{eqnarray*}
\mu(\psi) &=& \sum_{\omega' \in \Omega} \mu(\psi \mid \omega) \cdot \mu(\omega') \\
 &=& \mu( \psi \mid \omega ) \cdot \sum_{\omega' \in \Omega} \mu(\omega') \\
 &=& \mu(\psi \mid \omega) \cdot \mu(Y).
\end{eqnarray*}
But since $\psi \in Y$, we have $\mu(\psi) = \mu(\psi \mid Y) \cdot \mu(Y)$.
It follows that $\mu(\psi \mid \omega) = \mu(\psi \mid Y)$.

By symmetry, (b) implies (c). Trivially, (c) implies both (a) and (b).
\eprf
}
Suppose that $\mu$ is a probability on $W$, $X, Y \subseteq W$,
$Y_1, Y_2, \ldots$ is a 
countable
partition of $Y \subseteq W$, and $X, Y_1, Y_2,
\ldots$ are all measurable.  
The following are equivalent:
\ul
\item[(a)] $\mu(X \mid Y_i) = \mu(X \mid Y_j)$ for all $Y_i, Y_j$ such
that $\mu(Y_i) > 0$ and $\mu(Y_j) > 0$.
\item[(b)] $\mu(X \mid Y_i) = \mu(X \mid Y)$ for all 
$Y_i$ such that $\mu(Y_i) > 0$,
i.e., $Y_i$ is conditionally independent of $X$ given $Y$. 
\eul
\elem

\opro{pro:sync_independence}
If $(\R,\PR)$ is a probability system (resp., synchronous probability system) 
that satisfies the 
common prior assumption with prior probability $\mu_{\mathit{cp}}$, the
following are 
equivalent: 
\begin{itemize}
\item[(a)] Agent $j$ maintains probabilistic total (resp., synchronous) secrecy with respect
to $i$.
\item[(b)] Agent $i$ maintains probabilistic total (resp., synchronous) secrecy with respect
to $j$.
\item[(c)] For all points $(r,m)$ and $(r',m')$, 
$\mu_{\mathit{cp}}(\K_j(r',m') \mid \K_i(r,m)) =
\mu_{\mathit{cp}}(\K_j(r',m'))$ (resp., for all points
$(r,m)$ and $(r',m)$, $\mu_{\mathit{cp}}(\K_j(r',m) \mid \K_i(r,m)) 
= \mu_{\mathit{cp}}(\K_j(r',m) \mid \PT(m))$, where
$\PT(m)$ is the set of points occurring at time $m$;
that is, the events
$\K_i(r,m)$ and $\K_j(r',m)$ are conditionally independent with respect
to $\mu_{\mathit{cp}}$, given that the time is $m$).
\end{itemize}
\eopro

\prf
We prove the synchronous case here.  The proof 
for total secrecy
is almost
identical and left to the reader.
Recall that $j$ maintains probabilistic synchronous secrecy with respect
to $i$  
if, for all times $m$ and all runs $r,r',r''$, 
$$\mu_{r,m,i}(\K_j(r'',m) \inter \K_i(r,m)) = 
\mu_{r',m,i}(\K_j(r'',m) \inter \K_i(r',m)).$$
Because $(\R,\PR)$ satisfies the common prior assumption with 
prior probability
$\mu_{\mathit{cp}}$, this 
requirement can be restated as
$$
\mu_{\mathit{cp}}( \K_j(r'',m) \mid \K_i(r,m) ) = \mu_{\mathit{cp}}(
K_j(r'',m) \mid \K_i(r',m)). 
$$
By Lemma~\ref{lem:independence}, taking $Y = \PT(m)$ and the $Y_i$'s to
be the $i$-information sets at time $m$, it follows that $j$ maintains
probabilistic synchronous secrecy with respect to $i$ iff 
$\K_j(r'',m)$ is conditionally independent of $\K_i(r,m)$ conditional on
$\PT(m)$ for all runs $r$ and $r''$.  By the symmetry of conditional
independence, it immediately follows that this is true iff $i$ maintains
probabilistic synchronous secrecy with respect to $j$.
\eprf

\lem\label{lem:omegapartition}
If $\R$ is a system where agent $i$ has perfect recall and
$\Omega$ is an arbitrary set of $i$-information sets, then there
exists a set $\Omega' \subseteq \Omega$ such that 
$\{ R(\K) \mid \K \in \Omega' \}$
is a partition of $\bigcup_{\K \in \Omega} \R(\K).$
\elem

\prf
Define a set $\K \in \Omega$ to be {\em dominated} by a set $\K' \in \Omega$
if $\K \not= \K'$ and there exists a run $r$ and times $m' < m$ such that
$(r,m) \in \K$ and $(r,m') \in \K'$. Let $\Omega'$ consist of the information
sets in $\Omega$ that are not dominated by another set in $\Omega$. 
Note that if  $r \in \union_{\K \in \Omega} \R(\K)$, then $r \in
\R(\K')$ for some $\K' \in \Omega'$.  To see this, 
consider the set $\Omega(\K)$  
consisting of $\K$ and 
all information sets in $\Omega$ that dominate $\K$. By perfect recall,
$i$'s local state sequence at each information set in $\Omega(\K)$
is a (not necessarily strict) prefix of $i$'s local state sequence in
$\K$.  Let $\K'$ be the information set in $\Omega(\K)$ where $i$'s
local state sequence is shortest.  It follows 
that $\K'$ is not dominated by 
another information set in 
$\Omega(\K)$. Furthermore, 
if there exists an information set $\K'' \in \Omega - \Omega(\K)$ that
dominates $\K'$, then $\K''$ would dominate $\K$ as well, contradicting 
the construction of $\Omega(K)$.
Therefore, $\K' \in \Omega'$ and $r \in \K'$. Thus 
$\bigcup_{\K \in \Omega'} \R(\K) = \bigcup_{\K \in \Omega} \R(\K)$.
Moreover, if $\K$ and $\K'$ are different sets in $\Omega'$, then
$\R(\K)$ and $\R(\K')$ must be disjoint, for otherwise one of $\K$ or $\K'$ 
would dominate the other.
\eprf

\commentout{
\lem\label{lem:run_based_independence}
Let $(\R,\F,\mu)$ be a run-based probability system, and suppose that there
exist sets $\Psi$ and $\Omega$
of $i$-information sets and $j$-information sets such that
$\{ R(\K) \mid \K \in \Psi \}$ and $\{ R(\K) \mid \K \in \Omega \}$ both
partition $\R$. Then the following are equivalent:
\begin{itemize}
\item[(a)] Agent $j$ maintains run-based probabilistic secrecy with respect
to $i$.
\item[(b)] Agent $i$ maintains run-based probabilistic secrecy with respect
to $j$.
\item[(c)] For all points $(r,m), (r',m') \in \PT(\R)$,
$\R(\K_i(r,m))$ and $\R(\K_j(r',m'))$ are probabilistically independent
with respect to $\mu$.
\end{itemize}
\elem

\prf
Suppose that $j$ maintains run-based probabilistic secrecy with respect 
to $i$. By definition, 
$$\mu( \R(\K_j(r'',m'')) \mid \R(\K_i(r,m)) ) = \mu( \R(\K_j(r'',m'')) \mid \R(\K_i(r',m')) )$$
for all points $(r,m)$, $(r',m')$, and $(r'',m'')$.  In particular, for all
$\K, \K' \in \Psi$, and all points $(r',m')$,
$\mu( \R(\K_j(r',m') \mid \R(\K)) ) = \mu( \R(\K_j(r',m')) \mid \R(\K') )$.
By Lemma~\ref{lem:independence} it follows that
$\mu(\R(\K_j(r',m'')) \mid \R(\K)) = \mu(\R(\K_j(r',m'')))$ for all information sets
$\K \in \Psi$.  But then it follows by secrecy that 
$\mu(\R(\K_j(r',m')) \mid \R(\K_i(r,m))) = \mu(\R(\K_j(r',m'))$ for
all $i$-information sets $\R(\K_i(r,m))$.  Therefore $\R(\K_j(r',m'))$ and
$\R(\K_i(r,m))$ are independent for all information sets $\K_j(r',m')$ and $\K_i(r,m)$.  
Thus secrecy implies independence, and this holds if we reverse the roles of $i$ and $j$.

It is also clear that independence implies secrecy. Suppose that (c) holds and let
$(r,m), (r',m')$ and $(r'',m'')$ be given. We have
$$\mu( \R(\K_j(r'',m'')) \mid \R(\K_i(r,m)) ) = \mu( \R(\K_j(r'',m''))) = \mu( \R(\K_j(r'',m'')) \mid \R(\K_i(r',m')) )$$
so that $j$ maintains run-based probabilistic secrecy with respect to $i$. Similarly, $i$
maintains secrecy with respect to $j$.
\eprf
}

\opro{pro:run_based_independence}
If $(\R,\F,\mu)$ is a run-based probability system that is either
synchronous or one where agents $i$ and $j$ both have perfect recall, then
the following are
equivalent: 
\begin{itemize}
\item[(a)] Agent $j$ maintains run-based probabilistic secrecy with respect
to $i$.
\item[(b)] Agent $i$ maintains run-based probabilistic secrecy with respect
to $j$.
\item[(c)] For all points 
$(r,m)$ and $(r',m')$,
$\R(\K_i(r,m))$ and $\R(\K_j(r',m'))$ are probabilistically independent
with respect to $\mu$.
\end{itemize}
\eopro

\prf
First, note that if $\R$ is synchronous or if $i$ has perfect recall, then 
there exists a collection $\Omega$ of $i$-information sets such that the 
set $\{ R(\K) \mid \K \in \Omega \}$ is a partition of $\R$. In the case
of perfect recall, this follows by Lemma~\ref{lem:omegapartition}
applied to the set of all information sets (whose union is clearly $\R$).
With synchrony we can take $\Omega$ to consist of sets of the form
$\R(\K_i(r,m))$, for some fixed time $m$.

Now, suppose $j$ maintains run-based probabilistic secrecy with respect 
to $i$. By definition, 
$$\mu( \R(\K_j(r'',m'')) \mid \R(\K_i(r,m)) ) = \mu( \R(\K_j(r'',m'')) \mid \R(\K_i(r',m')) )$$
for all points $(r,m)$, $(r',m')$, and $(r'',m'')$.  In particular, for all
$\K, \K' \in \Omega$, and all points $(r',m')$,
$\mu( \R(\K_j(r',m') \mid \R(\K)) ) = \mu( \R(\K_j(r',m')) \mid \R(\K') )$.
By Lemma~\ref{lem:independence} it follows that
$\mu(\R(\K_j(r',m'')) \mid \R(\K)) = \mu(\R(\K_j(r',m'')))$ for all information sets
$\K \in \Omega$.  But then it follows by secrecy that 
$\mu(\R(\K_j(r',m')) \mid \R(\K_i(r,m))) = \mu(\R(\K_j(r',m'))$ for
all $i$-information sets $\R(\K_i(r,m))$.  Therefore $\R(\K_j(r',m'))$ and
$\R(\K_i(r,m))$ are independent for all information sets $\K_j(r',m')$ and $\K_i(r,m)$.  
Thus secrecy implies independence, and this holds if we reverse the roles of $i$ and $j$.

It is also clear that independence implies secrecy. For suppose that
(c) holds .
Then, for all points $(r,m)$, $(r',m')$, and $(r'',m'')$, we have
$$\mu( \R(\K_j(r'',m'')) \mid \R(\K_i(r,m)) ) = \mu( \R(\K_j(r'',m'')))
= \mu( \R(\K_j(r'',m'')) \mid \R(\K_i(r',m')) ),$$ 
so that $j$ maintains run-based probabilistic secrecy with respect to $i$. Similarly, $i$
maintains secrecy with respect to $j$.
\eprf

\opro{pro:sync_prob_sec}
If $(\R,\PR)$ is the standard system determined by the synchronous
run-based probability system $(\R,\F,\mu)$ and agents $i$ and $j$ have
perfect recall in $\R$, then agent
$j$ maintains run-based probabilistic secrecy with respect to $i$ in
$(\R,\F,\mu)$ iff  
$j$ maintains probabilistic synchronous secrecy with respect to $i$ in
$(\R,\PR)$. 
\eopro

\prf 
Clearly if $j$ maintains run-based probabilistic secrecy with respect to
$i$ in $(\R,\mu)$ and $(\R,\PR)$ is the standard system determined by
$(\R,\mu)$ then, at all times $m$, 
\begin{eqnarray*}
\mu_{r,m,i}(\K_j(r'',m) \inter \K_i(r,m)) &=& \mu(\K_j(r'',m) \mid \K_i(r,m)) \\
 &=& \mu(\K_j(r'',m) \mid \K_i(r',m)) \\
 &=& \mu_{r',m,i}(\K_j(r'',m) \inter \K_i(r',m)),
\end{eqnarray*}
so $j$ maintains probabilistic synchronous secrecy with respect to $i$ in
$(\R,\PR)$. 

For the converse, suppose
that $j$ maintains probabilistic synchronous secrecy with respect to $i$
in $(\R,\PR)$. 
We want to show that, for all points $(r,m)$, $(r',m')$, and $(r'',m'')$,
\begin{eqnarray}\label{toshow1}
\mu( \R(\K_j(r'',m'')) \mid \R(\K_i(r,m))) = \mu( \R(\K_j(r'',m'')) \mid
\R(\K_i(r',m'))).
\end{eqnarray}
We first show that,
for all runs $r$ and $r''$ and times $m$ and $m''$,
\begin{equation}\label{eqsymmetry}
\mu( \R(\K_j(r'',m'')) \mid \R(\K_i(r,m))) = \mu( \R(\K_j(r'',m'')) \mid
\R(\K_i(r,m''))).
\end{equation}
Since (\ref{eqsymmetry}) also holds with $r$ replaced by $r'$,
(\ref{toshow1}) easily follows from (\ref{eqsymmetry}) and the
assumption that $j$ maintains probabilistic synchronous
secrecy with respect to $i$.

To prove (\ref{eqsymmetry}), we consider two cases: $m \le m''$ and $m'' < m$.
If $m \leq m''$ then, by synchrony and
perfect recall,
we can partition the runs in $\R(\K_i(r,m))$ according to $i$'s local
state at time $m''$.
Let
$\Omega = \{ \K_i(r^*,m'') \mid r^* \in \R(\K_i(r,m)) \}$.
By perfect recall and synchrony, $\R(\K_i(r,m))$ is the disjoint union
of the sets in $\Omega$.
\commentout{
To see this, let $t \in \R(\K_i(r,m))$. We have $t_i(m) = r_i(m)$, so
$\K_i(t,m'') \in \Omega$, and 
$$t \in \R(\K_i(t,m'')) \subseteq \bigcup_{\K \in \Omega} \R(\K).$$
Conversely, suppose $t \in \bigcup_{\K \in \Omega} \R(\K)$, so that there
exists a run $t'$ such that $t'_i(m) = r_i(m)$, and $t \in \R(\K_i(t',m''))$.
Thus we also have $t'_i(m'') = t_i(m'')$. We need to show that $t_i(m) = r_i(m)$,
from which it follows that $t \in \R(\K_i(r,m))$.
By perfect recall, because 
$t'_i(m'') = t_i(m'')$ we have $t_i(m) = t'_i(m)$. It follows that 
$t_i(m) = r_i(m)$.
It follows that
$$  \mu( \R(\K_j(r'',m'')) \cap \R(\K_i(r,m)) ) = \sum_{\K \in \Omega}
\mu( \R(\K_j(r'',m'')) \cap \R(\K) ) $$
and
$$ \mu(\R(\K_i(r,m))) =  \sum_{\K \in \Omega} \mu(\R(\K)). $$
}
Thus,
$$\begin{array}{lll}
&\mu(\R(\K_j(r'',m'')) \mid \R(\K_i(r,m)) )\\
 =&
\sum_{\K \in \Omega} \mu( \R(\K_j(r'',m'')) \cap \R(\K) \mid \R(\K_i(r,m)) ) \\
= &\sum_{\K \in \Omega} \mu( \R(\K_j(r'',m'')) \mid \R(\K) ) \cdot
  \mu( \R(\K) \mid \R(\K_i(r,m)) ) \\ 
  =& \mu( \R(\K_j(r'',m'')) \mid \R(\K_i(r,m'')) ) \cdot \sum_{\K \in \Omega}  \mu(
\R(\K) \mid \R(\K_i(r,m)) )  &\mbox{[by synchronous secrecy]}\\
  =& \mu( \R(\K_j(r'',m'')) \mid \R(\K_i(r,m'')) ).
\end{array}$$
The argument is similar if $m'' < m$.  We now partition the
runs in $\R(\K_i(r,m''))$ according to $i$'s local state at time $m$
and  the runs in $\R(\K_j(r'',m''))$ according to $j$'s local
state at time $m$.
Define
$$\Omega_i = \{ \K_i(r^*,m) \mid r^* \in \R(\K_i(r,m'') \}.$$
and
$$\Omega_j = \{ \K_j(r^*,m) \mid r^* \in \R(\K_j(r'',m'')) \}.$$
We now have
$$\begin{array}{ll}
&\mu( \R(\K_j(r'',m'')) \mid \R(\K_i(r,m'')) )\\
=& \sum_{\K_j \in \Omega_j} \mu( \R(\K_j) \mid  \R(\K_i(r,m'')) )
\\
=& \sum_{\K_j \in \Omega_j} \sum_{\K_i \in \Omega_i} \mu( \R(\K_j) \mid
R(\K_i) ) \cdot \mu( \R(\K_i) \mid \R(\K_i(r,m'')) ) \\
=& \sum_{\K_j \in \Omega_j} \mu( \R(\K_j) \mid
R(\K_i) ) \cdot \sum_{\K_i \in \Omega_i} \mu( \R(\K_i) \mid \R(\K_i(r,m'')) ) \\
 =& \sum_{\K_j \in \Omega_j} \mu( \R(\K_j) \mid \R(\K_i(r,m)) ) \\
 =& \mu( \R(\K_j(r'',m'')) \mid \R(\K_i(r,m)) ),
\end{array}$$
as needed.
\eprf

\commentout{
\lem\label{lem:sync_prob_syntax}
In a synchronous probabilistic system $(R,\mu)$, $j$ maintains synchronous
probabilistic secrecy with respect to $i$ if and only if, for every time
$m$, and  
class $\Omega$ of $j$-information sets
(that is, sets of the form $\{ \K_j(r,m) \mid (r,m) \in P \}$
for some $P \subseteq \PT(\R)$), there exists a constant $\sigma_m$ such that for every run $r \in \R$,
$$ \mu_{r,m,i}( \bigcup_{\K \in \Omega} \K ) = \sigma_m. $$
\elem

\prf
Suppose that $j$ maintains synchronous probabilistic secrecy with respect to $i$
in $(R,\mu)$, and let $m$ and $\Omega$ be given. Let 
$\Psi = \bigcup_{\K \in \Omega} \K$. We must show that exists $\sigma_m$ such that
for all runs $r \in \R$, $\mu_{r,m,i}(\Psi) = \sigma_m$.
Let $S = \{ t \in \R \mid (t,m) \in \Psi \},$ and let
$\Omega(m) = \{ \K \in \Omega \mid (t,m) \in \K \mbox{~for some~} t \in \R \}.$
By secrecy, for every element $\K \in \Omega(m)$, there is a constant $\sigma(\K,m)$
such that $\mu( \R(\K) \mid \R(\K_i(r,m)) ) = \sigma(\K,m)$, for all runs $r \in \R$. 
Let $\sigma_m = \sum_{ \K \in \Omega(m) } \sigma(\K,m)$. Fix $r \in \R$.
Because $\Psi \cap \K_i(r,m) = \K_i(r,m)(S)$, we have
$\mu_{r,m,i}(\Psi) = \mu( S \mid \R(\K_i(r,m)) )$ (by definition).
But by synchrony, the set $\{ \R(\K) \mid \K \in \Omega(m) \}$ gives us a partition of $S$, and
$$\mu( S \mid \R(\K_i(r,m)) ) = \sum_{ \K \in \Omega(m) } \mu( \R(K)
\mid \R(\K_i(r,m)) ) = \sigma_m,$$ 
as required.

The converse is trivial.
\eprf
}%

\othm{thm:sync_prob_syntax}
\begin{itemize}
\item[(a)] If $(\R,\PR)$ is a probabilistic system, then 
agent $j$ maintains probabilistic total secrecy with respect to
agent $i$ iff,
for every interpretation $\pi$
and formula $\phi$ that is $j$-local in $\I = (\R,\PR,\pi)$, 
there exists a constant $\sigma$ such that 
$\I \sat \Pr_i(\phi) = \sigma$.
\item[(b)] 
If $(\R,\PR)$ is a synchronous probabilistic system, then 
agent $j$ maintains probabilistic synchronous secrecy with respect to
agent $i$ iff, 
for every interpretation $\pi$, time $m$, 
and formula $\phi$ that is $j$-local in $\I = (\R,\PR,\pi)$, 
there exists a constant $\sigma_m$ such that 
$(\I,r,m) \sat \Pr_i(\phi) = \sigma_m$ for all runs $r \in \R$.
\end{itemize}
\eothm

\prf
We prove part (b) here.  The proof of (a) is similar.

Suppose $\R$ is synchronous and 
that $j$ maintains synchronous probabilistic secrecy with
respect to $i$. 
Let $\pi$ be an interpretation, $m$ be an arbitrary time, and $\phi$ be a 
$j$-local formula in $\I = (\R,\mu,\pi)$.
Because $\phi$ is $j$-local,
by Proposition~\ref{pro:localsemantic},
there exists a set $\Omega$ of $j$-information sets such that 
$(\I,r,m) \sat \phi$
iff
$(r,m) \in \bigcup_{\K \in \Omega} \K$.
Let $\Psi = \bigcup_{\K \in \Omega} \K$.
Let $S = \{ r' \in \R \mid (r',m) \in \Psi \},$ and let
$\Omega(m) = \{ \K \in \Omega \mid (r',m) \in \K \mbox{~for some~} r' \in
\R \}.$ 
Since $j$ maintains synchronous probabilistic secrecy with respect to $i$,
for every element $\K \in \Omega(m)$, there is a constant $\sigma(\K,m)$
such that, for all runs $r \in \R$, $\mu( \R(\K) \mid \R(\K_i(r,m)) ) =
\sigma(\K,m)$.  
Let $\sigma_m = \sum_{ \K \in \Omega(m) } \sigma(\K,m)$, and fix $r \in \R$.
By synchrony, the set $\{ \R(\K) \mid \K \in \Omega(m) \}$ 
partitions $S$, and
$$\mu( S \mid \R(\K_i(r,m)) ) = \sum_{ \K \in \Omega(m) } \mu( \R(\K)
\mid \R(\K_i(r,m)) ) = \sigma_m.$$ 
Because $\Psi \cap \K_i(r,m) = \K_i(r,m)(S)$, we have
$\mu_{r,m,i}(\Psi) = \mu( S \mid \R(\K_i(r,m)) )$, and it follows that
$(\I,r,m) \sat \Pr_i(\phi) = \sigma_m$, as desired.

For the converse,
suppose that for every interpretation $\pi$ and time $m$, 
if $\phi$ is $j$-local in $\I = (\R,\mu,\pi)$, 
then
there exists a constant $\sigma_m$ such that 
$(\I,r,m) \sat \Pr_i(\phi) = \sigma_m$ for all runs $r \in \R$.
Fix a time $m$.
Suppose that $r, r', r'' \in \R$ and 
that
$\pi$ is an interpretation
such that $\pi(r^*,n)(p) = {\bf true}$ iff $(r^*,n) \in \K_j(r'',m)$.
The proposition $p$ is $j$-local, so there exists a constant $\sigma_m$
such that 
$(\I,r,m) \sat \Pr_i(p) = \sigma_m$ and
$(\I,r',m) \sat \Pr_i(p) = \sigma_m$. It follows that
$$\mu_{r,m,i}(\K_j(r'',m)) = \sigma_m = \mu_{r',m,i}(\K_j(r'',m)),$$
as desired.
\eprf

\commentout{
\lem\label{lem:probsec}
In a probabilistic system $(R,\mu)$ where agent $j$ has perfect recall, 
$j$ maintains probabilistic secrecy with respect to agent $i$ if and only if, for every
class $\Omega$ of $j$-information sets, there exists a constant $\sigma$ such that
for every point $(r,m) \in \PT(\R)$,
$$ \mu( \bigcup_{\K \in \Omega} \R(\K) \mid \R(\K_i(r,m))) = \sigma. $$
\elem

\prf
Let $\Psi = \bigcup_{\K \in \Omega} \R(\K).$ 
By Lemma~\ref{lem:omegapartition}, there exists a set $\Omega' \subseteq \Omega$
that partitions $\Psi$.
By probabilistic secrecy, for each $\K \in \Omega'$, there exists a constant
$\sigma(\K)$ such that 
$$\mu( \R(\K) \mid \R(\K_i(r,m)) ) = \sigma(\K)$$ 
for all
points $(r,m) \in \PT(\R)$. Let $\sigma = \sum_{\K \in \Omega'} \sigma(\K)$. 
Because $\{ \R(\K) \mid \K \in \Omega' \}$ is a partition of $\Psi$, for any
point $(r,m) \in \PT(\R)$,
$$ \mu( \Psi \mid \R(\K_i(r,m)) ) = \sum_{\K \in \Omega} \mu( \R(\K) \mid \R(\K_i(r,m)) ) = \sigma,$$
as required.

The converse is trivial. Given points $(r,m)$, $(r',m')$, $(r'',m'')$, let 
$\Omega = \{ \K_j(r'',m'') \}$.  By assumption, there exists $\sigma$ such that
$$ \mu( \R(\K_j(r'',m'')) \mid \R(\K_i(r,m)) ) = \mu( \R(\K_j(r'',m'')) \mid \R(\K_i(r',m'))) = \sigma, $$
as required.
\eprf
}

\othm{thm:runbased_prob_syntax}
If $(\R,\PR)$ is a standard probability system where agent $j$ has
perfect recall, then agent $j$ maintains 
run-based probabilistic secrecy with
respect to agent $i$ iff, for every interpretation $\pi$
and every formula $\phi$ that is $j$-local
in $\I = (\R, \PR, \pi)$, there exists a constant $\sigma$ such that
$\I \sat \Pr_i(\di \phi) = \sigma$. 
\eothm

\prf
Suppose that $j$ maintains probabilistic secrecy with
respect to agent $i$
in $(\R,\mu)$.
Given an interpretation $\pi$ and a formula $\phi$ 
that is $j$-local
in $\I = (\R, \mu, \pi)$, by Proposition~\ref{pro:localsemantic}
there exists a set $\Omega$ of $j$-information sets such that $(\I,r,m) \sat \phi$
whenever $(r,m) \in \bigcup_{\K \in \Omega} \K$. Let $\Psi = \bigcup_{\K \in \Omega} \R(\K).$
Note that
$(\I,r,m) \sat \di \phi$ iff
$r \in \R(\bigcup_{\K \in \Omega} \K) = \Psi.$
\commentout{
Now, there exists a set $\Omega' \subseteq \Omega$
that generates a partition of $\Psi$:
\ul
\item If $j$ has perfect recall, this follows by Lemma~\ref{lem:omegapartition}.
\item If the system is synchronous, let $\K \in \Omega'$ if $\K \in \Omega$ 
and if, for all runs $r$, $\K = \K_j(r,m)$ implies that that for all $m'
\leq m$, 
$\K_j(r,m') \not\in \Omega$. 
\eul
}
By Lemma~\ref{lem:omegapartition}, 
there exists a set $\Omega' \subseteq \Omega$ such that $\{\R(\K): 
\K \in \Omega\}$ is a partition of 
$\Psi$. 
By probabilistic secrecy, for each $\K \in \Omega'$, there exists a constant
$\sigma_{\K}$ such that 
$$\mu( \R(\K) \mid \R(\K_i(r,m)) ) = \sigma_{\K}$$ 
for all
points $(r,m)$. Let $\sigma = \sum_{\K \in \Omega'}
\sigma_{\K}$.  
Because $\{ \R(\K) \mid \K \in \Omega' \}$ is a partition of $\Psi$, for 
all points $(r,m)$,
$$ \mu( \Psi \mid \R(\K_i(r,m)) ) = \sum_{\K \in \Omega} \mu( \R(\K) \mid \R(\K_i(r,m)) ) = \sigma.$$
Because $\mu_{r,m,i}(\K_i(r,m)(\Psi)) = \mu(\Psi \mid \R(\K_i(r,m)))$, it follows that
$\I \sat \Pr_i(\di \phi) = \sigma$.

For the converse, suppose that for every interpretation $\pi$ and 
formula $\phi$ that is $j$-local in $\I = (\R,\mu,\pi)$, 
there exists a constant $\sigma$ such that
$\I \sat \Pr_i(\di \phi) = \sigma$. Given 
points $(r,m)$, $(r',m')$, and $(r'',m'')$,
let $\pi$ be an interpretation such that 
$\pi(r^*,n)(p) = {\bf true}$  iff $(r^*,n) \in \K_j(r'',m'')$. The
proposition $p$ is $j$-local,  
so $\I \sat \Pr_i(\di p) = \sigma$. It follows that
$$\mu( \R(\K_j(r'',m'')) \mid \R(\K_i(r,m)) ) =
\mu_{r,m,i}(\K_i(r,m)(\R(\K_j(r'',m'')))) = \sigma,$$ 
and the same holds if we replace $(r,m)$ with $(r',m')$, so
$$\mu( \R(\K_j(r'',m'')) \mid \R(\K_i(r,m)) ) = \mu( \R(\K_j(r'',m''))
\mid \R(\K_i(r',m'))).$$ 
This gives us probabilistic secrecy.
\eprf

\othm{thm:evidential}
Let $(\R, \D, \Delta)$ be the adversarial probability system determined
by $\INIT$ and suppose that $\R$ 
is either synchronous or a system where 
$i$ has perfect recall.
Agent $i$ obtains no
evidence for the initial choice in $(\R,\D,\Delta)$ iff agent $i^-$ maintains
generalized run-based probabilistic $f_{i^-}$-secrecy with respect to $i$ 
in $(\R,\M^{\INIT}_i(\Delta))$.
\eothm

\prf
For the forward direction, 
we want to show that $i^-$ maintains generalized run-based probabilistic
$f_{i^-}$-secrecy with respect to $i$ in $(\R,\M^{\INIT}_i(\Delta))$.
Suppose that $\mu \in \M^{\INIT}_i(\Delta)$.  The information function
$f_{i^-}$ maps an ${i^-}$-information set to the choices made by the
agents other than $i$.  Let an 
{\em $i^-$-choice set\/} 
be a set of runs of
the form $\inter_{j \ne i} D_{y_j}$.
We must show that for arbitrary points $(r,m)$ and $(r',m')$ and 
$i^-$-choice sets $D_{i^-}$, we have
\begin{equation}\label{ev_sec_eq}
\mu( D_{i^-} \mid \R(\K_i(r,m) )) =  \mu( D_{i^-} \mid \R(\K_i(r',m')) ). 
\end{equation}
\noindent
Since, by assumption,  $i$'s choice is encoded $i$'s local state, 
there exists a unique $y_i$ such that $\R(\K_i(r,m)) \subseteq D_{y_i}$. 
Since $i$ obtains no evidence for the initial choice, we have 
that for all $i^-$-choice sets $D_{i^-}$ and $D_{i^-}'$,
\begin{equation}\label{eqnoevidence}
\mu_{D_{y_i} \inter D_{i^-}}(\R(\K_i(r,m))) = 
\mu_{D_{y_i} \inter D_{i^-}'}(\R(\K_i(r,m))).
\end{equation}
Thus, whenever 
$\mu(D_{y_i} \inter D_{i^-}) > 0$ and  $\mu(D_{y_i} \inter D_{i^-}') > 0$, 
we have
\begin{eqnarray*}
\mu(\R(\K_i(r,m)) \mid D_{y_i} \inter D_{i^-}) 
&=& \mu_{D_{y_i} \inter D_{i^-}}(\R(\K_i(r,m))) \\
&=& \mu_{D_{y_i} \inter D_{i^-}'}(\R(\K_i(r,m))) \\
&=& \mu(\R(\K_i(r,m)) \mid D_{y_i} \inter D_{i^-}').
\end{eqnarray*}
It now follows by Lemma~\ref{lem:independence} that $\R(\K_i(r,m))$ is
conditionally independent of every $i^-$-choice set given $D_{y_i}$.
(Though Lemma~\ref{lem:independence} actually shows
only that $\R(\K_i(r,m))$ is conditionally independent of every $i^-$
choice set $D_{i^-}$ such that $\mu(D_{i^-} \inter D_{y_i}) > 0$,
conditional independence is immediate if $\mu(D_{i^-} \inter D_{y_i}) = 0$)   
Thus, for any $i^-$-choice set $D_{i^-}$, we have
$$\mu(D_{i^-} \mid \R(\K_i(r,m))) = \mu(D_{i^-} \mid \R(\K_i(r,m)) \inter D_{y_i}) =
\mu(D_{i^-} \mid D_{y_i}) = \mu(D_{i^-}),$$ 
where the last equality follows because
we have assumed that $i$'s choice is independent of the 
choices made by other agents.   
Similarly, $\mu(D_{i^-} \mid \R(\K_i(r',m'))) = \mu(D_{i^-})$, so
(\ref{ev_sec_eq}) follows, and 
$i^-$ does indeed maintain generalized run-based probabilistic
$f_{i^-}$-secrecy with respect to $i$.

For the converse, suppose that $i^-$ maintains generalized run-based
probabilistic $f_{i^-}$-secrecy with respect to $i$.  
Thus, for all points $(r,m)$, $i^-$-choice sets $D_{i^-}$,
and measures $\mu \in \M^{\INIT}_i(\Delta)$, we have 
(\ref{ev_sec_eq}).
Given two $i^-$-choice sets $D_{i^-}$ and $D_{i^-}'$ and an $i$-information set
$\K_i(r,m)$ such that $\R(\K_i(r,m)) \subseteq D_{y_i}$, we want to show
(\ref{eqnoevidence}).  
To do so we first show that 
there exists a measure $\mu \in \M^{\INIT}_i(\Delta)$
that places positive probability on all the cells.
(We will make use of this particular measure for the duration of the proof.)
Our strategy is to take a countable linear combination of the
cell-specific 
probability measures, such that the set of runs in each cell is assigned positive probability 
by $\mu$. Let $y_{i1}, y_{i2}, \ldots$ be a countable enumeration of $\INIT_i$, and let
$D_1, D_2, \ldots$ be a countable enumeration of the possible $i^-$-choice sets.
Define the function $\mu$ such that for $U \in \F$,
$$\mu(U) = \sum_{ j \geq 1, k \geq 1 } 
  \frac{\mu_{D_{y_{ij}} \inter D_k}(U \inter D_{y_{ij}} \inter D_k)}{2^{jk}}.$$
It is straightforward to check that $\mu \in \M^{\INIT}_i(\Delta)$ and that it places a
positive probability on 
all
the cells in $\D$. 
Furthermore, we have
$\mu_{D_{y_i} \inter D_{i^-}}(\R(\K_i(r,m))) = \mu(\R(\K_i(r,m)) \mid
D_{y_i} \inter D_{i^-})$, and the 
same holds if we replace $D_{i^-}$ with $D_{i^-}'$.

Given an $i$-information set $\K_i(r,m)$,
let $y_i$ be 
the
initial choice for $i$ such that $\R(\K_i(r,m))
\subseteq  D_{y_i}$.  
For all
$i^-$ choice sets $D_{i^-}$, we 
have
$$\mu_{D_{y_i} \inter D_{i^-}}(\R(\K_i(r,m))) = \mu(\R(\K_i(r,m) \mid D_{y_i} \inter D_{i^-}).$$
Thus, to prove (\ref{eqnoevidence}), it suffices to show that 
$$\mu(\R(\K_i(r,m) \mid D_{y_i} \inter D_{i^-}) = 
\mu(\R(\K_i(r,m) \mid D_{y_i} \inter D_{i^-}').$$
Standard probabilistic manipulations show that
\begin{equation}\label{eq.5}
\mu( \R(\K_i(r,m)) \mid D_{y_i} \inter D_{i^-}) \cdot \mu( D_{y_i} \mid D_{i^-} )
  = \mu( \R(\K_i(r,m)) \inter D_{y_i} \mid D_{i^-} );
\end{equation}
a similar equation holds if we replace $D_{i^-}$ by $D_{i^-}'$.
Since either $\R$ is synchronous or $i$ has perfect recall in $\R$,
there exists a set $\Omega$ of $i$-information sets 
such that $\{\R(\K): \K \in \Omega\}$ partitions $\R$.
By Lemma~\ref{lem:independence} 
and (\ref{ev_sec_eq}), 
it follows that $i^-$-choice sets are independent of 
the $i$-information sets in $\Omega$.  
Applying (\ref{ev_sec_eq})
again, it follows that $i^-$-choice sets are 
independent of {\em all}
$i$-information sets.  
Thus, $\mu(\R(\K_i(r,m)) \inter D_{y_i} \mid D_{i^-}) = 
\mu(\R(\K_i(r,m)) \mid D_{i^-}) = \mu(\R(\K_i(r,m)))$.
Since $D_{i^-}$ and $D_{y_i}$ are independent by assumption, it follows that
$\mu(D_{y_i} \mid D_{i^-}) = \mu(D_{y_i})$.  
Thus, (\ref{eq.5}) reduces to
$$\mu( \R(\K_i(r,m)) \mid D_{y_i} \inter D_{i^-}) \cdot \mu(D_{y_i}) =
\mu(\R(\K_i(r,m))).$$ 
The same is true for $D_{i^-}'$, so
because $\mu(D_{y_i}) > 0$ it follows that
$\mu( \R(\K_i(r,m)) \mid D_{y_i} \inter D_{i^-}) = \mu( \R(\K_i(r,m)) \mid D_{y_i} \inter D_{i^-}').$
(\ref{eqnoevidence}) is now immediate. 
\eprf

\commentout{

Note that because $\R(\K_i(r,m)) \subseteq D_p$, and by independence,
\begin{eqnarray*}
\mu( \R(\K_i(r,m)) \mid D_q ) &=& \mu( \R(\K_i(r,m)) \mid D_{(p,q)} ) \cdot \mu(D_p \mid D_q) \\
                            &=& \mu(  \R(\K_i(r,m)) \mid D_{(p,q)} ) \cdot \mu(D_p ).
\end{eqnarray*}
Multiplying both sides of (\ref{ev_sec_eq}) by $\mu(D_p)$, we 
get $\mu(\R(\K_i(r,m)) \mid D_q ) = \mu(\R(\K_i(r,m)) \mid D_{q'})$. By the symmetry of 
run-based secrecy,
$j$ maintains run-based probabilistic $f$-secrecy with respect to $i$.

For the converse, let $p \in \INIT_i$, $q,q' \in \INIT_j$, and $(r,m)$ be given. 
We want to show that
$$\mu_{(p,q)}(\R(\K_i(r,m)) \inter D_{(p,q)}) = \mu_{(p,q')}(\R(\K_i(r,m)) \inter D_{(p,q')}).$$
Without loss of generality, assume that $\R(\K_i(r,m)) \subseteq D_p$.
By symmetry, for any $\mu \in \M^{\INIT}(\Delta)$, we have 
$\mu( \R(\K_i(r,m)) \mid D_q) = \mu(\R(\K_i(r,m)) \mid D_{q'})$.
To prove that $i$ obtains no evidence for the initial choice 
it would be enough to show that there exists a 
measure $\mu \in \M^{\INIT}(\Delta)$
such that $\mu(\R(\K_i(r,m)) \mid D_{(p,q)}) = \mu(\R(\K_i(r,m)) \mid D_q)$ and 
$\mu(\R(\K_i(r,m)) \mid D_{(p,q')}) = \mu(\R(\K_i(r,m)) \mid D_q')$, but such a 
measure would violate the condition that $\mu(D_{(p,q)}) > 0$
for all $p \in \INIT_i, q \in \INIT_j$ because it would require $\mu$ to place all its weight
on $D_p$.

Therefore we shall show that for all $\epsilon > 0$, that there exists $\mu \in \M^{\INIT}(\Delta)$
such that
$$ \left| \mu_{(p,q)}(\R(\K_i(r,m)) \inter D_{(p,q)}) - \mu_{(p,q')}(\R(\K_i(r,m)) \inter D_{(p,q)}) \right| < \epsilon.$$
Let $\epsilon$ be given. Our strategy will be to pick a measure $\mu$
such that
$$ \left| \mu( \R(\K_i(r,m)) \mid D_{(p,q)}) -  \mu( \R(\K_i(r,m)) \mid D_q) \right| < \epsilon/2$$
and
$$ \left| \mu( \R(\K_i(r,m)) \mid D_{(p,q')}) -  \mu( \R(\K_i(r,m)) \mid D_{q'}) \right| < \epsilon/2,$$\
and the desired result follows by secrecy and the triangle inequality.

Givn $\delta > 0$, it will be useful if we can 
choose a measure $\mu_\delta$ such that $\mu_\delta(D_p) = 1 - \delta$,
and $\mu_\delta(D_q) = \mu_\delta(D_{q'}) = 1/4$. 
Our strategy for constructing $\mu$ will be to take a countable linear combination of the cell-specific
probability measures, such that the set of runs in each cell is assigned positive probability 
by $\mu$. Let $p_1, p_2, \ldots$ be
a countable enumeration of $\INIT_i - \{ p \}$, let $q_1, q_2, \ldots$ be
a countable enumeration of $\INIT_j - \{ q, q' \}$, 
and let $\mu_\delta$ be defined on $\R$ so that for $U \in \F$,
$\mu_\delta(U) = \sum_{(p^*,q^*) \in \INIT}
      g(p^*) \cdot h(q^*) \cdot \mu_{(p^*,q^*)}( U \inter D_{(p^*,q^*)} )$,
where 
\ul
\item $g(p^*) = (1-\delta)$ if $p^* = p$, and 
$g(p^*) = \delta/2^k$ if $p^* = p_k$; and
\item $h(q^*) = 1/4$ if $q^* = q$ or $q^* = q'$, and
$h(q^*) = 1/2^{k+1}$ if $q^* = q_k$.
\eul
It is straightforward to check that $\mu_\delta \in \M^{\INIT}(\Delta)$ and that it satisfies
the necessary properties mentioned above.

Now, because $\R(\K_i(r,m)) \subseteq D_p$, and by the construction of $\mu_\delta$,
we have 
$$\mu_\delta( \R(\K_i(r,m)) \mid D_q ) = \mu_\delta( \R(\K_i(r,m)) \mid D_{(p,q)} ) 
\cdot \mu_\delta( D_p ) = (1-\delta) \mu_\delta(\R(\K_i(r,m)) \mid D_{(p,q)})$$
and it follows that
$$ \left| \mu_\delta( \R(\K_i(r,m)) \mid D_{(p,q)}) -  \mu_\delta( \R(\K_i(r,m)) \mid D_q) \right| \leq \delta, $$
and similar reasoning shows that
$$ \left| \mu_\delta( \R(\K_i(r,m)) \mid D_{(p,q')}) -  \mu_\delta( \R(\K_i(r,m)) \mid D_q') \right| \leq \delta. $$
Because $\delta$ can be made arbitrarily small, this suffices to prove the result.
\commentout{
We have
\begin{eqnarray*}
 \mu(e \mid D_q) 
&=& \frac{\mu( e \cap D_q )}{\mu(D_q)} \\
&=& \frac{ \sum_{p' \in \INIT_i} \mu( e \cap D_{(p',q)}) }{ \sum_{p' \in \INIT_i} \mu( D_{(p',q)} ) } \\
&=& \frac{ \mu( e \cap D_{(p,q)}) + \sum_{p' \not= p} \mu( e \cap D_{(p',q)} ) }
         { \mu( D_{(p,q)} ) + \sum_{p' \not= p} \mu( D_{(p',q)} ) } \\
&=& \frac{ X + W }{ Y + Z }.
\end{eqnarray*}
Our goal is to show that 
$$ \left| \frac{ X + W }{ Y + Z } - \frac{X}{Y} \right| < \epsilon/2, $$
so we will show that
$$ \left| \frac{ X + W }{ Y + Z } - \frac{X}{Y+Z} \right| = \left| \frac{W}{Y+Z} \right| \leq  \left| \frac{Z}{Y+Z} \right|  < \epsilon/4, $$
and
$$ \left| \frac{ X }{ Y + Z } - \frac{X}{Y} \right| = \left|
\frac{XZ}{Y(Y+Z)} \right| < \epsilon/4, $$
and invoke the triangle inequality. Note that $Y = (1 - \delta)/4$ and $Z = \delta/4$, so 
$Z/(Y+Z) = \delta$ and $XZ/Y(Y+Z) = 4X \delta / (1-\delta)$. The first condition holds when 
$\delta < \epsilon/4$, and if $\epsilon < 4X$, 
the second condition holds when $\delta < \epsilon/(4X - \epsilon)$. The same results hold
if we replace $q$ with $q'$ throughout (replacing $X$ with $X' = \mu( e \mid D_{(p,q')})$), so
by choosing $\mu_{\delta}$ for an appropriately small value of $\delta$, the desired result
holds.
}
\eprf
}

\section{Generalizing from probability to plausibility}

In this section we give the details of the plausibilistic results
presented in Section \ref{sec:plaus}. 
All those results correspond to probabilistic results from the previous
section; in many 
cases the proofs are almost identical. For brevity we focus here on the
nontrivial subtleties 
that arise in the plausibilistic case.

To show that Proposition \ref{pro:sync_prob_sec} generalizes to run-based  
plausibility systems is straightforward.
We simply replace all occurrences of multiplication and
addition in the proof of Proposition~\ref{pro:sync_prob_sec}
with $\otimes$ and $\oplus$; all the resulting equations hold by the
properties of cacps's.

To define analogues of
Theorems \ref{thm:sync_prob_syntax} and \ref{thm:runbased_prob_syntax}, 
we need a language that allows statements of the form $\Pl_i(\phi) =
c$, where $c$ is a constant that is interpreted as a plausibility value.
Once we do this, the proofs of these results transfer to the
plausibilistic setting with almost no change.
We omit the straightforward details.
To prove Propositions~\ref{pro:sync_independence} and
\ref{pro:run_based_independence}, 
we first prove two results that generalize Lemma
\ref{lem:independence}. To do so, we 
need the following definition, taken from~\cite{Hal31}. 
Define a cacps to be {\em acceptable}
if $U \in \F'$ and $\Pl(V \mid U) \not= \bot$ implies that $V \inter U
\in \F'$. 
To understand the intuition behind this definition, consider the special
case where $U=W$.  Since $W \in \F'$ (this follows from the fact that
$\F'$ is a nonempty and is closed under supersets in $\F$), we get that 
if $\Pl(V) \ne \bot$, then $V \in \F'$. This is an analogue of the
situation in probability, where we can always condition on a set of
nonzero measure.
\lem\label{pro:plaus_independence}
Let $(W,\F,\F',\Pl)$ be an acceptable cacps. 
Suppose that $Y_1, Y_2, \ldots$ is a partition of $Y \in \F'$,
and that $Y_i \in \F$ for $i = 1, 2, 3, \ldots$.
For all $X \in \F$, the following are equivalent:
\ul
\item[(a)] $\Pl(X \mid Y_i) = \Pl(X \mid Y_j)$ for all $Y_i, Y_j \in \F'$.
\item[(b)] $\Pl(X \mid Y_i) = \Pl(X \mid Y)$ for all $Y_i \in \F'$.
\eul
\elem

\prf
Clearly (b) implies (a). To see that (a) implies (b), 
first note that since we are dealing with an acceptable cacps, if $Y_j
\notin \F'$, then $\Pl(Y_j \mid Y) = \bot$ and hence, for all $X$,
$\Pl(X \inter Y_j \mid Y) = \bot$.  Given $Y_i \in \F'$, 
it follows that 
\begin{eqnarray*}
\Pl(X \mid Y) &=& \oplus_{\{j: Y_j \in \F'\}} \Pl(X \inter
 Y_j \mid Y ) \\ 
 &=& \oplus_{\{j: Y_j \in \F'\}} ( \Pl(X \mid Y_{j}) \otimes \Pl(Y_{j}
 \mid Y) ) \\
 &=& \oplus_{\{j: Y_j \in \F'\}} ( \Pl(X \mid Y_{i}) \otimes \Pl(Y_{j} \mid Y) ) \\
 &=& \Pl( X \mid Y_i) \otimes (\oplus_{\{j: Y_j \in \F'\}} \Pl(Y_j \mid Y) )\\
 &=& \Pl( X \mid Y_i),
\end{eqnarray*}
as needed. 
\eprf
\commentout{
We have (b) implies (c) because 
if $Y_i \in \F'$ then, since $Y_i \inter Y = Y_i$, we have
$$ \Pl(X \inter Y_i \mid Y) = \Pl( X \mid Y_i \inter Y) \otimes \Pl( Y_i \mid Y )
  = \Pl(X \mid Y) \otimes \Pl(Y_i \mid Y). $$
For the converse, we have
\begin{eqnarray*}
\Pl( X \mid Y_i) \otimes \Pl(Y_i \mid Y) 
  &=& \Pl( X \mid Y_i \inter Y) \otimes \Pl(Y_i \mid Y) \\
  &=& \Pl( X \inter Y_i \mid Y) = \Pl( X \mid Y ) \otimes \Pl(Y_i \mid Y).
\end{eqnarray*}
Because $\Pl(Y_i \mid Y) \ne \bot$ we can cancel it from both sides of
the equation 
to derive $\Pl( X \mid Y_i ) = \Pl( X \mid Y)$, as needed.
}
In the probabilistic setting, if either part (a) or (b) of
Proposition~\ref{lem:independence} holds, we are able to conclude that
$Y_i$ is conditionally independent of $X$ given $Y$.  By the symmetry of
independence in the probabilistic setting, we can conclude that $X$ is
also conditionally independent of $Y_i$ given $Y$, that is, that
$\Pr(Y_i \mid X \inter Y) = \Pr(Y_i \mid Y)$.  
In the plausibilistic setting, independence is not 
symmetric in general
unless we make an additional 
assumption, namely that 
$\otimes$ is symmetric. 
We say that a cacps is {\em commutative\/} if its $\otimes$ operator is
commutative. 

\lem\label{cor:plaus_symmetry}
Suppose that $(W,\F,\F',\Pl)$ is a commutative acceptable cacps;
$Y_1, Y_2, \ldots$ is a partition of $Y \in \F'$; 
$X \in \F'$, $X \subseteq Y$, and $\Pl(X \mid Y) \not= \bot$; and 
for all $Y_i, Y_j \in \F'$, $\Pl(X \mid Y_i) = \Pl(X \mid Y_j)$.
Then, for all $Y_i \in \F$, $\Pl( Y_i \mid X ) = \Pl( Y_i \mid Y )$.
\elem

\prf
First, suppose that $Y_i \in \F'$.    By
Lemma~\ref{pro:plaus_independence}, we have that $\Pl(X \mid Y_i)
= \Pl(X \mid Y)$.  
Since $Y_i \inter Y = Y_i$ and $\otimes$ is commutative, we have
$$ \Pl(X \inter Y_i \mid Y) = \Pl( X \mid Y_i) \otimes \Pl( Y_i
\mid Y )   = \Pl(X \mid Y) \otimes \Pl(Y_i \mid Y) = 
\Pl(Y_i \mid Y) \otimes \Pl(X \mid Y).$$
Similarly, since $X \subseteq Y$, we have
$$ \Pl(X \inter Y_i \mid Y) = \Pl( Y_i \mid X) \otimes \Pl( X \mid Y).$$
Thus, $\Pl( Y_i \mid Y) \otimes \Pl( X \mid Y) =
\Pl( Y_i \mid X) \otimes \Pl( X \mid Y)$.
Since $\Pl(X \mid Y) \ne \bot$ by assumption, it follows from 
the definition of a cacps
that $\Pl(Y_i \mid Y) = \Pl(Y_i \mid X)$.  

If $Y_i \not\in \F'$ 
but $Y_i \in \F$, 
then $Y_i \inter X \notin \F'$ (since $\F'$ is closed under supersets in
$\F$).  Since we are working in an acceptable cacps, 
$\Pl(Y_i \mid Y) = \bot$ and $\Pl(Y_i \mid X) = \bot$,
so again $\Pl(Y_i \mid Y) = \Pl(Y_i \mid X)$.
\eprf
\commentout{
if $Y_i \in \F'$ then, since $Y_i \inter Y = Y_i$, we have
$$ \Pl(X \inter Y_i \mid Y) = \Pl( X \mid Y_i \inter Y) \otimes \Pl( Y_i \mid Y )
  = \Pl(X \mid Y) \otimes \Pl(Y_i \mid Y). $$

$$\Pl(X \inter Y_i \mid Y) = \Pl(X \mid Y) \otimes \Pl(Y_i \mid Y) = \Pl(Y_i \mid Y) \otimes \Pl(X \mid Y),$$
and so
From
$$\Pl(Y_i \mid X ) \otimes \Pl(X | Y) = \Pl(X \inter Y_i | Y) = \Pl(Y_i \mid Y) \otimes \Pl(X \mid Y).$$
Because $\Pl(X \mid Y) \not= \bot$ it follows that $\Pl( Y_i \mid X ) =
\Pl( Y_i | Y )$.

If $Y_i \not\in \F'$, then by acceptability $\Pl(Y_i \mid Y) = \bot$ and 
$\Pl(Y_i \mid X) = \Pl(Y_i \inter X \mid X) = \bot$, and equality follows.

}

With these results, plausibilistic versions of 
Propositions \ref{pro:sync_independence} and
\ref{pro:run_based_independence} can be proved 
with 
only
minor changes to the proof in the probabilistic case, provided
we make the additional assumptions stated in the main text.  We replace
the use of Lemma~\ref{lem:independence} by
Lemma~\ref{pro:plaus_independence}.  The appeal to the symmetry of
conditional independence is replaced by an appeal to
Lemma~\ref{cor:plaus_symmetry}.  However, to use this lemma, we need to
assume that $\otimes$ is commutative and that 
for all points $(r,m)$,
\begin{itemize}
\item $\Pl_{\mathit{cp}}(\K_i(r,m) \mid \PT(\R)) \ne \bot$ and 
$\Pl_{\mathit{cp}}(\K_j(r,m) \mid \PT(\R)) \ne \bot$ 
(in the proof of
total secrecy in the generalization of
Proposition~\ref{pro:sync_independence}); 
\item $\Pl_{\mathit{cp}}(\K_i(r,m) \mid \PT(m)) \ne \bot$ and 
$\Pl_{\mathit{cp}}(\K_j(r,m) \mid \PT(m)) \ne \bot$ (in the 
the proof of synchronous secrecy in the generalization of
Proposition~\ref{pro:sync_independence});  and 
\item $\Pl(\R(\K_i(r,m)) \mid \R) \ne \bot$ and $\Pl(\R(\K_j(r,m)) \mid \R)
\ne \bot$ (in the generalization of Proposition
\ref{pro:run_based_independence}).  
\end{itemize}
(We do not have to assume that the relevant cacps's are
acceptable for these propositions; it is enough that they are commutative.
We used acceptability in the proof of Lemma~\ref{pro:plaus_independence}
to show argue that if a set  $Y_i$ is not in $\F'$, then $\Pl(Y_i) =
\bot$.  Here, the sets $Y_i$ are of the form $\R(\K_i(r,m))$, and our
other assumptions guarantee that they are in $\F'$.)

Turning to the generalization of Theorem~\ref{thm:evidential}, the first
step is to define an {\em adversarial plausibility system}.  The definition
is completely analogous to 
that of that of an adversarial probability system,
except that now the set $\Delta$ consists of 
the 
acceptable conditional
plausibility spaces $(D,\F_D,\F_D',\Pl_D)$, for
each cell $D \in \D$.   
Again, we assume that $\R(\K_i(r,m)) \inter D \in \F_D$ and that, if
$\R(\K_i(r,m)) \inter D \ne \emptyset$, then 
$\R(\K_i(r,m)) \inter D \in \F_D'$ and 
$\Pl_D(\R(\K_i(r,m)) \inter D) \not= \bot$.
We say that an agent $i$ {\em obtains no plausibilistic evidence for the
initial choice} 
in $(\R,\D,\Delta)$ if for all $D,D' \in \D$ and all points $(r,m)$ such that
$\R(\K_i(r,m)) \inter D \not= \emptyset$ and $\R(\K_i(r,m)) \inter D' \not=
\emptyset$, 
we have 
$$\Pl_D(\R(\K_i(r,m)) \inter D) = \Pl_D(\R(\K_i(r,m)) \inter D').$$
Suppose that $\D$ is determined by $\INIT$ (as in the probabilistic case), 
and that the conditional 
plausibility spaces of $\Delta$ are all defined with respect to the
same domain $\mb{D}$ of plausibility values and with the same operations
$\oplus$ and $\otimes$,
where $\otimes$ is commutative.
Let $\F_\D$ be the $\sigma$-algebra generated by $\cup_{D \in \D} \F_D$.
Let $\M^{\INIT,\Pl}_i(\Delta)$ consist of all the 
acceptable
plausibility spaces $(\R,\F_\D,\F',\Pl)$ 
such that 
\ul
\item $\F'$ is a nonempty subset of $\F_\D$ that is closed under supersets;
\item if $A \in \F_D$ and $B \in \F' \inter \F_D'$ ,
then $\Pl(A \mid B) = \Pl_D(A \mid B)$;
\item for all agents $i$ and points $(r,m)$, there exists a cell $D$
such that $\Pl(D) \not= \bot$
and $\R(\K_i(r,m)) \inter D \ne \emptyset$; and
\item 
$\Pl(D_{(y_1, \ldots, y_n)}) = \Pl(D_{y_i}) \otimes
\Pl(\inter_{j \ne i} D_{y_j})$. 
\eul

\noindent We can now state and prove the plausibilistic analogue of 
Theorem~\ref{thm:evidential}.

\thm\label{thm:evidential1}
Let $(\R, \D, \Delta)$ be the adversarial plausibility system determined
by $\INIT$ and suppose that $\R$ 
is either synchronous or a system where $i$ has perfect recall.
Agent $i$ obtains no 
evidence for the initial choice in $(\R,\D,\Delta)$ iff agent $i^-$ maintains
generalized run-based plausibilistic $f_{i^-}$-secrecy with respect to $i$ 
in $(\R,\M^{\INIT,\Pl}_i(\Delta))$.
\ethm

\prf
The proof is basically the same as that of Theorem \ref{thm:evidential},
but some new subtleties arise because we are dealing with plausibility.
For the forward direction, we want to show that $i^-$ maintains
generalized run-based plausibilistic $f_{i^-}$-secrecy under the
assumption that $i$ obtains no evidence for the initial choice in
$(\R,\D,\Delta)$.  Much as in the proof of Theorem~\ref{thm:evidential},
we can show that
$\Pl(\R(\K_i(r,m)) \mid D_{y_i} \inter D_{i^-})  = 
\Pl(\R(\K_i(r,m)) \mid D_{y_i} \inter D_{i^-}')$ if 
$D_{i^-} \inter D_{y_i} \in \F'$ and $D_{i^-}' \inter D_{y_i} \in \F'$.
Continuing in the spirit of that proof, we now want to show that
$\Pl(D_{i^-} \mid \R(\K_i(r,m)) \inter D_{y_i}) = \Pl(D_{i^-} \mid
D_{y_i}) = \Pl(D_{i^-})$. 
For the second equality, note that,
by 
assumption, 
$\Pl(D_{i^-} \inter D_{y_i}) = \Pl(D_{i^-}) \otimes \Pl(D_{y_i})$.
Since the properties of acceptable conditional plausibility spaces
guarantee that $\Pl(D_{i^-} \mid D_{y_i}) \otimes \Pl(D_{y_i}) =
\Pl(D_{i^-} \inter D_{y_i})$, it follows that 
$\Pl(D_{i^-} \mid D_{y_i}) \otimes \Pl(D_{y_i}) = \Pl(D_{i^-}) \otimes
\Pl(D_{y_i})$. 
Since $\Pl(D_{y_i}) \ne \bot$, $\Pl(D_{i^-} \mid D_{y_i}) = \Pl(D_{i^-})$.

To prove the first equality, we want to apply
Lemma~\ref{cor:plaus_symmetry}.
To do so,  we must first show that
$\Pl( \R(\K_i(r,m)) \mid D_{y_i} ) \not= \bot$.
To see that this holds, recall that by assumption there exists a cell
$D$ such that $\Pl(D) \not= \bot$, $\Pl_{D}(\R(\K_i(r,m)) \inter D) \not= \bot$, and
$\R(\K_i(r,m)) \inter D \in \F'$.  
Since $\R(\K_i(r,m)) \inter D \ne
\emptyset$, we must have that $D \subseteq D_{y_i}$.  
Indeed, we must have $D = \widehat{D}_{i^-} \inter D_{y_i}$ for some
$i^-$-choice set $\widehat{D}_{i^-}$. 
Thus, 
we have
\begin{eqnarray*}
\Pl( \R(\K_i(r,m)) \mid D_{y_i} ) &\geq& \Pl( \R(\K_i(r,m)) \cap D \mid D_{y_i} ) \\
  &=& \Pl( \R(\K_i(r,m)) \mid D ) \otimes \Pl(\widehat{D}_{i^-} \mid
  D_{y_i}) \\  
 &=& \Pl_D( \R(\K_i(r,m)) \inter D) \otimes \Pl(\widehat{D}_{i^-})\\  
\end{eqnarray*}
By assumption $\Pl_D( \R(\K_i(r,m)) \inter D) \ne \bot$.
Since $\Pl(D) \ne \bot$ and $D \subseteq \widehat{D}_{i^-}$, it follows
that $\Pl(\widehat{D}_{i^-}) \ne \bot$.  Thus, 
$\Pl( \R(\K_i(r,m)) \mid D_{y_i} ) \ne \bot$.

For the converse, we must construct an
acceptable 
measure 
$\Pl$ and a set $\F'$ such that 
$(\R,\F_\D,\F',\Pl) \in \M^{\INIT,\mathrm{Pl}}_i(\Delta)$.
We take $\F'$ to consist of the sets $U$ such that $U \inter D \in
\F_D'$ for some cell $D$. For 
$\Pl$, we start
by taking some arbitrary total ordering $\prec$ of the cells in $\D$. 
Given $V \in \F$ and $U \in \F'$, let
$\Pl(V \mid U) = \Pl_D(V \cap D \mid U \cap D)$ where $D$ is the highest-ranked cell such that
$U \inter D \in \F_D$. By construction, $\Pl$ behaves identically to the cell-specific measures
when we condition on subsets of cells. It is easy to check that for all $y_i \in \INIT_i$ and
$i^{-}$-choice sets $D_{i^-}$,
we have $\Pl( D_{y_i} \cap D_{i^-} ) = \top$, $\Pl( D_{y_i} ) = \top$,
and $\Pl( D_{i^-} ) = \top$.  
The independence of the choices made by $i$ and $i^-$ follows immediately.

To see that the measure satisfies the conditioning axiom (in the definition of a cacps), 
suppose that $U_1, U_2, U_3 \in \F$ and 
$U_2 \inter U_3 \in \F'$. 
We must show that
$\Pl(U_1 \inter U_2 \mid U_3) = \Pl(U_1 \mid U_2 \inter U_3) \otimes \Pl(U_2 \mid U_3)$.
There are two cases. If the highest-ranked cell that intersects $U_3$ (call it $D$) also intersects $U_2$,
then all three terms in the equality are determined by $\Pl_D$, and the equality follows by applying
the conditioning axiom to $\Pl_D$ with $U_1 \inter D, U_2 \inter D,$ and $U_3 \inter D$.
If the highest-ranked cell $D$ that intersects $U_3$ does not intersect $U_2$, then the first and
third terms in the equality are both determined by $\Pl_D$ and must be $\bot$ because 
$U_2 \inter D = \emptyset$.

Finally, the measure $\Pl$ is acceptable (as required) because the
underlying cell-specific measures 
are acceptable.

The remainder of the proof is a relatively straightforward extension of
the probabilistic case. 
That $i^-$-choice sets are independent of $i$-information sets follows 
from Lemma~\ref{cor:plaus_symmetry},
using the facts that 
agent $i^-$ maintains generalized run-based 
plausibilistic $f_{i^-}$-secrecy, cells (and thus $i^-$-choice sets) 
have non-$\bot$ plausibility by construction,
and all information sets are in $\F'$.
\eprf

\commentout{
\othm{thm:plaus_evidential}
Let $(\R, \D, \Delta)$ be the adversarial plausibility system determined
by $\INIT$ and suppose that $\R$ 
is either synchronous or a system where $i$ has perfect recall.
Agent $i$ obtains no (plausibilistic)
evidence for initial choice in $(\R,\D,\Delta)$ iff agent 
$i^-$ maintains generalized run-based plausibilistic $f_{i^-}$-secrecy with
respect to $i$  in $(\R,\M^{\INIT}_i(\Delta))$.
\eothm

\prf
The proof has the same structure as the proof of Theorem~\ref{thm:evidential}.
The forward direction is identical, except that we replace probability measures
by conditional plausibility measures. 

For the converse, to show that secrecy
implies that $i$ obtains no evidence, 
let $p \in \INIT_i$, $q,q' \in \INIT_j$, and $(r,m)$ be given. 
We want to show that
$$\Pl_{(p,q)}(\R(\K_i(r,m)) \inter D_{(p,q)}) = \Pl_{(p,q')}(\R(\K_i(r,m)) \inter D_{(p,q')}).$$
Without loss of generality, assume that $\R(\K_i(r,m)) \subseteq D_p$.
By the symmetry of secrecy, for any $\Pl \in \M^{\INIT}(\Delta)$, we have 
$\Pl( \R(\K_i(r,m)) \mid D_q) = \Pl(\R(\K_i(r,m)) \mid D_{q'})$.
We will show that there exists a measure
$\Pl \in \M^{\INIT}(\Delta)$ such that $\Pl(D_p \mid \R) = \top$. Because
$\R(\K_i(r,m)) \subseteq D_p$, we have
$$ \Pl( \R(\K_i(r,m)) \mid D_q ) = \Pl( \R(\K_i(r,m)) \mid D_{(p,q)}) \cdot \Pl(D_p \mid \R) = 
   \Pl( \R(\K_i(r,m)) \mid D_{(p,q)},$$
and similarly,
$\Pl(\R(\K_i(r,m)) \mid D_{(p,q')}) = \Pl(\R(\K_i(r,m)) \mid D_{q'})$, and this suffices to complete the
proof.

To construct $\Pl$, we start by taking some arbitrary total ordering $\prec$ of the cells in $\D$ 
Given $V \in \F$ and $U \in \F'$, let
$\Pl(V \mid U) = \Pl_D(V \cap D \mid U \cap D)$ where $D$ is the highest-ranked cell such that
$U \inter D \not= \emptyset$. By construction, $\Pl$ behaves identically to the cell-specific measures
when we condition on subsets of cells. It is easy to check that for all $y_i \in \INIT_i$ and
$y_j \in \INIT_j$, we have $\Pl( D_{y_i} \cap D_{y_j} \mid \R ) = \top$, $\Pl( D_{y_i} \mid \R) = \top$,
and $\Pl( D_{y_j} \mid \R ) = \top$, and independence follows from this.

To see that the measure satisfies the conditioning axiom, suppose that
$U_1, U_2, U_3 \in \F$ and $U_1 \inter U_2 \in \F'$. We must show that
$\Pl(U_1 \inter U_2 \mid U_3) = \Pl(U_1 \mid U_2 \inter U_3) \otimes \Pl(U_2 \mid U_3)$.
There are two cases. If the highest-ranked cell that intersects $U_3$ (call it $D$) also intersects $U_2$,
then all three terms in the equality are determined by $\Pl_D$, and the equality follows by applying
the conditioning axiom to $\Pl_D$ with $U_1 \inter D, U_2 \inter D,$ and $U_3 \inter D$.
If the highest-ranked cell $D$ that intersects $U_3$ does not intersect $U_2$, then the first and
third terms in the equality are both determined by $\Pl_D$ and must be $\bot$ because 
$U_2 \inter D = \emptyset$.

Thus $\Pl$ is a valid conditional plausibility measure that is in $\M^{\INIT}(\Delta)$, as required.
\commentout{
For the forward direction, 
suppose that $\Pl \in \M^{\INIT}(\Delta)$, $q,q' \in \INIT_j$, and $(r,m) \in \PT(\R)$.
Without loss of generality suppose that $r \in D_{(p,q^*)}$ for some $p \in \INIT_i, q^* \in \INIT_j$. 
Because $i$ obtains no evidence and because $\Pl \in \M^{\INIT}(\Delta)$, we have
\begin{equation}\label{plaus_ev_sec_eq}
\Pl( \R(\K_i(r,m)) \mid D_{(p,q)} ) = \Pl( \R(\K_i(r,m)) \mid D_{(p,q')} ).
\end{equation}
Note that because $\R(\K_i(r,m)) \subseteq D_p$, and by plausibilistic independence,
\begin{eqnarray*}
\Pl( \R(\K_i(r,m)) \mid D_q ) &=& \Pl( \R(\K_i(r,m)) \mid D_{(p,q)} ) \cdot \Pl(D_p \mid D_q) \\
                            &=& \Pl(  \R(\K_i(r,m)) \mid D_{(p,q)} ) \cdot \Pl(D_p \mid \R ).
\end{eqnarray*}
Multiplying both sides of \ref{plaus_ev_sec_eq} by $\Pl(D_p \mid \R)$, we 
get $\Pl(\R(\K_i(r,m)) \mid D_q ) = \Pl(\R(\K_i(r,m)) \mid D_{q'})$. By the symmetry of 
run-based secrecy, $j$ maintains run-based plausibilistic $f$-secrecy with respect to $i$.
}
\eprf
} %

\section{Proofs for Section~\ref{sec:rw}}

\opro{pro:sync-sep->sep}
A limit-closed synchronous trace system $\Sigma$ satisfies separability
(resp. generalized noninterference)
iff
$H$ maintains synchronous secrecy 
(resp., synchronous $f_{hi}$-secrecy)
with respect to $L$ in $\R(\Sigma)$.
\eopro

\prf 
We give the argument for separability here; the argument for
generalized noninterference is similar and left to the reader.
The forward direction follows from Proposition~\ref{pro:sep->sync-sep}.  
For the converse, 
suppose that $H$ maintains synchronous secrecy with respect
to $L$ in $\R(\Sigma)$. 
Given $\tau, \tau' \in \Sigma$, let $\tau''$ be the trace such that 
$\tau'' |_L = \tau |_L$ and $\tau'' |_H = \tau' |_H$.
We must show that $\tau'' \in \Sigma$.
\commentout{
We start by constructing a 
tree $T$ consisting of trace prefixes in $\Sigma$.
The root node of $T$ represents time 0; for each time $i>0$, the nodes
represent state tuples 
$$t=\langle l_i, h_i, l_o, h_o \rangle.$$
Finite paths in $T$ correspond to finite trace prefixes of traces from
$\Sigma$,
and a node or state tuple $t$ is added to a path of the tree exactly if
doing
so will result in a trace prefix of a trace from $\Sigma$.

We now define the following sequence of subtrees of $T$.
Let $T_0 = T$, and let $T_{k+1}$ be the tree that results
from pruning all the nodes at level $k$ of $T_k$ exactly if those nodes
are on a path from the root corresponding to a trace prefix
$\sigma_{k+1}$
where either $\mc{L}(\sigma_{k+1}) \not= \mc{L}(\tau_{k+1})$ or
$\mc{H}(\sigma_{k+1}) \not= \mc{H}(\tau'_{k+1})$. Define $T_{\infty}$ to
be
the limit of this sequence of trees, so that at every depth $k$ of
$T_{\infty}$,
nodes are pruned if they make the corresponding trace prefixes
inconsistent
with either $\mc{L}(\tau)$ or $\mc{H}(\tau')$. Up to each depth $k$,
$T_{\infty}$
is equal to $T_k$. Furthermore, $T_{\infty}$ must be infinite, or else
it would
have only finitely many nodes, and would have some maximum depth
$j$. This
would mean that $T_{\infty}$ was equal to $T_j$. But if $T_j$ is
bounded, it
means that at every node at level $j+1$ can be pruned, so there is no
trace
$\sigma \in \Sigma$ that satisfies $\mc{L}(\sigma_{k+1}) =
\mc{L}(\tau_{k+1})$ and
$\mc{H}(\sigma_{k+1}) = \mc{H}(\tau'_{k+1})$. But this is a
contradiction, because
it implies that $H$ does not maintain synchronous secrecy with respect
to $L$ 
in $\R(\Sigma)$.

Since $T_{\infty}$ is infinite, and since each node of $\T_{\infty}$ has
bounded
degree (where the bound is given by the number of input/output values),
there
must exist an infinite path $\tau''$ in $T_{\infty}$, by K\"{o}nig's
Lemma.
This trace has the property that for every time $k$, 
$\mc{L}(\tau''_{k+1}) = \mc{L}(\tau_{k+1})$ and
$\mc{H}(\tau''_{k+1}) = \mc{H}(\tau'_{k+1})$. Furthermore, by limit
closure
we know that $\tau'' \in \Sigma$ since each finite trace prefix of
$\tau''$
is a trace prefix of some trace in $\Sigma$.
}
Since $H$ maintains synchronous secrecy with respect to $L$ in
$\R(\Sigma)$, for all $m$, there exists a run $r^m \in \R(\Sigma)$ such
that $r^m_L(m) = r^\tau_L(m)$ and $r^m_H(m) = r^{\tau'}_H(m)$.  Thus,
for all $m$, there exists a trace $\tau^m \in \Sigma$ such that
$\tau^m_m |_L = \tau |_L$ and $\tau^m_m |_H = \tau' |_H$.  
It follows that $\tau''_m = \tau^m_m$ for all $m$. 
Since $\tau^m \in \Sigma$ for all $m$, it follows by limit closure that
$\tau'' \in \Sigma$, as desired.
\eprf

\opro{pro:zl->totsec}
If $\Sigma$ is an asynchronous trace system that satisfies asynchronous
separability (resp. asynchronous generalized noninterference), then 
$H$ maintains total secrecy (resp. total $f_{hi}$-secrecy) with
respect to $L$ in $\R(\Sigma)$.
\eopro

\prf
Suppose that $\Sigma$ satisfies asynchronous separability, and
let $(r,m)$ and $(r',m')$ be arbitrary points. By the
construction of $\R(\Sigma)$,
there exist traces $\tau, \tau' \in T$ such that
$r_L(m) = \tau |_L$ and $r_H(m) = \tau' |_H$. 
Let $\tau''$ be an interleaving of $\tau |_L$ and $\tau' |_H$.  Since
$\Sigma$ satisfies asynchronous separability, $\tau'' \in \Sigma$.
Let $T''$ be a run-like set of traces that contains $\tau''$. (Such
a set must exist because $\Sigma$ is closed under trace prefixes.)
By definition, $r^{T''} \in \R(\Sigma)$.
Taking $m$ to be the length of $\tau''$, it follows that 
$r''_L(m'') = r_L(m)$ and
$r''_H(m'') = r'_H(m')$. 
Thus, $H$ maintains total secrecy with respect to $L$.

The proof for asynchronous generalized noninterference (and total
$f_{hi}$-secrecy) is analogous,
and left to the reader.
\eprf

\opro{pro:zl_iff_totsec}
If $\Sigma$ is an asynchronous trace system that is closed under
interleavings, then $\Sigma$ satisfies asynchronous separability iff
$H$ maintains total secrecy with respect to $L$ in $\R(\Sigma)$.
\eopro

\prf
We have already established the forward direction. For the converse, suppose
that $H$ maintains total secrecy with respect to $L$ in $\R(\Sigma)$, and that
$\Sigma$ is closed under interleavings.
Given $\tau, \tau' \in \Sigma$, 
there exist points $(r,m)$
and $(r',m')$ 
in $\PT(\R(\Sigma))$ such that $r_L(m) = \tau |_L$ and $r'_H(m') = \tau' |_H$.
Since $H$ maintains total secrecy with respect to $L$ in $\R(\Sigma)$,
there exists a point $(r'',m'')$ such that $r''_L(m'') = r_L(m)$
and $r''_H(m'') = r'_H(m')$.
By the construction of $\R(\Sigma)$,
there exists a run-like set $T$ of traces such that $r'' = r^T$.  Taking
$\tau''$ to be the traces of length $m''$ in $T$, it follows that
$\tau'' |_L = \tau |_L$ and $\tau''|_H = \tau' |_H$. 
Because $\Sigma$ is closed under interleavings, $\tau'' \in \Sigma$ as
required. 
\eprf

\commentout{
\opro{pro:ready_traces}
If $p$ and $q$ are processes, then $\RT(p) = \RT(q)$ if and only if 
$\R_{RT}(p) = \R_{RT}(q)$.
\eopro

\prf
Suppose that $\RT(p) = \RT(q)$, and let $r^T \in \R_{RT}(p)$, where 
$T \in \RT(p)$ is a run-like set of ready traces. It is immediate that
$T$ is a run-like subset of $\RT(q)$, so
$r^T \in \R_{RT}(p)$. Thus 
$\R_{RT}(p) \subseteq \R_{RT}(q)$.
Similarly, $\R_{RT}(q) \subseteq \R_{RT}(p)$.

If $\R_{RT}(p) = \R_{RT}(q)$, let $\tau \in \RT(p)$. We have
$\tau \in T$ for some run-like set $T \subseteq \RT(p)$, so
we have $r^T(m) = \tau$ for some time $m$, where $r^T \in \R_{RT}(p)$.
Because $r^T \in \R_{RT}(q)$, $\tau \in \RT(p)$. Thus
$\RT(p) \subseteq \RT(q)$, and similarly $\RT(q) \subseteq \RT(p)$.
\eprf
}

\bibliographystyle{chicagor}
\bibliography{z,joe}

\end{document}